\documentclass[journal]{IEEEtran}
\usepackage{graphicx}
\usepackage{amsmath,amssymb,dsfont,bbm,epstopdf,pgfplots,mathtools,enumitem,mathrsfs, bbm, algpseudocode , graphicx, dblfloatfix, physics} 
\usepackage[normalem]{ulem}
\usepackage{color}
\usepackage{cite}
\usepackage{graphics}
\usepackage{tikz}
\usepackage{pgfplots}
\usepackage[T1]{fontenc}
\usepackage{subcaption}
\usetikzlibrary{shapes, arrows, decorations.markings, arrows.meta}
\usetikzlibrary{patterns} 
\allowdisplaybreaks
\graphicspath{{.}{./Figures/}} 
\allowdisplaybreaks[4] 

\newtheorem{lemma}{Lemma}

\newtheorem{theorem}{Theorem}
\newtheorem{proposition}{Proposition}
\newtheorem{remark}{Remark}
\newtheorem{definition}{Definition}

\renewcommand{\Pr}{{\mathbb{P}}}

\renewcommand{\i}{{\iota}}

\newcommand{\vect}[1]{\boldsymbol{#1}}

\definecolor{ForestGreen}{rgb}{0.0, 0.5, 0.0}

\newcommand{\hn}[1]{{\color{black}#1}}
\newcommand{\hnr}[1]{{\color{black}#1}}

\renewcommand{\P}{\mathsf{P}}

\renewcommand{\P}{\mathsf{P}}

\newcommand{\syuf}{\sigma_{y_{1,2}|x_{1,2}}^2}
\newcommand{\syf}{\sigma_{y_{1,1}}^2}
\newcommand{\sys}{\sigma_{y_{1,2}}^2}

\newcommand{\ba}{\beta_{1,2}}
\newcommand{\bb}{\beta_{2,1}}
\newcommand{\yff}{\vect Y_{1,1}}
\newcommand{\yfs}{\vect Y_{1,2}}
\newcommand{\ysf}{\vect Y_{2,1}}
\newcommand{\yss}{\vect Y_{2,2}}
\newcommand{\zff}{\vect Z_{1,1}}
\newcommand{\zfs}{\vect Z_{1,2}}

\newcommand{\xf}{\vect X_{1,1}}
\newcommand{\xs}{\vect X_{\text c}}
\newcommand{\xt}{\vect X_{2,2}}
\newcommand{\xfs}{\vect X_{1,2}}
\newcommand{\xsf}{\vect X_{2,1}}
\newcommand{\nf}{n_{1,1}}
\newcommand{\ns}{n_{1,2}}
\newcommand{\nt}{n_{2,2}}
\newcommand{\bef}{\beta_{1,1}}
\newcommand{\bes}{\beta_{\text c}}
\newcommand{\bet}{\beta_{2,2}}

\newcommand{\sigff}{\sigma^2_{1,1}}
\newcommand{\sigfs}{\sigma^2_{1,2}}
\newcommand{\sigsf}{\sigma^2_{2,1}}
\newcommand{\sigss}{\sigma^2_{2,2}}

\title{Broadcast Channels with Heterogeneous Arrival and Decoding Deadlines: Second-Order Achievability}

\author{\hspace{0cm}\IEEEauthorblockN{Homa Nikbakht, \IEEEmembership{Member, IEEE},  Malcolm Egan,  Jean-Marie Gorce, \IEEEmembership{Senior Member, IEEE},  \\ and   H. Vincent Poor, \IEEEmembership{Life Fellow,~IEEE }}

\thanks{H.~Nikbakht and H.~V.~Poor are with the Department of Electrical and Computer Engineering, Princeton University, NJ, USA (email: {homa, poor}@princeton.edu). M.~Egan and J-M.~Gorce are with the INRIA and University of Lyon, CITI, INSA
Lyon, 69100 Villeurbanne, France (e-mail:  jean-marie.gorce@insa-lyon.fr, malcolm.egan@inria.fr).
The work of H.~Nikbakht and H.~V.~Poor has been supported by a  NextG Innovation Award from Princeton University.  The work of M. Egan has been supported by the program "PEPR Networks of the Future" of France 2030. The work of JM Gorce has been supported by the French National Research Agency (ANR-22-PEFT- 0005) as part of France 2030}
}
\begin{document}

\maketitle

\begin{abstract}
A standard assumption in the design of ultra-reliable low-latency communication systems is that the duration between message arrivals is larger than the number of channel uses before the decoding deadline. Nevertheless, this assumption fails when messages arrive rapidly and reliability constraints require that the number of channel uses exceed the time between arrivals. In this paper, we consider a broadcast setting in which a transmitter wishes to send two different messages to two receivers over Gaussian channels. Messages have different arrival times and decoding deadlines such that their transmission windows overlap. For this setting, we propose a coding scheme that exploits Marton's coding strategy.  We derive rigorous bounds on the achievable rate regions.  Those bounds can be easily employed in point-to-point settings with one or multiple parallel channels.  In the point-to-point setting with one  or multiple parallel channels, the proposed achievability scheme \hnr{is consistent with the normal approximation}. In the broadcast setting, our scheme agrees with  Marton's strategy for sufficiently large numbers of channel uses and  shows significant performance improvements over standard approaches based on time sharing for transmission of short packets. 

\end{abstract}

\begin{IEEEkeywords}
Ultra-reliable and low-latency communications, broadcast channels, Marton's coding strategy, heterogeneous arrival and decoding deadlines.
\end{IEEEkeywords}

\section{Introduction}

Mobile wireless networks in 5G and in 6G proposals are \hn{increasingly} intended for use in latency-critical and high-reliability systems, notably in industrial control applications, autonomous vehicles and remote surgery \cite{Bennis2018, Wang2022, HomaITW2020, Yao2022, HomaEntropy2022, Khan2022}. In such ultra-reliable low-latency communications (URLLC), packets are typically short. As a consequence, data transmission cannot  be made reliable by increasing channel code blocklength arbitrarily. 

A key challenge is, therefore, to design coding schemes that support high reliability requirements under finite blocklength. In recent years, a number of channel coding schemes have been proposed to address such requirements including short LDPC and polar codes \cite{Liu2018, Sharma2019}. At the same time, new characterizations of fundamental tradeoffs among the size of the message set, the probability of error, and the length of the code have been obtained via achievability and converse bounds,  thereby building on the work of \cite{Yuri2012, Strassen, Hayashi2009, Mahmood2023,Zhang2023, Fujiwara2024}.

A standard assumption in the design of coding schemes, even for URLLC, is that consecutive messages have distinct arrival times and decoding deadlines. As such, there is no choice but to encode the messages independently. However, this assumption is violated when a message arrives before the decoding deadline of a prior message. For example, a sensor in an unstable control system may send rapid measurements in order to stabilize the system \cite{Lucas2022}.  In order to ensure reliability of the sensor observations, the channel uses allocated to each observation of the speed may partially overlap.

It is therefore desirable to consider joint encoding of multiple sensor observations, albeit with heterogeneous decoding deadlines. That is, if the channel uses for two separate observations overlap, it is not possible to wait until the entire transmission for both sensor observations is received before decoding. In this situation, code design must account for two issues:
\begin{enumerate}
	\item[(i)]  messages with close arrival times; and
	\item[(ii)] messages with heterogeneous decoding deadlines.
\end{enumerate}
In addition to heterogeneous arrival times and decoding deadlines, a transmitter may seek to send messages to distinct receivers. While this setting has clear relevance for URLLC applications, there are currently no known designs for appropriate coding schemes.


\subsection{Related Work}

\subsubsection{Finite Blocklength Regime}
To capture both  reliability and latency, investigation of coding bounds in the finite blocklength regime is a requirement that dates back to the work of Shannon, Gallager, and Berlekamp \cite{Shannon1967}. Most of the existing literature focuses on identifying the limits of communication between a single transmitter and a single receiver for a coding block of size $n$. Among the proposed schemes, the widely used one is the \emph{normal approximation} that for a given $n$ approximates the point-to-point transmission rate $R$ by \cite{Yuri2012, Tan2015}:  
\begin{equation} \label{eq:1}
R \approx C - \sqrt{\frac{V}{n}} \log (e) \mathbb Q^{-1} (\epsilon) + \frac{\log n}{2n}, 
\end{equation}
where $C$ is the channel capacity, $V$ is the channel dispersion coefficient, $\epsilon$ is the average error probability and $\mathbb Q^{-1}(\cdot)$ is the inverse of the Gaussian cumulative distribution function. However this approximation has been proved to be a valid $O(n^{-1})$ asymptotic approximation for converse and achievability bounds \cite{Erseghe2016}, but an $O(n^{-1})$ bound is not necessarily reliable for small values of $n$ corresponding to URLLC applications. Therefore, exact bounds are of considerable interest. 
\subsubsection{Heterogeneous Decoding Deadlines}

The problem of designing a code to handle heterogeneous  decoding deadlines was first considered in the context of static broadcasting \cite{Shulman2000}, where a single message is decoded at multiple receivers under different relative decoding delay constraints. The work in \cite{Shulman2000} was recently generalized to multi-source and multi-terminal networks by Langberg and Effros in \cite{Langberg2021}. In particular, the notion of a time-rate region was introduced, which accounted for different decoding delay constraints for each message at each receiver. 

The work in both \cite{Shulman2000} and  \cite{Langberg2021} focused on the asymptotic regime. In the finite blocklength regime, a coding scheme for the Gaussian broadcast channel with heterogeneous blocklength constraints  was introduced in \cite{ Lin2021}, which decodes the messages at time-instances that depend on the realizations of the random channel fading. By employing an early decoding scheme, the authors showed that significant improvements  over standard successive interference cancellation are possible.  In \cite{Mross2022}  achievable rates and  latency of the early-decoding scheme  in \cite{Lin2021} are improved by introducing \emph{concatenated shell codes}.

\subsubsection{Heterogeneous Arrival Times}

The work in \cite{Shulman2000, Langberg2021,  Lin2021, Mross2022} focuses on the case where both messages are available at the time of encoding. In our previous works \cite{HomaWCNC2022, HomaISIT2022}, we introduced a coding scheme for the Gaussian point-to-point channel that encodes the first message before the second message arrives. The scheme proposed in \cite{HomaWCNC2022} exploited power sharing for symbols between the arrival time of the second message and the decoding deadline of the first message. Under a Gaussian interference assumption, bounds on the error probabilities for each message were established based on the message set size and finite decoding deadline constraints. In \cite{HomaISIT2022},  a coding scheme was proposed that exploits dirty-paper coding (DPC) \cite{Costa1983, Scarlett2015, Caire2003, Nikbakht2023Globecom}.We further derived  rigorous  bounds  on the achievable error probabilities of the messages.

\subsubsection{Broadcast Channels in the Finite Blocklength Regime}

Finite blocklength analysis of broadcast channels (BCs) was studied in \cite{Tajer2021, Sheldon2021, Tuninetti2022, Sheldon2022, Qiu2023}. The second-order Gaussian BC setting was investigated in \cite{Sheldon2021} where the authors studied the concatenate-and-code protocol \cite{Tuninetti2018} in which the transmitter concatenates the users' message bits into a single data packet and each user decodes the entire packet to extract its own bits. This scheme was shown to outperform superposition coding and time division multiplexing (TDM) schemes. The work in \cite{Tuninetti2022} is an extension of \cite{Sheldon2021} to $K$-user BCs.  The work in \cite{Sheldon2022} considered a two-user static BC and showed that under per-user reliability constraint,  superposition coding combined with a rate splitting technique in which the message intended for the user with the lowest signal-to-noise ratio (SNR) (the cloud center message) is allocated to either users gives the largest second-order rate region. 

\subsection{Main Contributions}

While coding schemes for heterogeneous arrival and decoding deadlines have been developed for point-to-point channels, adapting these codes to broadcast channels remains an open problem. In this paper, we address this question by developing and analyzing a coding scheme tailored to broadcast channels with heterogeneous arrival times and decoding deadlines. 

The main contributions of this work are: 
\begin{itemize}
\item We introduce a coding scheme for two-user Gaussian BCs with heterogeneous arrival times and decoding deadlines in \cite{HomaWCNC2022}, which exploits  Marton's coding strategy \cite{Marton1979}. Accounting for finite decoding deadline constraints (corresponding to fixed blocklengths), we first derive rigorous bounds on the achievable transmission rate for each of the messages. This is achieved by combining techniques to analyze the Gel'fand-Pinsker channel in the finite blocklength regime  \cite{Scarlett2015} and multiple parallel channels  \cite{Erseghe2016}. 
\item With the  developed  bounds, we  obtain further rigorous bounds for point-to-point settings with one or multiple parallel channels.  In the point-to-point setting with one channel, we show that our achievability scheme \hnr{is consistent with the  normal approximation in \eqref{eq:1}  proposed by Polyanskiy, Poor and Verdú in \cite{Yuri2012}.} Our results confirm the same statement for the point-to-point setting with multiple parallel channels when comparing  our achievability bound with  the normal approximation proposed by Erseghe in \cite{Erseghe2016}. 
\item  In the  broadcast setup, we provide a second-order analysis for the achievable rate regions. We show that our scheme agrees  with Marton's bound in \cite{Marton1979} for a sufficiently large number of channel uses. 

\item Finally, we show that our scheme outperforms the time-sharing scheme that transmits each message independently but over fewer number of channel uses.  
\end{itemize}

\subsection{Organization}
The rest of this paper is organized as follows. We end this section with some remarks on notation. Sections~\ref{sec:setup} and \ref{sec:coding} describe the problem setup and the proposed coding scheme. Section~\ref{sec:main} and \ref{sec:discussions} present our main results and discussions on the related works.   Section~\ref{sec:conclusion} concludes the paper. Some technical proofs are referred to in appendices. 

\subsection{Notation}
 The set of all integers is denoted by  $\mathbb Z$, the set of positive integers by $\mathbb Z ^{+}$ and the set of real numbers by $\mathbb R$. For other sets we use calligraphic letters, e.g., $\mathcal{X}$.  Random variables are denoted by uppercase letters, e.g., $X$, and their realizations by lowercase letters, e.g., $x$. For vectors we use boldface notation, i.e., upper case boldface letters such as $\mathbf{X}$  for random vectors and lower case boldface letters such as $\mathbf{x}$ for deterministic vectors. Matrices are depicted with sans serif font, e.g., $\mathsf{H}$. We also write $X^n$ for the tuple of random variables $(X_1, \ldots, X_n)$ and  ${\mathbf X}^n$ for the tuple of random vectors   $( \mathbf{X}_1, \ldots,\mathbf{X}_n)$. 

\section{Problem setup} \label{sec:setup}
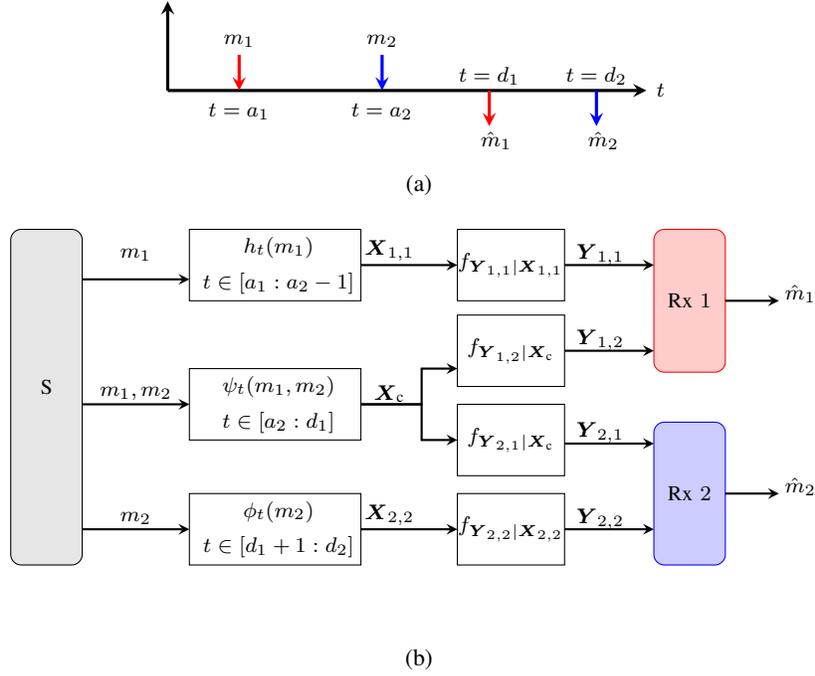
\begin{figure*}[t!]
  \centering
\begin{subfigure}{1 \textwidth}
\centering
\begin{tikzpicture}[scale=0.95, >=stealth]
\centering
\footnotesize
\draw[very thick, ->] (-1.5,-4.5)--(5.2,-4.5);
\draw[very thick, ->] (-1.5,-4.5)--(-1.5,-3.25);
\draw[very thick, ->, red] (-0.5,-4)--(-0.5,-4.5);
\draw[very thick, ->, red] (3,-4.5)--(3,-5);
\draw [ very thick, ->, blue] (1.5,-4)--(1.5,-4.5);
\draw [ very thick, ->, blue] (4.5,-4.5)--(4.5,-5);

\node[draw =none] at (-0.5,-3.8) {$m_1$};
\node[draw =none] at (-0.5,-4.8) {$t = a_1$};
\node[draw =none] at (1.5,-3.8) {$m_2$};
\node[draw =none] at (1.5,-4.8) {$t = a_2$};
\node[draw =none] at (3.1,-3.8-1.3-0.1) {$\hat m_1$};
\node[draw =none] at (3,-4.3) {$t = d_1$};
\node[draw =none] at (4.6,-3.8-1.3-0.1) {$\hat m_2$};
\node[draw =none] at (4.5,-4.3) {$t = d_2$};
\node[draw =none] at (5.4,-4.5) {$t$};
\end{tikzpicture}
\caption{}
\label{fig1a}
\end{subfigure}

  \begin{subfigure}{1 \textwidth}
\centering
\begin{tikzpicture}[scale=0.95, >=stealth]
\centering
\footnotesize
\tikzstyle{every node}=[draw,shape=circle, node distance=0.5cm];
\draw [ rounded corners, fill = gray!20](0,0) rectangle (1,4.5+0.2);
\node[draw =none] at (0.5,1.75+0.75) {S};
\draw (2.5,0) rectangle (4.9,1);
\draw (2.5,0+1.25+0.5) rectangle (4.9,1+1.25+0.5);
\draw (2.5,0+2.5+1+0.2) rectangle (4.9,1+2.5+1+0.2);
\draw[thick, ->](1,0.5)--(2.5,0.5);
\node[draw =none] at (1.75,0.65) {$m_2$};
\node[draw =none] at (3.75,0.75) {$\phi_t(m_2)$};
\node[draw =none] at (3.75,0.25) {$t \in [d_1+1:d_2]$};
\draw[thick, ->](1,0.5+1.25+0.5)--(2.5,0.5+1.25+0.5);
\node[draw =none] at (1.75,0.65+1.25+0.5) {$m_1, m_2$};
\node[draw =none] at (3.75,0.75+1.25+0.5) {$\psi_t(m_1,m_2)$};
\node[draw =none] at (3.75,0.25+1.25+0.5) {$t \in [a_2:d_1]$};
\draw[thick, ->](1,0.5+2.5+1)--(2.5,0.5+2.5+1);
\node[draw =none] at (1.75,0.65+2.5+1+0.2) {$m_1$};
\node[draw =none] at (3.75,0.75+2.5+1+0.2) {$h_t(m_1)$};
\node[draw =none] at (3.75,0.25+2.5+1+0.2) {$t \in [a_1:a_2-1]$};
\draw[thick, ->](4.9,0.5)--(5.75+0.5,0.5);
\draw (5.75+0.5,0) rectangle (7.25+0.5,1);
\node[draw =none] at (5.3,0.7) {$\xt$};
\node[draw =none] at (6.5+0.5,0.5) {$f_{\yss| \xt}$};
\draw[thick, ->](4.9,0.5+1.25+0.5)--(5.75,0.5+1.25+0.5)--(5.75, 1.75)--(5.75+0.5, 1.75);
\draw[thick, ->](4.9,0.5+1.25+0.5)--(5.75,0.5+1.25+0.5)--(5.75, 1.75+1)--(5.75+0.5, 1.75+1);
\draw (5.75+0.5,0+1.25) rectangle (7.25+0.5,1+1.25);
\node[draw =none] at (5.3,0.65+1.25+0.5) {$\xs$};
\node[draw =none] at (6.5+0.5,0.5+1.25) {$f_{\ysf| \xs}$};
\draw (5.75+0.5,0+2.5) rectangle (7.25+0.5,1+2.5);
\node[draw =none] at (6.5+0.5,0.5+2.5) {$f_{\yfs| \xs}$};
\draw[thick, ->](4.9,0.5+2.5+1+0.2)--(5.75+0.5,0.5+2.5+1+0.2);
\node[draw =none] at (5.3,0.7+2.5+1+0.2) {$\xf$};
\draw (5.75+0.5,0+2.5+1.2) rectangle (7.25+0.5,1+2.5+1.2);
\node[draw =none] at (6.5+0.5,0.5+2.5+1.2) {$f_{\yff| \xf}$};

\draw [ rounded corners, blue, fill = blue!20](8.25+0.75,0) rectangle (9.25+0.75,2);
\draw[thick, ->](7.25+0.5,0.5)--(8.25+0.75,0.5);
\node[draw =none] at (7.75+0.5,0.65) {$\yss$};
\node[draw =none] at (8.75+0.75,0.5+0.5) {Rx~$2$};
\draw[thick, ->](9.25+0.75,1)--(10+0.75,1);
\node[draw =none] at (10.25+0.8,0.5+0.6) {$\hat m_2$};
\draw[thick, ->](7.25+0.5,0.5+1.2)--(8.25+0.75,0.5+1.2);
\node[draw =none] at (7.75+0.5,0.65+1.2) {$\ysf$};

\draw[rounded corners, red, fill= red!20] (8.25+0.75,0+2.7) rectangle (9.25+0.75,2+2.7);
\draw[thick, ->](7.25+0.5,0.5+2.5)--(8.25+0.75,0.5+2.5);
\node[draw =none] at (7.75+0.5,0.65+2.5) {$\yfs$};
\node[draw =none] at (8.75+0.75,0.5+0.6+2.6) {Rx~$1$};
\draw[thick, ->](9.25+0.75,1+2.7)--(10+0.75,1+2.7);
\node[draw =none] at (10.25+0.8,0.5+0.6+2.7) {$\hat m_1$};
\draw[thick, ->](7.25+0.5,0.5+1.2+2.5)--(8.25+0.75,0.5+1.2+2.5);
\node[draw =none] at (7.75+0.5,0.65+1.2+2.5) {$\yff$};
\end{tikzpicture}
\caption{}
\label{fig1b}
\end{subfigure}
\vspace{-0.4cm}
  \caption{System model. (a) Arrival times and decoding deadlines of $m_1$ and $m_2$, (b)  Problem setup with one transmitter and two receivers.}
   \label{fig1} 
\vspace{-0.4cm}
  \end{figure*}
Consider a transmitter S that seeks to communicate with two receivers Rx~$1$ and Rx~$2$. It wishes to transmit message $m_1$ to receiver Rx~$1$ and message $m_2 $ to receiver Rx~$2$. At time $t = a_1$, transmission commences for the first message $m_1$. At time $t = a_2$, transmission commences for the second message $m_2 $. The two messages $m_1,m_2$ are assumed to be drawn independently and uniformly on $\{1,\ldots,M_1\}$ and $\{1, \ldots, M_2\}$, respectively. 

Each message is subject to different decoding delay constraints. \footnote{In this work we do not consider the encoding/decoding time delays, but rather the delay due to transmission, i.e., in terms of number of channel uses.} In particular, at time $d_1$, Rx~$1$ attempts to reconstruct the message $m_1$. Similarly, at time $d_2 > d_1$, Rx~$2$ attempts to reconstruct the message $m_2$. See Fig.~\ref{fig1a}.

Under the assumption that $a_1 < a_2$ and $a_2 <d_1< d_2$, the encoder outputs symbols at time $t \in \{a_1,\ldots,d_2\}$ as 
\begin{align}
X_t = \begin{cases} {h}_t ( m_1), & \quad t \in \{a_1, \ldots,a_2-1\}\\
\psi_t(m_1, m_2), & \quad t \in \{a_2, \ldots, d_1\} \\
\phi_t(\textcolor{black}{m_2}), & \quad t \in \{d_1 + 1, \ldots, d_2\},
\end{cases} 
\end{align}
where $h,\psi,\phi$ are the encoding functions corresponding to the channel uses where only message $m_1$ has arrived but not $m_2$, where both $m_1,m_2$ are present, and $m_1$ has been decoded at Rx~$1$. We highlight that $m_2$ is not known before time $t = a_2$; i.e., encoding is causal. Define 
\begin{equation}
n: = d_2 - a_1 + 1, \quad \nf := a_2 - a_1, \quad \ns := d_1 - a_2 + 1 \quad \text{and} \quad \nt := d_2 - d_1.
\end{equation}
We assume that the encoding functions satisfy an average block power constraint; namely,
\begin{equation} \label{eq:power}
\frac{1}{n} \sum_{i = a_1}^{d_2}X_i^2 \le \P
\end{equation}

\hn{To share the power among the codewords of the first $\nf$ channel uses, second $\ns$ channel uses and the last $\nt$ channel uses, we introduce the power sharing parameters $\bef, \bes, \bet \in [0,1]$ such that 
\begin{equation}
\nf \bef + \ns \bes + \nt \bet= n.
\end{equation}
}

Consequently
\begin{IEEEeqnarray}{rCl}
\frac{1}{\nf} \sum_{i = a_1}^{a_2-1}X_i^2 &\le& \bef \P, \\
\frac{1}{\ns} \sum_{i = a_2}^{d_1}X_i^2 &\le& \bes \P, \\
\frac{1}{\nt} \sum_{i = d_1+1}^{d_2}X_i^2 &\le& \bet \P.
\end{IEEEeqnarray}

Denote the channel inputs by 
\begin{align}
\xf &= \{X_{a_1},\ldots,X_{a_2-1}\},\notag\\
\xs &= \{X_{a_2},\ldots,X_{d_1}\}, \notag\\
\xt &= \{X_{d_1 +1},\ldots,X_{d_2}\},
\end{align}

and the  corresponding channel outputs at Rx~$1$ by $\yff$ and $\yfs$ and the channel outputs at Rx~$2$ by $\ysf$ and $\yss$.
The conditional distributions governing the four channels are then denoted by $f_{\vect{Y}_{1,1}|\xf}$, $f_{\vect{Y}_{1,2}|\xs}$, $f_{\vect{Y}_{2,1}|\xs}$ and $f_{\yss|\xt}$. We assume that each channel is additive, memoryless, stationary, and Gaussian; that is,
\begin{IEEEeqnarray}{rCl}
\vect Y_{1,1} &=& \xf + \vect{Z}_{1,1}, \\
\vect Y_{1,2} &=& \xs+ \vect{Z}_{1,2}, \\
\vect Y_{2,1} &=& \xs + \vect{Z}_{2,1}, \\
\vect Y_{2,2} &=& \xt + \vect{Z}_{2,2}, 
\end{IEEEeqnarray}
where $\vect{Z}_{i,j} \sim \mathcal{N}(\vect{0},\sigma_{i,j}^2\mathbf{I})$ for $i = 1,2$ and $j = 1,2$. 

This setup is illustrated in Fig.~\ref{fig1b}.

Receiver Rx~$1$ attempts to reconstruct message $m_1$ based on the channel outputs $\yff$ and $\yfs$ via the decoding function $g_1$; i.e.,
\begin{equation}
\hat{m}_1 = g_1(\yff,\yfs).
\end{equation}
Receiver Rx~$2$ attempts to reconstruct message $m_2$ based on the channel outputs $\ysf$ and $\yss$ via the decoding function $g_2$; i.e.,
\begin{equation}
\hat{m}_2 = g_2(\ysf,\yss).
\end{equation}
Observe that both receivers are causal. 

The average probability of error for each of the messages is then 
\begin{align}
\Pr[\hat{m}_1 \neq m_1],~~~\Pr[\hat{m}_2 \neq m_2].
\end{align}
The focus of the remainder of this paper is on characterizing the tradeoff among the size of the message sets $M_1$ and $M_2$, the error probabilities, and the decoding deadlines $d_1,d_2$. Formally, we study the achievable region defined as follows.

\begin{definition}
	Given the power constraint $\P$, a tuple $(a_1,a_2,d_1,d_2,M_1, M_2,\epsilon_1,\epsilon_2)$ is achievable if the messages $m_1$ and $m_2$ of cardinality $M_1$ and $M_2$, respectively, arriving at the $a_1$-th and $a_2$-th channel uses can be decoded by the $d_1$-th and $d_2$-th channel uses with an average probability of error satisfying 
	\begin{equation}
	\Pr[\hat{m}_1 \neq m_1] \le \epsilon_1, \quad  \Pr[\hat{m}_2 \neq m_2]\le \epsilon_2.
	\end{equation}
\end{definition}

\section{Random Coding Scheme}\label{sec:coding}

In this section, we introduce our random coding scheme. Notice that instead of employing Gaussian codebooks, our analysis relies on the use of power-shell codebooks. A power-shell codebook of length $n$ consists of codewords that are uniformly distributed on the centered $(n-1)$-dimensional sphere with radius $\sqrt{n\P}$ where $\P$ is the average input power constraint.  

\subsection{Encoding}

 The encoding process consists of three phases. 

\subsubsection{Transmitting only $m_1$}
In the first channel, consisting of $\nf$ channel uses, only $m_1$ is known to the encoder. The channel input $\xf$ corresponding to message $m_1$ is a codeword $\xf (m_1) \in \mathbb{R}^{\nf}$, which  is independently distributed on the sphere $\mathbb{S}^{\nf-1}$ with power $\nf \bef \P$.  That is, the probability density function of $\xf$ is given by
\begin{IEEEeqnarray}{rCl}\label{eq:fu}
	f_{\xf} (\vect x_{1,1}) = \frac{\delta \left (||\vect x_{1,1}||^2 - \nf \bef \P \right )}{S_{\nf} (\sqrt{\nf \bef \P })},
\end{IEEEeqnarray}
where $\delta(\cdot)$ is the Dirac delta function, and $S_{n}(r) $ is the surface area of a sphere of radius $r$ in $n$-dimensional space.

\subsubsection{Transmitting both $m_1$ and $m_2$}

Over the next $\ns$ channel uses, the encoder exploits Marton's coding strategy \cite{Marton1979} to jointly encode $m_1$ and $m_2$. To this end, we choose $\bb \in [0,1]$ and $\ba \in [0,1]$ such that for a given $\rho \in [0,1]$,
\begin{equation}
\ba + \bb + 2\rho \sqrt{\ba \bb} = \bes
\end{equation}
The parameter $\rho$ is defined shortly.

The following two codebooks are then generated.

\underline{\textit{Codebook Generation:}}
\begin{itemize}
\item Denote by $L_1$ the random coding parameter illustrating  the number of auxiliary codewords 
for each message $m_1 \in \{1, \ldots, M_1\}$.  A random codebook $C_{\xfs}$ containing $M_1 L_1$ auxiliary codewords $\{\xfs (m_1, \ell_1)\}$ with $\ell_1 \in \{ 1, \ldots, L_1\}$ and $m_1 \in \{1, \ldots, M_1\}$ is generated  where each codeword is independently distributed on the sphere $\mathbb{S}^{\ns-1}$ with power $\ns \ba \P$. 
\item Denote by $L_2$ the random coding parameter illustrating  the number of auxiliary codewords for each message $m_2 \in \{1, \ldots, M_2\}$.  A random codebook $C_{\xsf}$ containing $M_2 L_2$ auxiliary codewords $\{\xsf (m_2, \ell_2)\}$ with $\ell_2\in \{ 1, \ldots, L_2\}$ and $m_2 \in \{1, \ldots, M_2\}$ is generated where each codeword is independently distributed on the sphere $\mathbb{S}^{\ns -1}$ with power $\ns \bb  \P$. 
\end{itemize}
The codebooks are revealed to all terminals.

The transmitter chooses the pair ($\ell_1,\ell_2$) such that 
$\xfs (m_1, \ell_1) \in C_{\xfs}$ and  $\xsf (m_2, \ell_2) \in C_{\xsf}$ satisfy
\begin{equation} \label{eq:con}
\langle \xfs(m_1, \ell_1), \xsf (m_2, \ell_2) \rangle \in \mathcal D
\end{equation}
where
\begin{IEEEeqnarray}{rCl}\label{eq:D}
\mathcal D \triangleq \left [\ns  \sqrt{\ba \bb  }\P \rho :  \ns  \sqrt{\ba \bb  }\P  \right ]
\end{IEEEeqnarray}
 and $\rho$ is the correlation parameter.  If more that one such a pair exists, then one pair is selected arbitrarily.

The channel input over the $\ns$ symbols allocated for the joint transmission of $m_1$ and $m_2$ is given by 
\begin{align}
\xs =\alpha\left ( \xfs (m_1, \ell_1) + \xsf (m_2, \ell_2) \right),
\end{align}
where 
\begin{IEEEeqnarray}{rCl}\label{eq:alphan}
\alpha : = \sqrt \frac{\bes}{\bes^{\star}}.
\end{IEEEeqnarray}
with 
\begin{equation}
\bes^{\star}: = \ba + \bb +2 \rho^{\star} \sqrt{\ba \bb},
\end{equation}
and
$\rho^\star \in [\rho, \; \; 1]$ is the correlation coefficient  between the chosen codewords $\xfs(m_1, \ell_1)$ and $ \xsf (m_2, \ell_2)$. Notice that $\alpha$ is a power normalization coefficient that ensures that the transmit signal satisfies the power constraint  in \eqref{eq:power}.



We have the encoding error event:
\begin{IEEEeqnarray}{rCl} \label{eq:e12}
\mathcal E_{1,2} &\triangleq&  \{\text{no}\; (\ell_1, \ell_2) \; \text{exists such that \eqref{eq:con} is satisfied}\}.
\end{IEEEeqnarray} 
In our analysis, we set $\epsilon_{1,2} \in [0,1]$ as the  threshold on $\Pr[\mathcal E_{1,2}]$, i.e., given $\ns, \ba, \bb, \P$ and $\rho$ we choose $L_1$ and $L_2$  such that $\Pr[\mathcal E_{1,2}] \le \epsilon_{1,2}$. 
\subsubsection{Transmitting only $m_2$}
Over the last $\nt$ channel uses, the transmitter encodes only  $m_2$ with a codeword $\xt (m_2) $ which  is independently distributed on the sphere $\mathbb{S}^{\nt-1}$ with power $\nt \bet \P $.  

\subsection{Decoding}
Given the structure of the encoding functions, the output sequences at each receiver can be viewed as arising from two parallel channels: Rx~$1$ observes the two channel outputs $\yff$ and $\yfs$ of $\nf$ and $\ns$ blocks, respectively; and Rx~$2$ observes the two channel outputs $\ysf$ and $\yss$ of $\ns$ and $\nt$ blocks, respectively.

\subsubsection{Decoding $m_1$}
Given observations $\yff$ and $\yfs$, Rx~$1$  estimates $m_1$ according to the pair $(\hat m_1, \hat \ell_1)$, such that the corresponding sequences $\xfs (\hat m_1, \hat \ell_1)$ and $\xf (\hat m_1)$ maximize
\begin{IEEEeqnarray}{rCl}
\lefteqn{i_1\left ( \{\vect x_{1,j}\}_{j \in \{1,2\}}; \{\vect y_{1,j}\}_{j \in \{1,2\}}\right )} \notag \\
& :=& \log \prod_{j = 1,2} \frac{  f_{\vect Y_{1,j} | \vect X_{1,j}} (\vect y_{1,j} | \vect x_{1,j}) }{f_{\vect Y_{1,j}}(\vect y_{1,j})}
\end{IEEEeqnarray} 
over all pairs of $\vect x_{1,1}$ and $\vect x_{1,2} \in C_{\xfs}$. 
We have the following error event while decoding $m_1$: 
\begin{IEEEeqnarray}{rCl}\label{eq:e2}
\mathcal E_{1} &\triangleq&  \{ \text{Rx~$1$ chooses a message} \; \hat m_1 \neq m_1 \}.
\end{IEEEeqnarray}
Thus the average error of decoding $m_1$ is  bounded by 
\begin{IEEEeqnarray}{rCl}
\Pr[\hat{m}_1 \neq m_1] \le \Pr[\mathcal E_{1,2}] + \Pr [\mathcal E_{1}| \mathcal E_{1,2}^c].
\end{IEEEeqnarray}

\subsubsection{Decoding $m_2$}
 Given observations  $\ysf$ and $\yss$,  Rx~$2$ estimates $m_2$ according to the pair $(\hat m_2, \hat \ell_2)$, such that the corresponding sequences $\xsf (\hat m_2, \hat \ell_2)$ and $\xt (\hat m_2)$ maximize
\begin{IEEEeqnarray}{rCl}
\lefteqn{i_2\left ( \{\vect x_{2,j}\}_{j \in \{1,2\}}; \{\vect y_{2,j}\}_{j \in \{1,2\}}\right ) } \notag \\
&:=& \log \prod_{j = 1,2} \frac{  f_{\vect Y_{2,j} | \vect X_{2,j}} (\vect y_{2,j} | \vect x_{2,j}) }{f_{\vect Y_{2,j}}(\vect y_{2,j})}
\end{IEEEeqnarray}
over all pairs of $\vect x_{2,1} \in C_{\xsf}$ and $\vect x_{2,2}$.  
We have the following error event while decoding $m_2$: 
\begin{IEEEeqnarray}{rCl} \label{eq:e22}
\mathcal E_2 &\triangleq&  \{ \text{Rx~$2$ chooses a message} \; \hat m_2 \neq m_2 \}.
\end{IEEEeqnarray}
Thus the average error of decoding $m_2$ is bounded by 
\begin{IEEEeqnarray}{rCl}
\Pr[\hat{m}_2 \neq m_2]\le \Pr[\mathcal E_{1,2}] + \Pr [\mathcal E_{2}| \mathcal E_{1,2}^c]. 
\end{IEEEeqnarray}

\begin{remark} \label{rem1}
To improve the finite blocklength performance,  all the codewords are uniformly distributed on the \emph{power shell}. According to Shannon’s observation,  the optimal decay of the probability of error near capacity of the point-to-point Gaussian channel is achieved by codewords on the power-shell \cite{Shannon1959}. 
  As a result of this code construction,  the induced output distributions $f_{\vect Y_{i,j}}(\vect y_{i,j})$, with $i = 1,2$, $j = 1,2$, $f_{\yfs|\xfs}(\vect y_{1,2}| \vect x_{1,2})$ and $f_{\ysf|\xsf}(\vect y_{2,1}| \vect x_{2,1})$ are non-i.i.d.; thus we propose to bound  the corresponding information density measure $i_i\left ( \{\vect x_{i,j}\}_{j \in \{1,2\}}; \{\vect y_{i,j}\}_{j \in \{1,2\}}\right ) $, for each $i \in \{1,2\}$, by 
\begin{IEEEeqnarray}{rCl}
\lefteqn{\tilde i_i\left ( \{\vect x_{i,j}\}_{j \in \{1,2\}}; \{\vect y_{i,j}\}_{j \in \{1,2\}}\right )} \notag \\
& :=& \log  \frac{f_{\vect Y_{i,i} | \vect X_{i,i}} (\vect y_{i, i} | \vect x_{i, i})  Q_{\vect Y_{i,j\neq i} | \vect X_{i,j\neq i}} (\vect y_{i,j\neq i} | \vect x_{i,j \neq i}) }{\prod_{j =1}^2 Q_{\vect Y_{i,j}}(\vect y_{i,j})}, \IEEEeqnarraynumspace
\end{IEEEeqnarray} 
where the $Q$s are i.i.d Gaussian distributions. We then work with this modified information density throughout the analysis. 
\end{remark}
\section{Main Results} \label{sec:main}
Define $n_{2,1} = n_{1,2}$. For each $i \in \{1,2\}$ let $j \in \{1,2\} \backslash i$, and define
\begin{subequations} \label{eq:def-omega}
\begin{IEEEeqnarray}{rCl}
\Omega_{i,i}&:=& \frac{\beta_{i,i} \P}{\sigma_{i,i}^2}, \\ 
\Omega_{i,j} &:=& \frac{(\beta_{i,j} + \rho^ 2 \beta_{j,i} + 2 \rho \sqrt{\beta_{i,j} \beta_{j,i}}) \P }{\sigma_{i,j}^2 + (1 - \rho^2) \beta_{j,i} \P}, \\
J_i & :=&  \frac{4}{\sqrt{\pi}} \frac{\sqrt{\beta_{i,j} \beta_{j,i} (1+ 2 \Omega_{i,i})}}{(\beta_{i,j}+ \beta_{j,i}) (1+ \Omega_{i,i})} \prod_{\ell =1}^2 \left ( \frac{n_{i,\ell} - 2}{n_{i,\ell}}\right)^{\frac{n_{i,\ell}+1}{2}} \notag \\
&& \left ( \frac{n_{i,j} - 1}{n_{i,j}}\right)^{\frac{n_{i,j}-2}{2}}  \exp \left (-\frac{1}{6n_{i,i}}-\frac{1}{6n_{i,j}} + \frac{1}{2} \right). \IEEEeqnarraynumspace \label{eq:Ji}
\end{IEEEeqnarray}
\end{subequations}

Our first result gives upper bounds on the achievable rates of the first and the second message. 
\begin{theorem} \label{Th1:bounds}
Given $\nf, \ns, \nt$ and $\P$, for  each $i \in \{1,2\}$, let $j \in \{1,2\} \backslash i$,  $\log L_i := n_{i,j} R_{L_i}$ and    $ \log M_i := R_i \sum_{j =1}^ 2n_{i,j}$. For  each $i \in \{1,2\}$, we then have the following upper bound: 
\begin{IEEEeqnarray}{rCl}
\lefteqn{ \log M_i + \log L_i } \notag \\
&\le&  \sum_{j =1}^ 2n_{i,j} C\left (\Omega_{i,j}\right) -  \sqrt{ \sum_{j =1}^ 2n_{i,j} V_{i,j}\left (\Omega_{i,j}\right)}\mathbb Q^{-1} \left ( \epsilon_i - \Delta_i\right ) \notag \\
&& +K_i \log \left (\sum_{j =1}^ 2 n_{i,j} \right) \IEEEeqnarraynumspace  \label{eq:upp1}
\end{IEEEeqnarray}
subject to
\begin{IEEEeqnarray}{rCl}
\sum_{i = 1}^ 2 \log L_i &\ge& \log \frac{ \log \left( \epsilon_{1,2} \right )}{ \log \left(1  - (1- \rho^2)^{n_{1,2}-1}\right )},\label{eq:Ls}\IEEEeqnarraynumspace
\end{IEEEeqnarray}
where $C(x): = \frac{1}{2} \log (1 + x)$, $V_{i,i}(x) := \frac{x(2+x)}{2(1+x)^2} $, $V_{i,j}(x) = \frac{x(2+x)}{2(1+x)^2} + \tilde V_i$, with $\tilde V_i$ defined in  \eqref{eq:Vj},  $K_i$ is a constant, and 
\begin{IEEEeqnarray}{rCl}
 \Delta_{i}& : =& \frac{6 T_{\max,i} }{\sqrt{(\sum_{j =1}^ 2n_{i,j}V(\Omega_{i,j}))^3}} \notag \\
 && +  \left (1 - \left (1-\rho^2 \right )^{n_{i,j}-1} \right )^{L_1 \cdot L_2} \notag \\
 && + \frac{J_i 2^{\delta_i}}{2^{\delta_i}-1}\Bigg(\frac{\delta_i}{\sqrt{2 \pi \sum_{j = 1}^2 n_{i,j}V(\Omega_{i,j})}} \notag \\
 &&\hspace{0.5cm}+\frac{6 T_{\max,i} }{\sqrt{(\sum_{j =1}^ 2n_{i,j}V(\Omega_{i,j}))^3}} \Bigg) \left (\sum_{j =1}^ 2n_{i,j}\right) ^{K_i}, \IEEEeqnarraynumspace \label{eq:Deltai}
\end{IEEEeqnarray}
for any $\delta_i >0 $ and where $V(x) := \frac{x(2+x)}{2(1+x)^2} $. The parameter  $T_{\max,i}$ is defined in \eqref{eq:Tmax} with   $\Phi (\cdot, \cdot, \cdot)$ is the Hurwitz Lerch transcendent,  $\kappa_i>1$, $\zeta_i > 1$ and $\tilde \zeta_i >1$ are constants.  
\end{theorem}
\begin{IEEEproof}
See Appendix~\ref{sec:proofTh1}.
\end{IEEEproof}
\begin{figure*}[b!] 
\hrule
\begin{subequations}\label{eq:37}
\begin{IEEEeqnarray}{rCl}
T_{\max,i} & : = &2^3 \kappa_i + 2^4 \max \left \{ \frac{1}{\Gamma \left (\frac{n_{i,i}}{2} \right)} \zeta_i e^{-c_i}A(n_{i,i}, k_i, b_i, c_i), \frac{1}{\Gamma \left (\frac{n_{i,j}}{2} \right)}\tilde \zeta_i e^{-\tilde c_i} A(n_{i,j}, \tilde k_i, \tilde b_i,\tilde c_i)\right\},   \IEEEeqnarraynumspace \label{eq:Tmax}\\ 
A (n, k, b, c) &:=& \left ( (\tilde \kappa + 1)^3 - \frac{k^3 b^6}{8} - \frac{3}{2} (\tilde \kappa + 1) k b^2  (\tilde \kappa + 1- \frac{1}{2} k b^2 ) \right ) \Phi (e^{-1} , -\frac{n}{2} + 1, c) \notag \\
&-&  3k\left (5k^2b^4 + 4(\tilde \kappa + 1) ((\tilde \kappa + 1) - 3kb^2) \right) \Phi (e^{-1} , -\frac{n}{2}, c) - 96\sqrt{2}k^3b \Phi (e^{-1} , -\frac{n}{2} - \frac{3}{2}, c) \notag \\
&-&  3kb \left ( \frac{k^2b^4}{\sqrt{2}} + 2\sqrt{2} (\tilde \kappa + 1) b ((\tilde \kappa + 1) - b^2 k)\right )\Phi (e^{-1} , -\frac{n}{2} + \frac{1}{2}, c)  - 64k^3 \Phi (e^{-1} , -\frac{n}{2} -2, c) \notag \\
&+&  8\sqrt{2} k^2 b (6(\tilde \kappa + 1) -5kb^2) \Phi (e^{-1} , -\frac{n}{2} - \frac{1}{2}, c) +24 k (2(\tilde \kappa + 1)^2 - 5k^2 b^2 )\Phi (e^{-1} , -\frac{n}{2} -1, c), \\
\tilde \kappa_i &: =& \frac{k_i b_i^2}{4} +\frac{\tilde k_i \tilde b_i^2}{4} -n_{i,i}C( \Omega_{i,i}) -n_{i,j}C(\Omega_{i,j})  - \frac{n_{i,i} \Omega_{i,i}}{2(1+ \Omega_{i,i})} - n_{i,j} (1- \sigma_{i,j}^2 \tilde k_i), \\
c_i  & := & \frac{1}{2\sigma_{i,i}^2} \left ( \sqrt{\frac{\tilde \kappa_i + \kappa_i}{2k_i}} - \frac{b_i}{2}\right )^2, \; \tilde c_i  :=  \frac{1}{2\sigma_{i,j}^2} \left ( \sqrt{\frac{\tilde \kappa_i + \kappa_i }{2\tilde k_i}} - \frac{\tilde b_i}{2}\right )^2, \; k_i:=\frac{2+\Omega_{i,i}}{2\sigma_{i,i}^2(1+\Omega_{i,i})} , \; b_i: = \frac{\sigma_{i,i} \sqrt{n_{i,i} \Omega_{i,i}}}{k_i  (1+\Omega_{i,i})}, \IEEEeqnarraynumspace \\
\tilde k_i &: =& \frac{(2+ \Omega_{i,j})}{2((1-\rho^2)\beta_{j,i} \P + \sigma_{i,j}^2)(1+ \Omega_{i,j})},  \quad \tilde b_i:=  \sqrt{n_{i,j} \beta_{j,i} \P} + \frac{2}{(2+\Omega_{i,j})} \left ( (1+\Omega_{i,j}) \rho \sqrt{\beta_{j,i}} + \sqrt{\beta_{i,j}}\right), \\
\tilde V_i&: =& \frac{\sigma_{i,j}^2 \P}{(\sigma_{i,j}^2 + \bes \P)^2 } \left (\frac{\P(1+ 2\beta_{j,i} \P )(\rho \sqrt{\beta_{j,i} }+ \sqrt{\beta_{i,j}})^4}{\left ((1-\rho^2) \beta_{j,i} \P + \sigma_{i,j}^2\right)^2} + 2 \beta_{i,j} \right) - \frac{1}{2}\left(1- \left(\frac{(1-\rho^2)\beta_{j,i} \P + \sigma_{i,j}^2}{\sigma_{i,j}^2 + \bes \P}\right)^2\right).\label{eq:Vj}
\end{IEEEeqnarray}
\end{subequations}
\end{figure*}

\hn{
\begin{remark}
Note that, for each $i \in \{1,2\}$,  $K_i$ and $\delta_i$ should be chosen such that the argument inside $\mathbb Q^{-1}(\cdot)$ stays positive. In our numerical analysis, given other parameters, we choose these parameters such that the effect of $\Delta_i$ is negligible. 
\end{remark}
}

\begin{proposition} \label{prop1}
For sufficiently large $\nf, \ns, \nt$, and for $\rho >0$,  we have  
\begin{IEEEeqnarray}{rCl}
\lefteqn{\log M_i + \log L_i } \notag \\ &\le&  \sum_{j =1}^ 2n_{i,j} C_{i,j}\left (\Omega_{i,j}\right) \notag \\
&&  -  \sqrt{ \sum_{j =1}^ 2n_{i,j} V_{i,j}\left (\Omega_{i,j}\right)}\mathbb Q^{-1} \left ( \epsilon_i \right )  + O\left (\log \left (\sum_{j =1}^ 2 n_{i,j} \right) \right )  \notag \\\IEEEeqnarraynumspace  \label{eq:upp1asym}
\end{IEEEeqnarray}
subject to
\begin{IEEEeqnarray}{rCl}
\sum_{i = 1}^ 2 \log  L_i &\ge&  \log (-\log  \epsilon_{1,2}) - \ns \log(1- \rho^2).\label{eq:Lsasym}\IEEEeqnarraynumspace
\end{IEEEeqnarray}

\end{proposition}
\begin{IEEEproof}
See Appendix~\ref{sec:prop1}. 
\end{IEEEproof}


\section{Discussion on the Main Results and Related Works}\label{sec:discussions}
In this section, we review related settings that can be covered by Theorem~\ref{Th1:bounds}. 

\begin{figure}[t]
\center
				\includegraphics[width=0.45\textwidth]{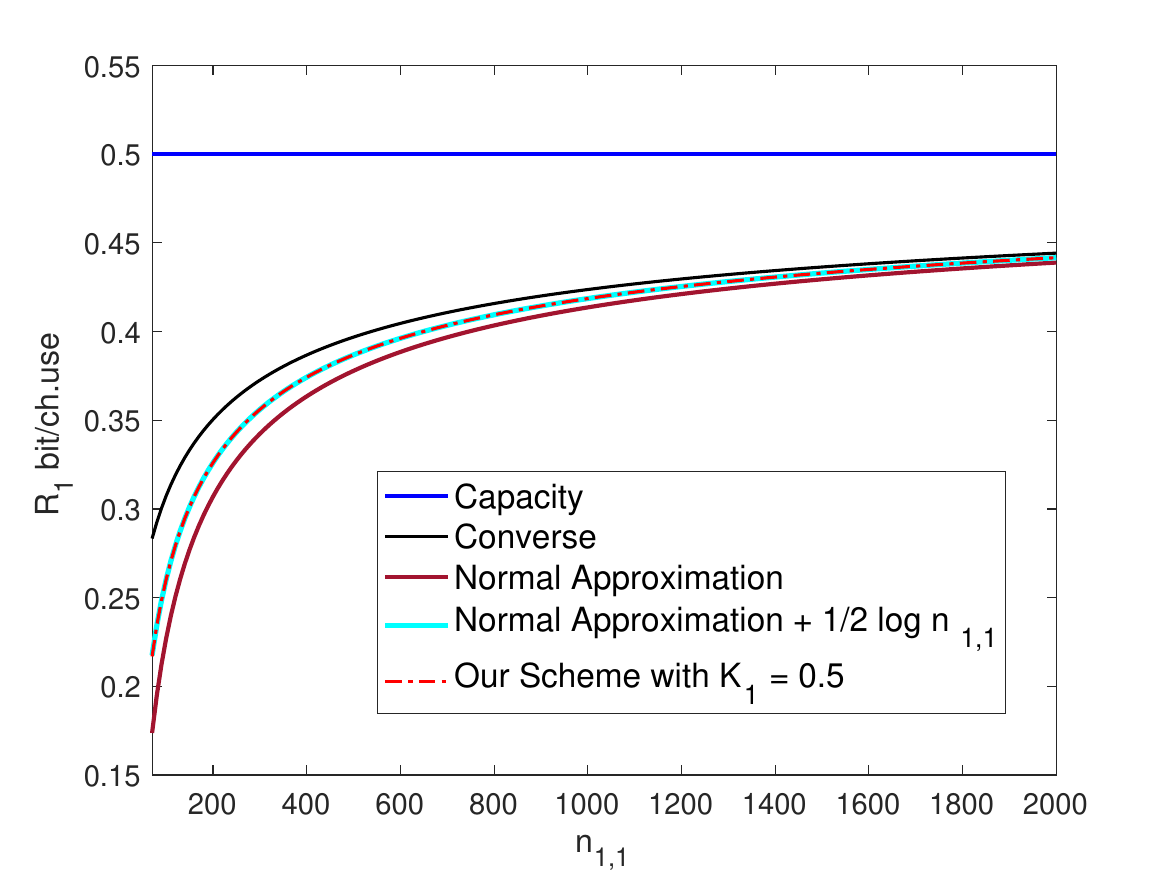}
\caption{\hnr{Converse and achievability bounds on $R_1$ versus $\nf$ in the  point-to-point case with $\Omega_{1,1} = 0$ dB,  and $\epsilon_1 = 10^{-3}$. The value of $\nf$ varies from $70$ to $2000$ with a step size of $10$. }}
\label{fig2new1}
			\end{figure}
\subsection{Point-to-Point Settings}
\subsubsection{Transmission of only  $m_1$ over $\nf$ channel uses}
 In this setting, we have only the following channel outputs: 
\begin{equation}
\yff = \xf + \vect Z_{1,1}
\end{equation}
with $||\xf||^ 2= \nf \P$ and $\vect Z_{1,1} \sim \mathcal N(0, \sigff I_{\nf})$. Set
\begin{equation}
 \Omega_{1,1} = \frac{\P}{\sigff}.
\end{equation}
\hnr{
By \cite[Appendix B]{Molavianjazi}:
\begin{IEEEeqnarray}{rCl}
J_1: =\left (\frac{\nf-2}{\nf}\right)^{\frac{\nf+1}{2}}\cdot  e^{-\frac{1}{6\nf}}\cdot \frac{2\sqrt{1 + 2 \Omega_{1,1}}}{\pi  (1 + \Omega_{1,1})} .
\end{IEEEeqnarray}

}

\begin{theorem}[P2P, Only $m_1$] \label{Th2:bounds}
Given $\nf$ and $\P$,   $ \log M_1:=  \nf R_1$ is upper bounded as
\begin{IEEEeqnarray}{rCl}
\log M_1  &\le&  \nf C\left (\Omega_{1,1}\right)\notag \\
&& -   \sqrt{ \nf V\left (\Omega_{1,1}\right)}\mathbb Q^{-1} \left ( \epsilon_1  -\Delta_1 \right)+ K_1 \log \left (\nf \right) , \IEEEeqnarraynumspace \label{eq:41}
\end{IEEEeqnarray}
with
\begin{IEEEeqnarray}{rCl}
\Delta_1&: =&\frac{6 T_{\max,1} }{\sqrt{(n_{1,1}V(\Omega_{1,1}))^3}} + \frac{J_12^{\delta_1}}{2^{\delta_1}-1}\Bigg(\frac{\delta_1}{\sqrt{2 \pi n_{1,1} V(\Omega_{1,1})}} \notag \\
&& +\frac{6 T_{\max,1} }{\sqrt{(n_{1,1}V(\Omega_{1,1}))^3}} \Bigg)  \left (n_{1,1}\right) ^{K_1}, \label{eq:43} \IEEEeqnarraynumspace
\end{IEEEeqnarray}
for  any $\delta_1$ and with $K_1$ a constant.

Note that $T_{\max,1}$ can be easily calculated from \eqref{eq:Tmax}  by setting  $ i = 1$ and $n_{i,j} = 0$. \end{theorem}

\begin{IEEEproof}
The proof of this result follows the proof of Theorem~\ref{Th1:bounds} by setting 
\begin{IEEEeqnarray}{rCl}
\ns = n_{2,1} =\nt = 0 \quad \text{and} \quad \bef = 1 .
\end{IEEEeqnarray}
\end{IEEEproof}

\hnr{
\begin{remark}
For sufficiently large $\nf$ and sufficiently small $\delta_1$, by setting  $K_1 = 0.5$ the bound in \eqref{eq:41} matches the normal approximation in \cite[Eq. 223]{Yuri2012} which is for  Gaussian channels. 
\end{remark}
}

We now make a more specific comparison with the work of Polyanskiy, Poor and Verdú in \cite{Yuri2012}. To this end,  note that, the authors in \cite{Yuri2012} show for both discrete memoryless channels and Gaussian channels, the normal approximation achieves

\begin{IEEEeqnarray}{rCl}
\lefteqn{\log M_1} \notag \\
& =& \nf C\left (\Omega_{1,1}\right) -   \sqrt{ \nf V\left (\Omega_{1,1}\right)}\mathbb Q^{-1} \left ( \epsilon_1  \right)+ O\left( \log \left (\nf \right) \right). \notag \\
\end{IEEEeqnarray}
See \cite[Eq.223]{Yuri2012}. The authors then show that for the Gaussian channels with equal-power and maximal-power constraints, the $O\left( \log \left (\nf \right) \right)$ can be bounded as
\begin{equation}\label{eq:47}
O\left( \log \left (\nf \right) \right) \le \frac{1}{2}\log \nf + O(1),
\end{equation}
and with the average power constraint as
\begin{equation}
O\left( \log \left (\nf \right) \right) \le \frac{3}{2} \log \nf + O(1).
\end{equation}
See \cite[Eq.294 and Eq.295]{Yuri2012}. The achievability of \eqref{eq:47}  for Gaussian channels is also shown in \cite{Tan2015}.

\hnr{
Our achievability bound in \eqref{eq:41} shows that 
\begin{equation}
O\left( \log \left (\nf \right) \right) =K_1 \log \nf,
\end{equation}
with $K_1 \le 0.5$. 
}

 \hnr{To compare our achievability bound in \eqref{eq:41} with the normal approximation,  Fig.~\ref{fig2new1} illustrates the bound in \eqref{eq:41} for $\epsilon_1 = 10^{-3}$, $\Omega_{1,1} = 0$dB, and  $K_1 = 0.5$ as well as the normal approximation with $O\left( \log \left (\nf \right)\right) $ equal to $0$ and $0.5 \log \nf$. The converse bound is the meta-converse bound proposed in \cite[Theorem 41]{Yuri2012}. As can be seen from this figure, our achievability scheme with $K_1 = 0.5$ matches the normal approximation with  $O\left( \log \left (\nf \right)\right)  = 0.5 \log \nf$.  }
%
%

\subsubsection{Transmission of only  $m_1$ over  two parallel channels of $\nf$ and $\ns$ channel uses}
In this setting, we have the following two channel outputs: 
\begin{equation}
\yff = \xf + \vect Z_{1,1}, \quad \yfs = \xfs + \vect Z_{1,2}, 
\end{equation}
with $||\xf||^ 2= \nf \bef \P$, $||\xfs||^ 2= \ns \ba \P$,  and $\vect Z_{1,1} \sim \mathcal N(0, \sigff I_{\nf})$ and $\vect Z_{1,2} \sim \mathcal N(0, \sigfs I_{\ns})$. The power sharing parameters $\bef \in [0,1]$ and $\ba \in [0,1]$ are chosen such that 
\begin{equation}
\bef \nf + \ba \ns = n.
\end{equation}
Set
\begin{equation}
 \Omega_{1,1} = \frac{\bef \P}{\sigff} \quad \text{and} \quad \Omega_{1,2} = \frac{\ba \P}{\sigfs}.
\end{equation}
\begin{theorem}[P2P, Parallel Channels, Only $m_1$] \label{Th3:bounds}
Given $\nf, \ns, \bef, \ba$ and $\P$,   $ \log M_1:= R_1 (\nf + \ns)$ is upper bounded as
\begin{IEEEeqnarray}{rCl}
\lefteqn{\log M_1} \notag \\ &\le& \sum_{j =1}^ 2n_{1,j} C\left (\Omega_{1,j}\right) -   \sqrt{ \sum_{j =1}^ 2n_{1,j} V\left (\Omega_{1,j}\right)}\mathbb Q^{-1} \left ( \epsilon_1 - \Delta_1\right )\notag \\
&& \hspace{0.5cm}+K_1 \log \left (\sum_{j =1}^ 2 n_{1,j} \right), \IEEEeqnarraynumspace
\label{eq:49}
\end{IEEEeqnarray}
where
\begin{IEEEeqnarray}{rCl}
\Delta_1&: =&\frac{6 T_{\max,1} }{\sqrt{(\sum_{j =1}^ 2n_{1,j}V(\Omega_{1,j}))^3}} \notag \\
&& + \frac{J_1 2^{\delta_1}}{2^{\delta_1}-1}\Bigg(\frac{\delta_1}{\sqrt{2 \pi \sum_{j = 1}^2 n_{1,j}V(\Omega_{1,j})}} \notag \\
&& \hspace{0.5cm}+\frac{6 T_{\max,1} }{\sqrt{(\sum_{j =1}^ 2n_{1,j}V(\Omega_{1,j}))^3}} \Bigg) \left (\sum_{j =1}^ 2n_{i,j}\right) ^{K_1} 
\end{IEEEeqnarray}
with 
\begin{IEEEeqnarray}{rCl}
J_1: =\prod_{j = 1}^2 \left (\frac{n_{1,j}-2}{n_{1,j}}\right)^{\frac{n_{1,j}+1}{2}}\cdot  e^{-\frac{1}{6n_{1,j}}}\cdot \frac{2\sqrt{1 + 2 \Omega_{1,j}}}{\pi (1 + \Omega_{1,j})} \IEEEeqnarraynumspace
\end{IEEEeqnarray}
and $K_1$ a constant.
Notice that $T_{\max,1}$  can be easily calculated from \eqref{eq:Tmax} by setting  $ i = 1$, $\bb = 0$, $\rho = 1$ and  $\bes = \ba$. 
\end{theorem}
\begin{IEEEproof}
The proof follows the proof of Theorem~\ref{Th1:bounds} by setting 
\begin{IEEEeqnarray}{rCl}
\nt = 0, \quad \bb = 0, \quad \text{and} \quad  \bet = 0.
\end{IEEEeqnarray}
\end{IEEEproof}

\hnr{
\begin{remark}
Let  $\ns = \nf$. For sufficiently large $\nf$ and $\ns$ and for sufficiently small $\delta_1$,  by setting  \hnr{$K_1 =0.5$} the bound in \eqref{eq:49} matches the normal approximation in \cite[Eq. 54]{Erseghe2016} that is for a two-parallel additive white Gaussian noise (AWGN) channels. This setting can be easily extended to a setting with more than two parallel channels. 
\end{remark}
}
\begin{figure}[t]
\centering
				\includegraphics[width=0.45\textwidth]{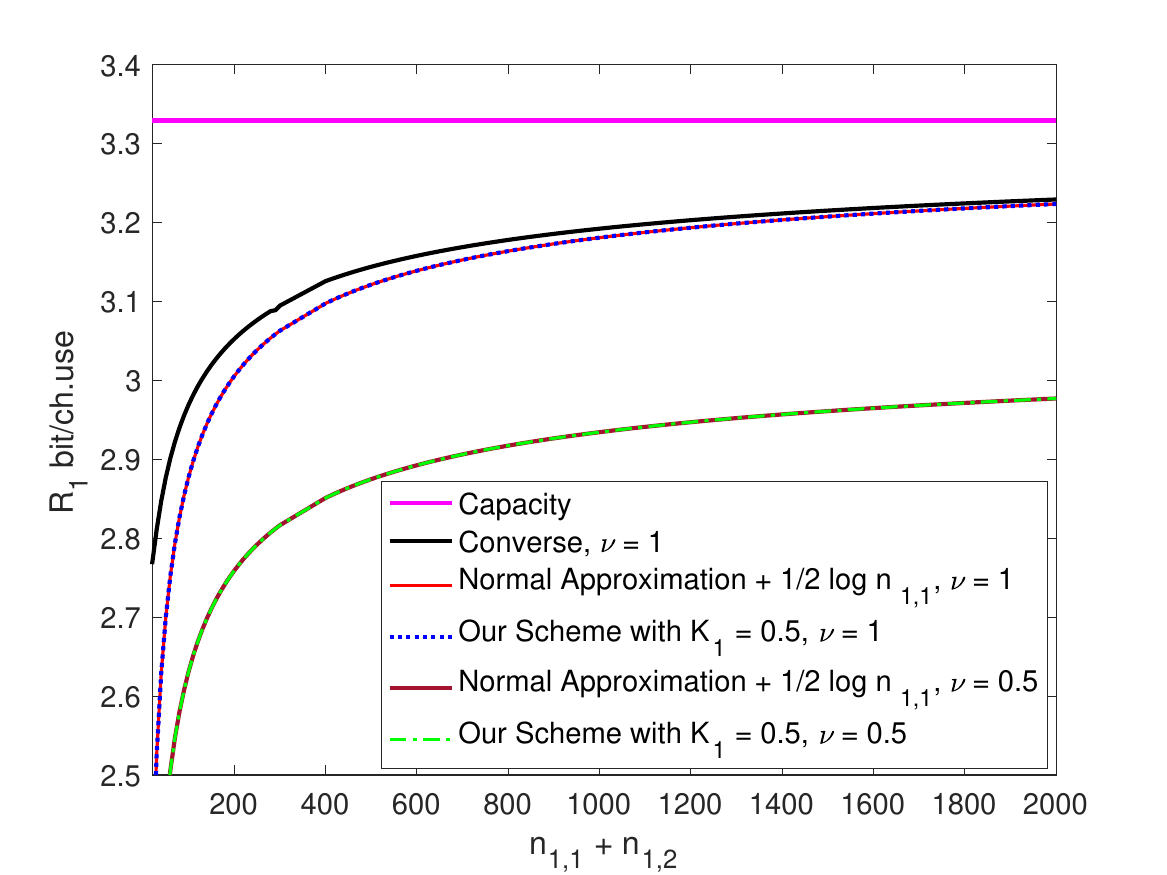}
\caption{\hnr{Converse and achievability bounds on $R_1$ versus $\nf + \ns$ in the  two parallel Gaussian channel case  with $\Omega_{1,1} = 20$ dB, $\Omega_{1,1} = \nu \Omega_{1,2}$ and $\epsilon_1 = 10^{-6}$. The value of $n = \nf + \ns$ varies from $20$ to $2000$ with a step size of $10$. }}
\label{fig3new1}
			\end{figure}
We now make a more detailed comparison with the work of Erseghe in \cite{Erseghe2016}. The leading idea of \cite{Erseghe2016} is based on the fact that a probability of the form $\Pr [\sum_{i = 1}^n u_i \ge n \tilde \lambda]$ where $u_i$s are i.i.d continuous  random variables can be written and numerically evaluated using standard Laplace transforms. The author further employs this idea to propose new asymptotic approximations for the meta-converse and random union coding (RCU) achievability bounds for the parallel AWGN channels. In  \cite[Theorem 10]{Erseghe2016} the author shows that in a $K$-parallel AWGN channels with each channel  of $\frac{n}{K}$ channel uses, the meta-converse bound, proposed by Polyanskiy, Poor and Verdú in \cite[Theorem 41]{Yuri2012}, is consistent with the normal approximation \eqref{eq:1} with 
\begin{IEEEeqnarray}{rCl}
V = \frac{1}{K} \sum_{k = 1}^K \frac{\Omega_k (2 + \Omega_k)}{2(1+ \Omega_k)^2} \quad \text{and} \quad C = \frac{1}{K} \sum_{k = 1}^K \frac{1}{2} \log (1 + \Omega_k), \notag \\
\end{IEEEeqnarray}
where $\Omega_k$ is the SNR of the $k$-th channel. We now  numerically compare our bound in \eqref{eq:49} with the normal approximation bound in \cite{Erseghe2016} for the case of $2$-parallel channels.  Since in \cite{Erseghe2016} all $K$ channels are of equal blocklength thus we assume $\nf = \ns$  and $\Omega_{1,1} = \nu \cdot \hn{\Omega_{1,2}}$ for some $\nu >0$. Fig.~\ref{fig3new1} illustrates this comparison for $\Omega_{1,1} = 20$ dB, $\nu = 0.5$ and $\nu = 1$, $\epsilon_1 = 10^{-6}$ and $K_1 = 0.5$.  \hnr{Our achievability bound in \eqref{eq:49} with $K_1 = 0.5$ is consistent with the normal approximation. }

\subsection{Broadcast Setting: Marton's Bounds} \label{sec:Marton}
In this section, we compare our results with Marton's  inner bound in \cite{Marton1979} proposed for a general two-receiver broadcast channel. In \cite{Marton1979}, the Tx wishes to transmit message $m_1 $ to Rx~$1$ and message $m_2$ to Rx~$2$ each over $n$ channel uses. The Tx thus encodes $m_1$ using the codeword $\vect U_1^n (m_1)$ and encodes $m_2$ using the codeword $\vect U_2^n (m_2)$ where $\vect U_1$ and $\vect U_2$ are distributed according to $f_{\vect U_1 \vect U_2} (\vect u_1, \vect u_2)$ and are correlated with a correlation parameter $ \rho$ such that 
\begin{equation}
\langle \vect U_1, \vect U_2 \rangle = n \rho \sqrt{\ba \bb} \P.
\end{equation}
With the codewords $\vect U_1$ and $\vect U_2$, the following codeword is formed:
\begin{equation}
\vect X = \vect U_1 + \vect U_2,
\end{equation}
 and  is transmitted to Rx~$1$ over the channel  $f_{\vect Y_1 | \vect X}$ and to Rx~$2$ over the channel $f_{\vect Y_2| \vect X}$. Here, $\vect Y_1$ and $\vect Y_2$ denote the output sequences of the first and second channels, respectively.

Let $R_1 : = \log M_1/n$ and $R_2 : = \log M_2/n$, then the following inner bounds hold for this setting: 
\begin{theorem} [Marton's bound] \label{th5}
The capacity region $\mathcal C$ of this setting is a set of rate pairs $(R_1, R_2)$ satisfying
\begin{subequations}
\begin{IEEEeqnarray}{rCl}
R_1 &\le& I(\vect U_1; \vect Y_1), \\
R_2 &\le& I(\vect U_2; \vect Y_2), \\
R_1+ R_2 &\le& I(\vect U_1; \vect Y_1) + I(\vect U_2; \vect Y_2) - I(\vect U_1; \vect U_2).
\end{IEEEeqnarray}
\end{subequations}
\end{theorem}
To compare this setup with ours, let $m_1$ and $m_2$ to be jointly sent over  the entire $n$ channel uses. Rx~$1$ observes the channel outputs  $\yff$ and Rx~$2$ observes the channel outputs  $\yss$ given by
\begin{IEEEeqnarray}{rCl}
\yff &=& \alpha(\xfs + \xsf) + \vect Z_{1,1}, \\
 \yss &=& \alpha(\xfs + \xsf )+ \vect Z_{2,2},
\end{IEEEeqnarray}
where $\xfs \sim \mathcal N(0, \ba \P I_n)$ and $\xsf \sim \mathcal N(0, \bb I_n)$ where $\ba$ and $\bb$ are chosen such that 
\begin{equation}
\ba + \bb + 2\rho \sqrt{\ba \bb} = 1.
\end{equation}
 For this Gaussian case, Theorem~\ref{th5} can be written as the  following proposition.   
\begin{proposition}\label{prop2}
Let $R_1 : = \frac{\log M_1}{n}$, $R_2 : = \frac{\log M_2}{n}$, $R_{L_1} := \frac{\log L_1}{n}$, $R_{L_2} := \frac{\log L_2}{n}$, $\xfs \sim \mathcal N(0, \ba \P \mathrm I_n)$ and $\xsf \sim \mathcal N(0, \bb \P \mathrm I_n)$,  we have the following inner bounds:
\begin{subequations} \label{eq:Marton}
\begin{IEEEeqnarray}{rCl}
R_1 + R_{L_1} &\le& \frac{1}{2} \log \left (\frac{\sigff + \P}{\sigff + (1- \rho^2) \bb \P}\right)
\end{IEEEeqnarray}
and
\begin{IEEEeqnarray}{rCl}
R_2 + R_{L_2} &\le& \frac{1}{2} \log \left (\frac{\sigss + \P}{\sigss + (1- \rho^2) \ba \P}\right)
\end{IEEEeqnarray}
\end{subequations}
subject to 
\begin{IEEEeqnarray}{rCl}  \label{eq:Marton2}
R_{L_1} + R_{L_2} &\ge& - \frac{1}{2}\log (1- \rho^2). 
\end{IEEEeqnarray}
\end{proposition}
\begin{remark}
By setting $\nf = \nt  = 0$, $\ns = n$ and 
\begin{IEEEeqnarray}{rCl}\label{eq:deltaMarton}
\epsilon_{1,2}&=& \frac{1}{2^{(1-\rho^2)^{\frac{\ns}{2}}}}.
\end{IEEEeqnarray} 
Proposition~\ref{prop1} matches Proposition~\ref{prop2} in the asymptotic regime (i.e., when $\ns \to \infty$) and when all the codewords are i.i.d Gaussian. 
\end{remark}

\begin{remark}
If all the codewords are i.i.d Gaussian, by setting either $L_1$ or $L_2$ to $0$,  in the asymptotic regime Proposition~\ref{prop1} matches the DPC results of Costa  \cite{Costa1983} and in the finite blocklength regime matches the results of Scarlett \cite[Theorem 2]{Scarlett2015}.
\end{remark}
\begin{figure}[t]
\begin{subfigure}{0.5\textwidth}
\centering
				\includegraphics[width=0.8\textwidth]{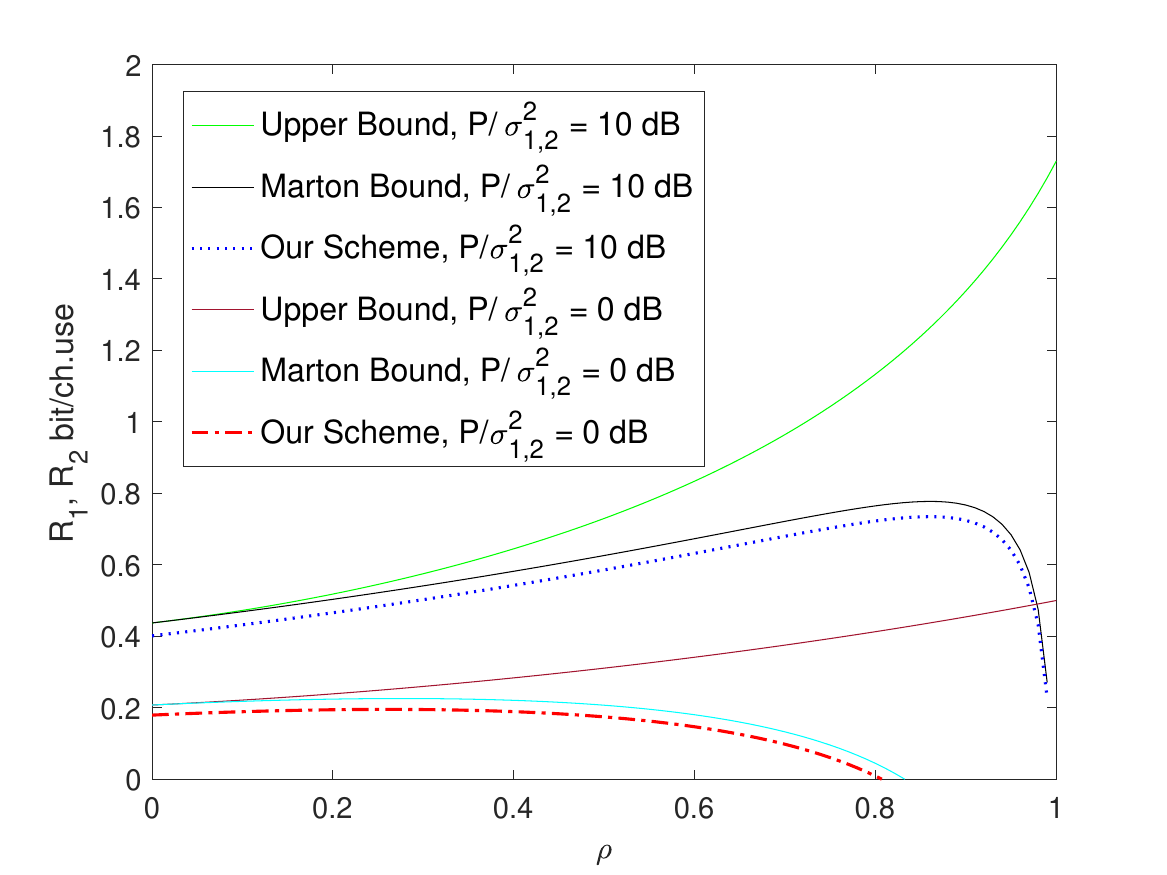}
\caption{$n = 5000$}
\label{fig4new1a}
\end{subfigure}
\begin{subfigure}{0.5\textwidth}
\centering
				\includegraphics[width=0.8\textwidth]{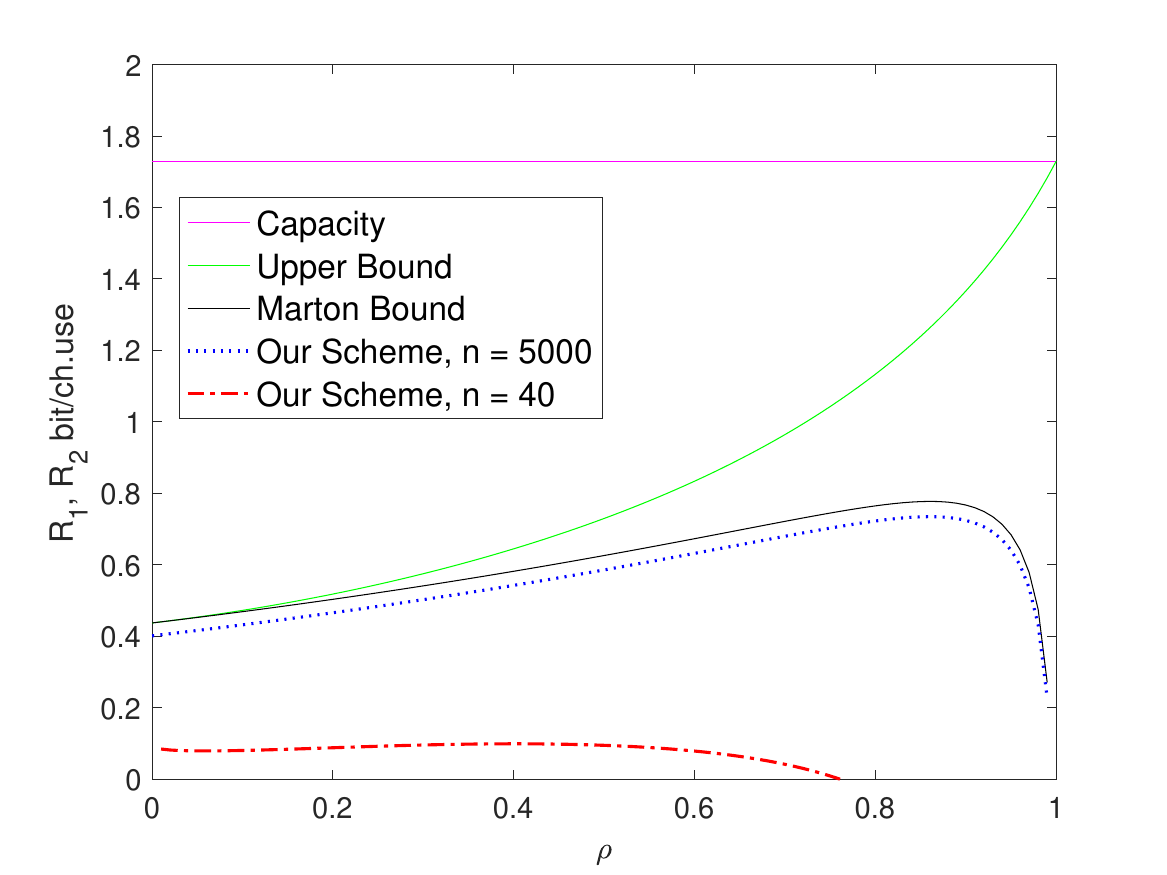}
\caption{$\P/\sigfs = 10$ dB }
\label{fig4new1b}
\end{subfigure}
\caption{\hnr{Effect of the correlation parameter $\rho$ on the bounds in Theorem~\ref{Th1:bounds} and Proposition~\ref{prop2}.}}
\label{fig4new1}
			\end{figure}

The correlation parameter $\rho$ is of a vital importance in both our scheme and Marton's scheme.  This is due to the fact that  increasing  $\rho$ increases the right-hand side of  \eqref{eq:upp1} and \eqref{eq:Marton}  as well as the right-hand side of  \eqref{eq:Ls} and \eqref{eq:Marton2} of Theorem~\ref{Th1:bounds} and Proposition~\ref{prop2}, respectively. More specifically, increasing $\rho$ increases the upper bound on $\log M_i + \log L_i$ as well as the lower bound on $\sum_{i =1}^2 \log L_i$. Hence, increasing $\rho$ will not always increase the upper bound on $R_i$.  Fig.~\ref{fig4new1} illustrates the effects of increasing $\rho$ on upper bounds on $R_1$ and $R_2$ in \eqref{eq:upp1} and Marton's bounds in Proposition~\ref{prop2}. To make a comparison with Marton's bounds, in this figure, we set $n = \ns$, $\nf = \nt =0$, $\epsilon_{1,2} = 10^{-3}$, $K_1 = K_2 = 0.5$, $\ba = \bb$, $L_1 = L_2$, $R_1 = R_2$, and $\sigfs = \sigsf$. Fig.~\ref{fig4new1a} is for the case where $n = 5000$ and $\P /\sigfs$ is equal to $10$ dB and $0$ dB. As can be seen from Fig.~\ref{fig4new1a},  when $\P /\sigfs$ is equal to $10$ dB, the transmission rate under our scheme and Marton's scheme maximizes around $\rho = 0.9$ and decays for the values of $\rho$ larger than $0.9$. Whereas, for $\P /\sigfs$ equal to $0$ dB, the transmission rate under our scheme and Marton's scheme maximizes at $\rho = 0$.  A similar trend can be seen for very small values of $n$. See Fig.~\ref{fig4new1b}. Notice that the upper bound is the case where $R_{L_1}$ ( or  $R_{L_2} $) is set at $0$ thus $R_1$ (or $R_2$) can be maximized.


\subsection{Time-sharing: Transmission of both $m_1$ and $m_2$} \label{sec:TS}
In this setting, we divide the $\ns$ channel uses into two parts $\eta \ns$ and $(1-\eta) \ns$ for $\eta \in [0,1]$. Therefore, $m_1$ is transmitted over $\nf + \eta \ns$ channel uses and $m_2$ is transmitted over $(1-\eta)\ns + \nt$ channel uses. Transmissions of $m_1$ and $m_2$ are thus independent. In this setting, we have the following two channel outputs: 
\begin{equation}
\yff = \xf + \vect Z_{1,1}, \quad \yss = \xt + \vect Z_{2,2}, 
\end{equation}
with $||\xf||^ 2= (\nf  + \eta \ns) \bef \P$, $||\xt||^ 2= ((1- \eta )\ns + \nt) \bet \P$,  and $\vect Z_{1,1} \sim \mathcal N(0, \sigff I_{\nf  + \eta \ns})$ and $\vect Z_{2,2} \sim \mathcal N(0, \sigss I_{(1- \eta )\ns + \nt})$. The power sharing parameters $\bef \in [0,1]$ and $\bet \in [0,1]$ are chosen such that 
\begin{equation}
(\nf  + \eta \ns) \bef + ((1- \eta )\ns + \nt)\bet  = n.
\end{equation}
Set
\begin{equation}
 \Omega_{1,1} = \frac{\bef \P}{\sigff} \quad \text{and} \quad \Omega_{2,2} = \frac{\bet \P}{\sigss}.
\end{equation}

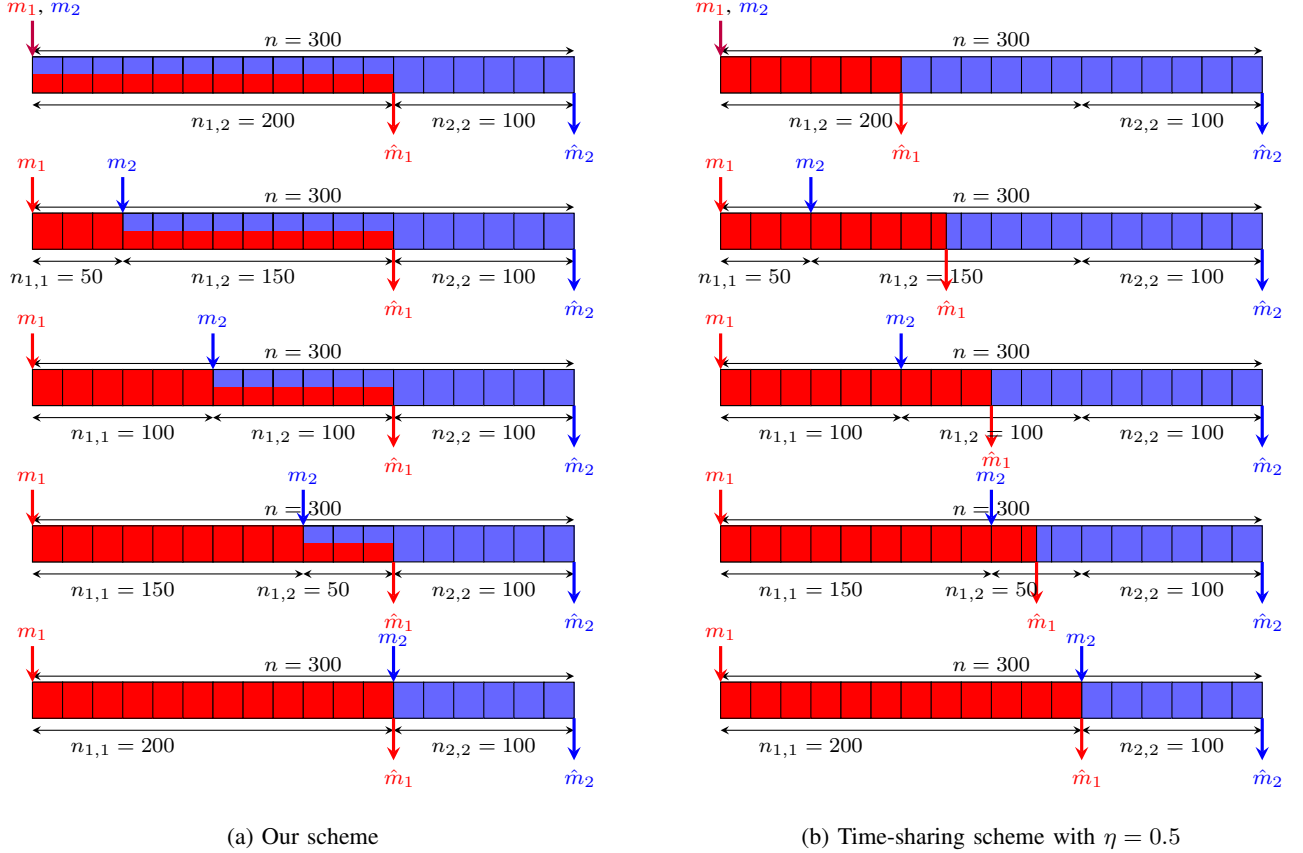
\begin{figure}[t!]
\begin{subfigure}{0.5\textwidth}
\centering
\begin{tikzpicture}[scale=1.4, >=stealth]
\centering
\tikzstyle{every node}=[draw,shape=circle, node distance=0cm];
\footnotesize
\foreach \i in {0,1,2,...,17}{
\draw [pattern=north west lines, pattern color=blue]  (0+\i*0.25,-0.1) rectangle (0.25+0.25*\i,0.2);
\draw  (0+\i*0.25,-0.1) rectangle (0.25+0.25*\i,0.2);
}
\foreach \i in {0,1,2,...,11}{
\draw [draw = none, fill = red]  (0+\i*0.25,-0.1) rectangle (0.25+0.25*\i,0.05);
\draw  (0+\i*0.25,-0.1) rectangle (0.25+0.25*\i,0.2);
}

\draw [<->] (0,0.25)--(4.5,0.25);
\draw [<->] (0,-0.2)--(3,-0.2);
\draw [<->] (3,-0.2)--(4.5,-0.2);
\draw [->, very thick, purple] (0,0.5)--(0,0.2);
\draw [->, very thick, red] (3,-0.1)--(3,-0.45);
\draw [->, very thick, blue] (4.5,-0.1)--(4.5,-0.45);
\node [draw = none] at (2.25,0.35) {$n = 300$};
\node [draw = none] at (1.75,-0.35) {$\ns = 200$};
\node [draw = none] at (3.75,-0.35) {$\nt = 100$};
\node [draw = none] at (0.1,0.6) {{\color{red}$m_1$}, {\color{blue}$m_2$}};
\node [draw = none] at (3.05,-0.6) {{\color{red}$\hat m_1$}};
\node [draw = none] at (4.55,-0.6) {{\color{blue}$\hat m_2$}};


\foreach \i in {0,1,2,...,17}{
\draw [pattern=north west lines, pattern color=blue]  (0+\i*0.25,-0.1-1.3) rectangle (0.25+0.25*\i,0.2-1.3);
\draw  (0+\i*0.25,-0.1-1.3) rectangle (0.25+0.25*\i,0.2-1.3);
}
\foreach \i in {3,...,11}{
\draw [draw = none, fill = red]  (0+\i*0.25,-0.1-1.3) rectangle (0.25+0.25*\i,0.05-1.3);
\draw  (0+\i*0.25,-0.1-1.3) rectangle (0.25+0.25*\i,0.2-1.3);
}
\foreach \i in {0,1,2}{
\draw [draw = none, fill = red]  (0+\i*0.25,-0.1-1.3) rectangle (0.25+0.25*\i,0.2-1.3);
\draw  (0+\i*0.25,-0.1-1.3) rectangle (0.25+0.25*\i,0.2-1.3);
}

\draw [<->] (0,0.25-1.3)--(4.5,0.25-1.3);
\draw [<->] (0,-0.2-1.3)--(0.75,-0.2-1.3);
\draw [<->] (0.75,-0.2-1.3)--(3,-0.2-1.3);
\draw [<->] (3,-0.2-1.3)--(4.5,-0.2-1.3);
\draw [->, very thick, red] (0,0.5-1.3)--(0,0.2-1.3);
\draw [->, very thick, blue] (0.75,0.5-1.3)--(0.75,0.2-1.3);
\draw [->, very thick, red] (3,-0.1-1.3)--(3,-0.45-1.3);
\draw [->, very thick, blue] (4.5,-0.1-1.3)--(4.5,-0.45-1.3);
\node [draw = none] at (2.25,0.35-1.3) {$n = 300$};
\node [draw = none] at (1.75,-0.35-1.3) {$\ns = 150$};
\node [draw = none] at (0.2,-0.35-1.3) {$\nf = 50$};
\node [draw = none] at (3.75,-0.35-1.3) {$\nt = 100$};
\node [draw = none] at (0,0.6-1.3) {{\color{red}$m_1$}};
\node [draw = none] at (0.75,0.6-1.3) {{\color{blue}$m_2$}};
\node [draw = none] at (3.05,-0.6-1.3) {{\color{red}$\hat m_1$}};
\node [draw = none] at (4.55,-0.6-1.3) {{\color{blue}$\hat m_2$}};
\foreach \i in {0,1,2,...,17}{
\draw [pattern=north west lines, pattern color=blue]  (0+\i*0.25,-0.1-1.3-1.3) rectangle (0.25+0.25*\i,0.2-1.3-1.3);
\draw  (0+\i*0.25,-0.1-1.3-1.3) rectangle (0.25+0.25*\i,0.2-1.3-1.3);
}
\foreach \i in {6,...,11}{
\draw [draw = none, fill = red]  (0+\i*0.25,-0.1-1.3-1.3) rectangle (0.25+0.25*\i,0.05-1.3-1.3);
\draw  (0+\i*0.25,-0.1-1.3-1.3) rectangle (0.25+0.25*\i,0.2-1.3-1.3);
}
\foreach \i in {0,...,5}{
\draw [draw = none, fill = red]  (0+\i*0.25,-0.1-1.3-1.3) rectangle (0.25+0.25*\i,0.2-1.3-1.3);
\draw  (0+\i*0.25,-0.1-1.3-1.3) rectangle (0.25+0.25*\i,0.2-1.3-1.3);
}

\draw [<->] (0,0.25-1.3-1.3)--(4.5,0.25-1.3-1.3);
\draw [<->] (0,-0.2-1.3-1.3)--(1.5,-0.2-1.3-1.3);
\draw [<->] (1.5,-0.2-1.3-1.3)--(3,-0.2-1.3-1.3);
\draw [<->] (3,-0.2-1.3-1.3)--(4.5,-0.2-1.3-1.3);
\draw [->, very thick, red] (0,0.5-1.3-1.3)--(0,0.2-1.3-1.3);
\draw [->, very thick, blue] (1.5,0.5-1.3-1.3)--(1.5,0.2-1.3-1.3);
\draw [->, very thick, red] (3,-0.1-1.3-1.3)--(3,-0.45-1.3-1.3);
\draw [->, very thick, blue] (4.5,-0.1-1.3-1.3)--(4.5,-0.45-1.3-1.3);
\node [draw = none] at (2.25,0.35-1.3-1.3) {$n = 300$};
\node [draw = none] at (2.25,-0.35-1.3-1.3) {$\ns = 100$};
\node [draw = none] at (0.75,-0.35-1.3-1.3) {$\nf = 100$};
\node [draw = none] at (3.75,-0.35-1.3-1.3) {$\nt = 100$};
\node [draw = none] at (0,0.6-1.3-1.3) {{\color{red}$m_1$}};
\node [draw = none] at (1.5,0.6-1.3-1.3) {{\color{blue}$m_2$}};
\node [draw = none] at (3.05,-0.6-1.3-1.3) {{\color{red}$\hat m_1$}};
\node [draw = none] at (4.55,-0.6-1.3-1.3) {{\color{blue}$\hat m_2$}};

\foreach \i in {0,1,2,...,17}{
\draw [pattern=north west lines, pattern color=blue]  (0+\i*0.25,-0.1-1.3-1.3-1.3) rectangle (0.25+0.25*\i,0.2-1.3-1.3-1.3);
\draw  (0+\i*0.25,-0.1-1.3-1.3-1.3) rectangle (0.25+0.25*\i,0.2-1.3-1.3-1.3);
}
\foreach \i in {9,...,11}{
\draw [draw = none, fill = red]  (0+\i*0.25,-0.1-1.3-1.3-1.3) rectangle (0.25+0.25*\i,0.05-1.3-1.3-1.3);
\draw  (0+\i*0.25,-0.1-1.3-1.3-1.3) rectangle (0.25+0.25*\i,0.2-1.3-1.3-1.3);
}
\foreach \i in {0,...,8}{
\draw [draw = none, fill = red]  (0+\i*0.25,-0.1-1.3-1.3-1.3) rectangle (0.25+0.25*\i,0.2-1.3-1.3-1.3);
\draw  (0+\i*0.25,-0.1-1.3-1.3-1.3) rectangle (0.25+0.25*\i,0.2-1.3-1.3-1.3);
}

\draw [<->] (0,0.25-1.3-1.3-1.3)--(4.5,0.25-1.3-1.3-1.3);
\draw [<->] (0,-0.2-1.3-1.3-1.3)--(2.25,-0.2-1.3-1.3-1.3);
\draw [<->] (2.25,-0.2-1.3-1.3-1.3)--(3,-0.2-1.3-1.3-1.3);
\draw [<->] (3,-0.2-1.3-1.3-1.3)--(4.5,-0.2-1.3-1.3-1.3);
\draw [->, very thick, red] (0,0.5-1.3-1.3-1.3)--(0,0.2-1.3-1.3-1.3);
\draw [->, very thick, blue] (2.25,0.5-1.3-1.3-1.3)--(2.25,0.2-1.3-1.3-1.3);
\draw [->, very thick, red] (3,-0.1-1.3-1.3-1.3)--(3,-0.45-1.3-1.3-1.3);
\draw [->, very thick, blue] (4.5,-0.1-1.3-1.3-1.3)--(4.5,-0.45-1.3-1.3-1.3);
\node [draw = none] at (2.25,0.35-1.3-1.3-1.3) {$n = 300$};
\node [draw = none] at (2.25,-0.35-1.3-1.3-1.3) {$\ns = 50$};
\node [draw = none] at (0.75,-0.35-1.3-1.3-1.3) {$\nf = 150$};
\node [draw = none] at (3.75,-0.35-1.3-1.3-1.3) {$\nt = 100$};
\node [draw = none] at (0,0.6-1.3-1.3-1.3) {{\color{red}$m_1$}};
\node [draw = none] at (2.25,0.6-1.3-1.3-1.3) {{\color{blue}$m_2$}};
\node [draw = none] at (3.05,-0.6-1.3-1.3-1.3) {{\color{red}$\hat m_1$}};
\node [draw = none] at (4.55,-0.6-1.3-1.3-1.3) {{\color{blue}$\hat m_2$}};

\foreach \i in {0,1,2,...,17}{
\draw [pattern=north west lines, pattern color=blue]  (0+\i*0.25,-0.1-1.3-1.3-1.3-1.3) rectangle (0.25+0.25*\i,0.2-1.3-1.3-1.3-1.3);
\draw  (0+\i*0.25,-0.1-1.3-1.3-1.3-1.3) rectangle (0.25+0.25*\i,0.2-1.3-1.3-1.3-1.3);
}

\foreach \i in {0,...,11}{
\draw [draw = none, fill = red]  (0+\i*0.25,-0.1-1.3-1.3-1.3-1.3) rectangle (0.25+0.25*\i,0.2-1.3-1.3-1.3-1.3);
\draw  (0+\i*0.25,-0.1-1.3-1.3-1.3-1.3) rectangle (0.25+0.25*\i,0.2-1.3-1.3-1.3-1.3);
}

\draw [<->] (0,0.25-1.3-1.3-1.3-1.3)--(4.5,0.25-1.3-1.3-1.3-1.3);
\draw [<->] (0,-0.2-1.3-1.3-1.3-1.3)--(3,-0.2-1.3-1.3-1.3-1.3);
\draw [<->] (3,-0.2-1.3-1.3-1.3-1.3)--(4.5,-0.2-1.3-1.3-1.3-1.3);
\draw [->, very thick, red] (0,0.5-1.3-1.3-1.3-1.3)--(0,0.2-1.3-1.3-1.3-1.3);
\draw [->, very thick, blue] (3,0.5-1.3-1.3-1.3-1.3)--(3,0.2-1.3-1.3-1.3-1.3);
\draw [->, very thick, red] (3,-0.1-1.3-1.3-1.3-1.3)--(3,-0.45-1.3-1.3-1.3-1.3);
\draw [->, very thick, blue] (4.5,-0.1-1.3-1.3-1.3-1.3)--(4.5,-0.45-1.3-1.3-1.3-1.3);
\node [draw = none] at (2.25,0.35-1.3-1.3-1.3-1.3) {$n = 300$};
\node [draw = none] at (0.75,-0.35-1.3-1.3-1.3-1.3) {$\nf = 200$};
\node [draw = none] at (3.75,-0.35-1.3-1.3-1.3-1.3) {$\nt = 100$};
\node [draw = none] at (0,0.6-1.3-1.3-1.3-1.3) {{\color{red}$m_1$}};
\node [draw = none] at (3,0.5-1.3-1.3-1.3-1.25) {{\color{blue}$m_2$}};
\node [draw = none] at (3.05,-0.6-1.3-1.3-1.3-1.3) {{\color{red}$\hat m_1$}};
\node [draw = none] at (4.55,-0.6-1.3-1.3-1.3-1.3) {{\color{blue}$\hat m_2$}};
\end{tikzpicture}
\caption{Our scheme}
\end{subfigure}
\begin{subfigure}{0.5\textwidth}
\centering
 \begin{tikzpicture}[scale=1.4, >=stealth]
\centering
\tikzstyle{every node}=[draw,shape=circle, node distance=0cm];
\footnotesize
\foreach \i in {0,1,2,...,17}{
\draw [pattern=north west lines, pattern color=blue]  (0+\i*0.25,-0.1) rectangle (0.25+0.25*\i,0.2);
\draw  (0+\i*0.25,-0.1) rectangle (0.25+0.25*\i,0.2);
}

\foreach \i in {0,...,5}{
\draw [draw = none, fill = red]  (0+\i*0.25,-0.1) rectangle (0.25+0.25*\i,0.2);
\draw  (0+\i*0.25,-0.1) rectangle (0.25+0.25*\i,0.2);
}

\draw [<->] (0,0.25)--(4.5,0.25);
\draw [<->] (0,-0.2)--(3,-0.2);
\draw [<->] (3,-0.2)--(4.5,-0.2);
\draw [->, very thick, purple] (0,0.5)--(0,0.2);
\draw [->, very thick, red] (3-1.5,-0.1)--(3-1.5,-0.45);
\draw [->, very thick, blue] (4.5,-0.1)--(4.5,-0.45);
\node [draw = none] at (2.25,0.35) {$n = 300$};
\node [draw = none] at (1,-0.35) {$\ns = 200$};
\node [draw = none] at (3.75,-0.35) {$\nt = 100$};
\node [draw = none] at (0.1,0.6) {{\color{red}$m_1$}, {\color{blue}$m_2$}};
\node [draw = none] at (3.05-1.5,-0.6) {{\color{red}$\hat m_1$}};
\node [draw = none] at (4.55,-0.6) {{\color{blue}$\hat m_2$}};


\foreach \i in {0,1,2,...,17}{
\draw [pattern=north west lines, pattern color=blue]  (0+\i*0.25,-0.1-1.3) rectangle (0.25+0.25*\i,0.2-1.3);
\draw  (0+\i*0.25,-0.1-1.3) rectangle (0.25+0.25*\i,0.2-1.3);
}

\foreach \i in {0,...,6}{
\draw [draw = none, fill = red]  (0+\i*0.25,-0.1-1.3) rectangle (0.25+0.25*\i,0.2-1.3);
\draw  (0+\i*0.25,-0.1-1.3) rectangle (0.25+0.25*\i,0.2-1.3);
}

\draw [draw = none, fill = red]  (0+7*0.25,-0.1-1.3) rectangle (0.5*0.25+0.25*7,0.2-1.3);
\draw  (0+7*0.25,-0.1-1.3) rectangle (0.5*0.25+0.25*7,0.2-1.3);

\draw [<->] (0,0.25-1.3)--(4.5,0.25-1.3);
\draw [<->] (0,-0.2-1.3)--(0.75,-0.2-1.3);
\draw [<->] (0.75,-0.2-1.3)--(3,-0.2-1.3);
\draw [<->] (3,-0.2-1.3)--(4.5,-0.2-1.3);
\draw [->, very thick, red] (0,0.5-1.3)--(0,0.2-1.3);
\draw [->, very thick, blue] (0.75,0.5-1.3)--(0.75,0.2-1.3);
\draw [->, very thick, red] (3-1.125,-0.1-1.3)--(3-1.125,-0.45-1.3);
\draw [->, very thick, blue] (4.5,-0.1-1.3)--(4.5,-0.45-1.3);
\node [draw = none] at (2.25,0.35-1.3) {$n = 300$};
\node [draw = none] at (1.75,-0.35-1.3) {$\ns = 150$};
\node [draw = none] at (0.2,-0.35-1.3) {$\nf = 50$};
\node [draw = none] at (3.75,-0.35-1.3) {$\nt = 100$};
\node [draw = none] at (0,0.6-1.3) {{\color{red}$m_1$}};
\node [draw = none] at (0.75,0.6-1.3) {{\color{blue}$m_2$}};
\node [draw = none] at (3.05-1.125,-0.6-1.3) {{\color{red}$\hat m_1$}};
\node [draw = none] at (4.55,-0.6-1.3) {{\color{blue}$\hat m_2$}};
\foreach \i in {0,1,2,...,17}{
\draw [pattern=north west lines, pattern color=blue]  (0+\i*0.25,-0.1-1.3-1.3) rectangle (0.25+0.25*\i,0.2-1.3-1.3);
\draw  (0+\i*0.25,-0.1-1.3-1.3) rectangle (0.25+0.25*\i,0.2-1.3-1.3);
}

\foreach \i in {0,...,8}{
\draw [draw = none, fill = red]  (0+\i*0.25,-0.1-1.3-1.3) rectangle (0.25+0.25*\i,0.2-1.3-1.3);
\draw  (0+\i*0.25,-0.1-1.3-1.3) rectangle (0.25+0.25*\i,0.2-1.3-1.3);
}

\draw [<->] (0,0.25-1.3-1.3)--(4.5,0.25-1.3-1.3);
\draw [<->] (0,-0.2-1.3-1.3)--(1.5,-0.2-1.3-1.3);
\draw [<->] (1.5,-0.2-1.3-1.3)--(3,-0.2-1.3-1.3);
\draw [<->] (3,-0.2-1.3-1.3)--(4.5,-0.2-1.3-1.3);
\draw [->, very thick, red] (0,0.5-1.3-1.3)--(0,0.2-1.3-1.3);
\draw [->, very thick, blue] (1.5,0.5-1.3-1.3)--(1.5,0.2-1.3-1.3);
\draw [->, very thick, red] (3-0.75,-0.1-1.3-1.3)--(3-0.75,-0.45-1.3-1.3);
\draw [->, very thick, blue] (4.5,-0.1-1.3-1.3)--(4.5,-0.45-1.3-1.3);
\node [draw = none] at (2.25,0.35-1.3-1.3) {$n = 300$};
\node [draw = none] at (2.25,-0.35-1.3-1.3) {$\ns = 100$};
\node [draw = none] at (0.75,-0.35-1.3-1.3) {$\nf = 100$};
\node [draw = none] at (3.75,-0.35-1.3-1.3) {$\nt = 100$};
\node [draw = none] at (0,0.6-1.3-1.3) {{\color{red}$m_1$}};
\node [draw = none] at (1.5,0.6-1.3-1.3) {{\color{blue}$m_2$}};
\node [draw = none] at (3.05-0.75,-0.6-1.3-1.25) {{\color{red}$\hat m_1$}};
\node [draw = none] at (4.55,-0.6-1.3-1.3) {{\color{blue}$\hat m_2$}};

\foreach \i in {0,1,2,...,17}{
\draw [pattern=north west lines, pattern color=blue]  (0+\i*0.25,-0.1-1.3-1.3-1.3) rectangle (0.25+0.25*\i,0.2-1.3-1.3-1.3);
\draw  (0+\i*0.25,-0.1-1.3-1.3-1.3) rectangle (0.25+0.25*\i,0.2-1.3-1.3-1.3);
}
\foreach \i in {0,...,9}{
\draw [draw = none, fill = red]  (0+\i*0.25,-0.1-1.3-1.3-1.3) rectangle (0.25+0.25*\i,0.2-1.3-1.3-1.3);
\draw  (0+\i*0.25,-0.1-1.3-1.3-1.3) rectangle (0.25+0.25*\i,0.2-1.3-1.3-1.3);
}

\draw [draw = none, fill = red]  (0+10*0.25,-0.1-1.3-1.3-1.3) rectangle (0.5*0.25+0.25*10,0.2-1.3-1.3-1.3);
\draw  (0+10*0.25,-0.1-1.3-1.3-1.3) rectangle (0.5*0.25+0.25*10,0.2-1.3-1.3-1.3);

\draw [<->] (0,0.25-1.3-1.3-1.3)--(4.5,0.25-1.3-1.3-1.3);
\draw [<->] (0,-0.2-1.3-1.3-1.3)--(2.25,-0.2-1.3-1.3-1.3);
\draw [<->] (2.25,-0.2-1.3-1.3-1.3)--(3,-0.2-1.3-1.3-1.3);
\draw [<->] (3,-0.2-1.3-1.3-1.3)--(4.5,-0.2-1.3-1.3-1.3);
\draw [->, very thick, red] (0,0.5-1.3-1.3-1.3)--(0,0.2-1.3-1.3-1.3);
\draw [->, very thick, blue] (2.25,0.5-1.3-1.3-1.3)--(2.25,0.2-1.3-1.3-1.3);
\draw [->, very thick, red] (3-0.375,-0.1-1.3-1.3-1.3)--(3-0.375,-0.45-1.3-1.3-1.3);
\draw [->, very thick, blue] (4.5,-0.1-1.3-1.3-1.3)--(4.5,-0.45-1.3-1.3-1.3);
\node [draw = none] at (2.25,0.35-1.3-1.3-1.3) {$n = 300$};
\node [draw = none] at (2.25,-0.35-1.3-1.3-1.3) {$\ns = 50$};
\node [draw = none] at (0.75,-0.35-1.3-1.3-1.3) {$\nf = 150$};
\node [draw = none] at (3.75,-0.35-1.3-1.3-1.3) {$\nt = 100$};
\node [draw = none] at (0,0.6-1.3-1.3-1.3) {{\color{red}$m_1$}};
\node [draw = none] at (2.25,0.6-1.3-1.3-1.3) {{\color{blue}$m_2$}};
\node [draw = none] at (3.05-0.375,-0.6-1.3-1.3-1.3) {{\color{red}$\hat m_1$}};
\node [draw = none] at (4.55,-0.6-1.3-1.3-1.3) {{\color{blue}$\hat m_2$}};

\foreach \i in {0,1,2,...,17}{
\draw [pattern=north west lines, pattern color=blue]  (0+\i*0.25,-0.1-1.3-1.3-1.3-1.3) rectangle (0.25+0.25*\i,0.2-1.3-1.3-1.3-1.3);
\draw  (0+\i*0.25,-0.1-1.3-1.3-1.3-1.3) rectangle (0.25+0.25*\i,0.2-1.3-1.3-1.3-1.3);
}

\foreach \i in {0,...,11}{
\draw [draw = none, fill = red]  (0+\i*0.25,-0.1-1.3-1.3-1.3-1.3) rectangle (0.25+0.25*\i,0.2-1.3-1.3-1.3-1.3);
\draw  (0+\i*0.25,-0.1-1.3-1.3-1.3-1.3) rectangle (0.25+0.25*\i,0.2-1.3-1.3-1.3-1.3);
}

\draw [<->] (0,0.25-1.3-1.3-1.3-1.3)--(4.5,0.25-1.3-1.3-1.3-1.3);
\draw [<->] (0,-0.2-1.3-1.3-1.3-1.3)--(3,-0.2-1.3-1.3-1.3-1.3);
\draw [<->] (3,-0.2-1.3-1.3-1.3-1.3)--(4.5,-0.2-1.3-1.3-1.3-1.3);
\draw [->, very thick, red] (0,0.5-1.3-1.3-1.3-1.3)--(0,0.2-1.3-1.3-1.3-1.3);
\draw [->, very thick, blue] (3,0.5-1.3-1.3-1.3-1.3)--(3,0.2-1.3-1.3-1.3-1.3);
\draw [->, very thick, red] (3,-0.1-1.3-1.3-1.3-1.3)--(3,-0.45-1.3-1.3-1.3-1.3);
\draw [->, very thick, blue] (4.5,-0.1-1.3-1.3-1.3-1.3)--(4.5,-0.45-1.3-1.3-1.3-1.3);
\node [draw = none] at (2.25,0.35-1.3-1.3-1.3-1.3) {$n = 300$};
\node [draw = none] at (0.75,-0.35-1.3-1.3-1.3-1.3) {$\nf = 200$};
\node [draw = none] at (3.75,-0.35-1.3-1.3-1.3-1.3) {$\nt = 100$};
\node [draw = none] at (0,0.6-1.3-1.3-1.3-1.3) {{\color{red}$m_1$}};
\node [draw = none] at (3,0.5-1.3-1.3-1.3-1.25) {{\color{blue}$m_2$}};
\node [draw = none] at (3.05,-0.6-1.3-1.3-1.3-1.3) {{\color{red}$\hat m_1$}};
\node [draw = none] at (4.55,-0.6-1.3-1.3-1.3-1.3) {{\color{blue}$\hat m_2$}};
\end{tikzpicture}
\caption{Time-sharing scheme with $\eta = 0.5$}
\end{subfigure}
\caption{Example of comparing our scheme with the time-sharing scheme for $n = 300$ and $\ns$ taking values  from $200$ to $0$ with a step size of $50$. }
\label{fig3}
\vspace{-0.5cm}
\end{figure}
\begin{theorem}[Time-Sharing, Both $m_1$ and $m_2$] \label{Th4:bounds}
Given $\nf$, $\nt$, $\bef$, $\bet$ and $\P$,   $ \log M_1:=  (\nf + \eta \ns) R_1$ and $\log M_2:=  (\nt + (1-\eta) \ns) R_2$ are upper bounded as
\begin{IEEEeqnarray}{rCl}
\log M_1
 &\le&   (\nf + \eta \ns) C\left (\Omega_{1,1}\right) \notag \\ 
 &&-   \sqrt{  (\nf + \eta \ns) V\left (\Omega_{1,1}\right)}\mathbb Q^{-1} \left ( \epsilon_1  -\Delta_1 \right) \notag \\
 && \hspace{0.5cm}+ \tilde K_1 \log \left ( (\nf + \eta \ns)\right) , \label{eq:55} 
\end{IEEEeqnarray}
and
\begin{IEEEeqnarray}{rCl}
\log M_2 &\le&  (\nt + (1-\eta) \ns) C\left (\Omega_{2,2}\right) \notag \\
&&  -   \sqrt{ (\nt + (1-\eta) \ns) V\left (\Omega_{2,2}\right)}\mathbb Q^{-1} \left ( \epsilon_2 -\Delta_2 \right) \notag \\
&& \hspace{0.5cm} + \tilde K_2 \log \left ((\nt + (1-\eta) \ns) \right) , \label{eq:56} \IEEEeqnarraynumspace
\end{IEEEeqnarray}
where 
\begin{IEEEeqnarray}{rCl}
\Delta_1&: =&\frac{6 T_{\max,1} }{\sqrt{ (\nf + \eta \ns)  V(\Omega_{1,1}))^3}} \notag \\
&& + \frac{J_12^{\delta_1}}{2^{\delta_1}-1}\Bigg(\frac{\delta_1}{\sqrt{2 \pi (\nf + \eta \ns) V(\Omega_{1,1})}} \notag \\
&& \hspace{1.5cm}+\frac{6 T_{\max,1} }{\sqrt{((\nf + \eta \ns)V(\Omega_{1,1}))^3}} \Bigg)\notag \\
&&\hspace{2cm} \cdot \left (\nf + \eta \ns \right) ^{K_1},\IEEEeqnarraynumspace \\
\Delta_2&: =&\frac{6 T_{\max,2} }{\sqrt{ (\nt + (1-\eta) \ns)  V(\Omega_{2,2}))^3}} \notag \\
&& +  \frac{J_2 2^{\delta_2}}{2^{\delta_2}-1}\Bigg(\frac{\delta_2}{\sqrt{2 \pi (\nt + (1-\eta) \ns) V(\Omega_{2,2})}} \notag \\
&& \hspace{1cm} +\frac{6 T_{\max,2} }{\sqrt{((\nt + (1-\eta) \ns)V(\Omega_{2,2}))^3}} \Bigg)\notag \\
&& \hspace{1.5cm} \cdot \left (\nt + (1-\eta) \ns \right) ^{K_2}, \IEEEeqnarraynumspace
\end{IEEEeqnarray}
with 
\begin{IEEEeqnarray}{rCl}
J_1&: =& \left (\frac{\nf +\eta \ns -2}{\nf + \eta \ns}\right)^{\frac{\nf+ \eta \ns+1}{2}}\notag \\
&& \cdot  e^{-\frac{1}{6(\nf + \eta \ns)}}\cdot \frac{2\sqrt{1 + 2 \Omega_{1,1}}}{\pi  (1 + \Omega_{1,1})}, \\
J_2&: =& \left (\frac{\nt +(1-\eta) \ns -2}{\nt + (1-\eta) \ns}\right)^{\frac{\nt+ (1-\eta) \ns+1}{2}} \notag \\
&& \cdot  e^{-\frac{1}{6(\nt + (1-\eta) \ns)}}\cdot \frac{2\sqrt{1 + 2 \Omega_{2,2}}}{\pi  (1 + \Omega_{2,2})},
\end{IEEEeqnarray}
with $\tilde K_1$ and $\tilde K_2$  being constants. 
Notice that $T_{\max,1}$  can be easily calculated from \eqref{eq:Tmax} by setting  $ i = 1$, $\ba = \bb = 0$ and replacing $\nf$ by $\nf + \eta \ns$.  Similarly, $T_{\max,2}$ can be calculated from \eqref{eq:Tmax}  by setting  $ i = 2$, $n_{2,1} = 0$ and replacing $\nt$ by $\nt + (1-\eta) \ns$.
\end{theorem}
\begin{IEEEproof}
Follows the proof of Theorem~\ref{Th1:bounds} by setting 
\begin{IEEEeqnarray}{rCl}
  \ba= 0, \quad  \bb =0, 
\end{IEEEeqnarray}
and replacing $\nf$ by $\nf + \eta \ns$ and $\nt$ by $\nt + (1-\eta) \ns$.
\end{IEEEproof}

\begin{figure}[t!]
\begin{subfigure}{0.5\textwidth}
\centering
\begin{tikzpicture}[scale=0.9]
\begin{axis}[
    xlabel={\small {$R_1$ }},
    ylabel={\small {$R_2$ }},
     xlabel style={yshift=.5em},
     ylabel style={yshift=0em},
    xmin= 0.3245, xmax=0.52,
    ymin= 0.3374, ymax=0.55,
    xtick={0.1,0.15,0.2,0.25,0.3,0.35, 0.4,0.45, 0.5,0.55,0.6,0.7,0.8,0.9},
    ytick={0.1,0.2,0.3,0.35,0.4,0.45,0.5},
    yticklabel style = {font=\small,xshift=0.25ex},
    xticklabel style = {font=\small,yshift=0.25ex},
    legend pos=north east,
    legend pos=north east,
]


\addplot[ color=black, very thick, mark = star] coordinates { (0.3419-0.0121,0.5336-0.0121)(0.3826-0.0134,0.5020-0.0134)(0.4347-0.014,0.4347-0.014)(0.4567-0.01051,0.3821-0.01051)(0.4820, 0.3374)};

\addplot[ color=blue, very thick, mark = halfcircle] coordinates { ( 0.3374,0.4820)(0.3682,0.4172)(0.3904,0.3904)(0.4172,0.3682)(0.4820, 0.3374)};

%

\legend{{Our Scheme},  {Time-Sharing}}  
\end{axis}
\end{tikzpicture}
\caption{}
\label{fig4a}
\end{subfigure}
\begin{subfigure}{0.5\textwidth}
\centering
\begin{tikzpicture}[scale=0.9]
\begin{axis}[
    xlabel={\small {$R_1$ }},
    ylabel={\small {$R_2$ }},
     xlabel style={yshift=.5em},
     ylabel style={yshift=0em},
    xmin= 0.3374/3, xmax=0.4,
    ymin= 0.3374/3, ymax=0.55,
   xtick={0.1,0.15,0.2,0.25,0.3,0.35, 0.4,0.45, 0.5,0.55,0.6,0.7,0.8,0.9},
    ytick={0.1,0.15,0.2,0.25, 0.3,0.35,0.4,0.45,0.5},
    yticklabel style = {font=\small,xshift=0.25ex},
    xticklabel style = {font=\small,yshift=0.25ex},
    legend pos=north east,
    legend pos=north east,
]


\addplot[ color=black, very thick, mark = star] coordinates { (0.3419*2/3 -0.0111,0.5336-0.0111)(0.3826*2/3 - 0.01224,0.5020*2.5/3-0.01244)(0.4347*2/3-0.01311,0.4347*2/3-0.01311)(0.4567*2/3-0.01435,0.3821*1.5/3-0.01435)(0.4820*2/3, 0.3374/3)};

\addplot[ color=blue, very thick, mark = halfcircle] coordinates { ( 0.3374/3,0.4820*2/3)(0.3682*1.25/3,0.4172*1.75/3)(0.3904/2,0.3904/2)(0.4172*1.75/3,0.3682*1.25/3)(0.4820*2/3, 0.3374/3)};

%

\legend{{Our Scheme},  {Time-Sharing}}  
\end{axis}
\end{tikzpicture}
\caption{}
\label{fig4b}
\end{subfigure}

\caption{\hnr{Comparison between the transmission rate pairs $(R_1,R_2)$ obtained in our scheme and the time-sharing scheme under the example shown in Fig.~\ref{fig3}. (a) In our scheme: $R_1 = \log M_1/(\nf + \ns)$ and $R_2 = \log M_2/(\nf + \nt)$, in the time-sharing scheme: $R_1 = \log M_1/(\nf + \eta \ns)$ and $R_2 = \log M_2/(\nf + (1-\eta)\ns)$ with $\eta = 0.5$, (b) In both schemes: $R_1 = \log M_1/n$ and $R_2 = \log M_2/n$. }}
\label{fig4}
\end{figure}
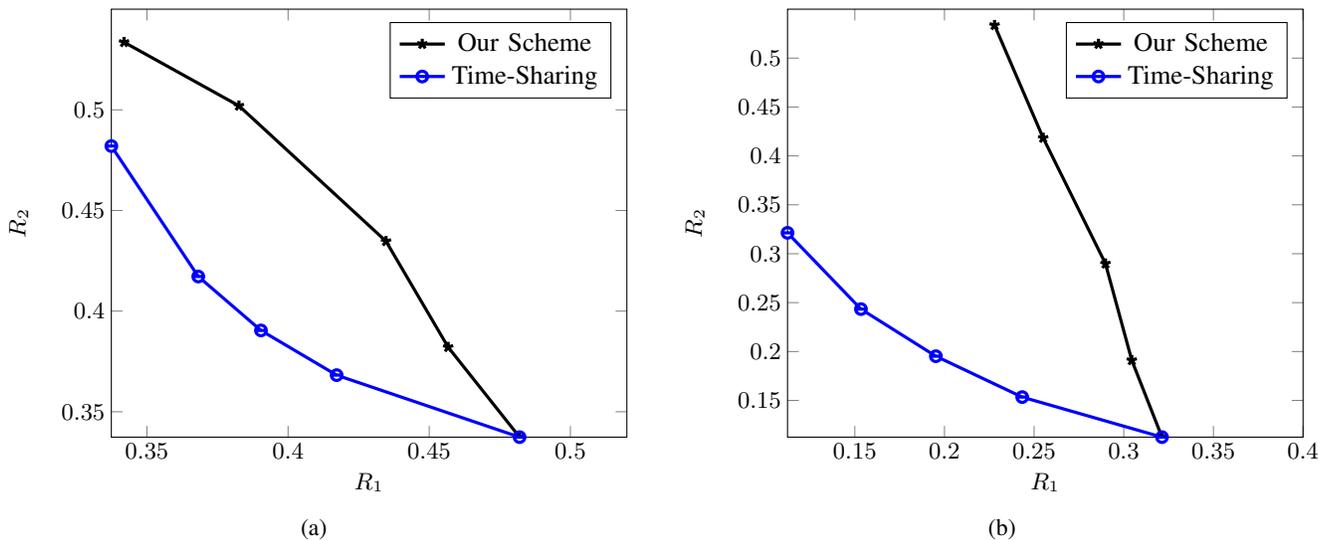

To compare our scheme with the time-sharing scheme  we consider a scenario where the total number of available channel uses is equal to $n =300$. The first message $m_1$ arrives at the beginning of the first channel use (i.e., $a_1 = 1$) and is sent over $200$ channel uses (i.e., $d_1 = 200$). The second message $m_2$ arrives at the time $a_2 \in [1:200]$ and has to be decoded at the end of $300$ channel uses (i.e., $d_2 = 300$). Fig.~\ref{fig3} illustrates a schematic representation of this example under our scheme (Fig.~\ref{fig3}.a) and the time sharing scheme with $\eta = 0.5$ (Fig.~\ref{fig3}.b). 

In Fig.~\ref{fig4}, we plot  the transmission rates $R_1$ and $R_2$ of our scheme and the time-sharing scheme for this example. Each point corresponds to a different value of $\ns$ shown in the example starting from $\ns = 200$ to $\ns = 0$ with a step size of 50. At $\ns = 0$, our scheme coincides with the time-sharing scheme.    In this figure, $\P = 5$, $\rho = 0.9$, $\sigff = \sigfs = \sigsf = \sigss = 1$, $K_1 = K_2 = \tilde K_1 = \tilde K_2 = 0.5$ and the values of the parameters $\bef, \bes, \bet, \ba, \bb$ are optimized to obtain the maximum sum transmission rates. In Fig.~\ref{fig4a}, in our scheme  $R_1 = \log M_1/(\nf + \ns)$ and $R_2 = \log M_2/(\nf + \nt)$ and in the time sharing scheme  $R_1 = \log M_1/(\nf + \eta \ns)$ and $R_2 = \log M_2/(\nf + (1-\eta)\ns)$. In Fig.~\ref{fig4b}, in both schemes  $R_1 = \log M_1/n$ and $R_2 = \log M_2/n$. As can be seen from this figure, our scheme significantly outperforms the time-sharing scheme.

\begin{figure}[t!]
\centering
\begin{subfigure}{0.5\textwidth}
\centering
\begin{tikzpicture}[scale=0.9]
\begin{axis}[
    xlabel={\small {$\lambda$ }},
    ylabel={\small {$R_1,R_2$ }},
     xlabel style={yshift=.5em},
     ylabel style={yshift=0em},
    xmin=0, xmax=1,
    ymin=0, ymax=0.65,
    xtick={0,0.2,0.4,0.6,0.8,1},
    ytick={0.1,0.2,0.3,0.4,0.5, 0.6},
    yticklabel style = {font=\small,xshift=0.25ex},
    xticklabel style = {font=\small,yshift=0.25ex},
    legend pos=north east,
    legend pos=south east,
]


\addplot[ color=black, very thick, mark =star] coordinates { (0,0.0415)    (0.2,0.0915)    (0.4,0.1415)    (0.6,0.1915)    (0.8,0.2415)    (1,0.2915)};

\addplot[ color=red, very thick, mark = halfcircle] coordinates { (0,0.5079)    (0.2,0.4579)    (0.4,0.4079)    (0.6,0.3579)    (0.8,0.3079)    (1,0.2579)};

\addplot[ color=green, very thick, mark = square] coordinates { (0,0.0415+0.5079)    (0.2,0.0915+0.4579)    (0.4,0.1415+0.4079)    (0.6,0.1915+0.3579)    (0.8,0.2415+0.3079)    (1,0.2915+0.2579)};


\footnotesize
\legend{ {$R_1, \sigff = \sigfs = 1$}, {$R_2, \sigsf = \sigss = 0.1$}, {$R_1 + R_2$}}  
\end{axis}

\vspace{-0.8cm}
\end{tikzpicture}

\caption{$\nf = \ns = \nt = 60$}
\label{fig3a}
\end{subfigure}
\begin{subfigure}{0.5\textwidth}
\centering
\begin{tikzpicture}[scale=0.9]
\begin{axis}[
    xlabel={\small {$\lambda$ }},
    ylabel={\small {$R_1,R_2$ }},
     xlabel style={yshift=.5em},
     ylabel style={yshift=0em},
    xmin=0, xmax=1,
    ymin=0, ymax=0.9,
    xtick={0,0.2,0.4,0.6,0.8,1},
    ytick={0.1,0.2,0.3,0.4,0.5,0.6,0.7,0.8},
    yticklabel style = {font=\small,xshift=0.25ex},
    xticklabel style = {font=\small,yshift=0.25ex},
    legend pos=north east,
    legend pos = south east,
]


\addplot[ color=black, very thick, mark = star] coordinates { (0,0.1802)    (0.2,0.2302)    (0.4,0.2802)    (0.6,0.3302)    (0.8,0.3802)    (1,0.4302)};

\addplot[ color=red, very thick, mark = halfcircle] coordinates { (0,0.6409)    (0.2,0.59009)    (0.4,0.5409)    (0.6,0.49009)    (0.8,0.4409)    (1,0.39009)};

\addplot[ color=green, very thick, mark = square] coordinates { (0,0.6409+0.1802)    (0.2,0.59009+0.2302)    (0.4,0.5409+0.2802)    (0.6,0.49009+0.3302)    (0.8,0.4409+0.3802)    (1,0.39009+0.4302)};


\footnotesize
\legend{ {$R_1, \sigff = \sigfs = 1$}, {$R_2, \sigsf = \sigss = 0.1$}, {$R_1 + R_2$}}  
\end{axis}

\vspace{-0.8cm}
\end{tikzpicture}

\caption{$\nf = \ns = \nt = 180$}
\label{fig3b}
\end{subfigure}
\vspace{-0.5cm}
\caption{\hnr{The figure illustrates the effect of changing $\lambda =\frac{1}{\ns}\log \left (\frac{L_2}{L_1}\right ) $ on the achievable rates $R_1$ and $R_2$ in a scenario where $\sigff = \sigfs = 1$, and  $\sigsf = \sigss = 0.1$ and (a) $\nf = \ns =\nt = 60$, (b)  $\nf = \ns =\nt = 180$. In this figure,  $\P = 1$, $\epsilon_1 =\epsilon_2 = 10^{-5}$, $\epsilon_{12} = 10^{-7}$ and $\rho = 0.8$.}  }
\label{fig3new}
\end{figure}
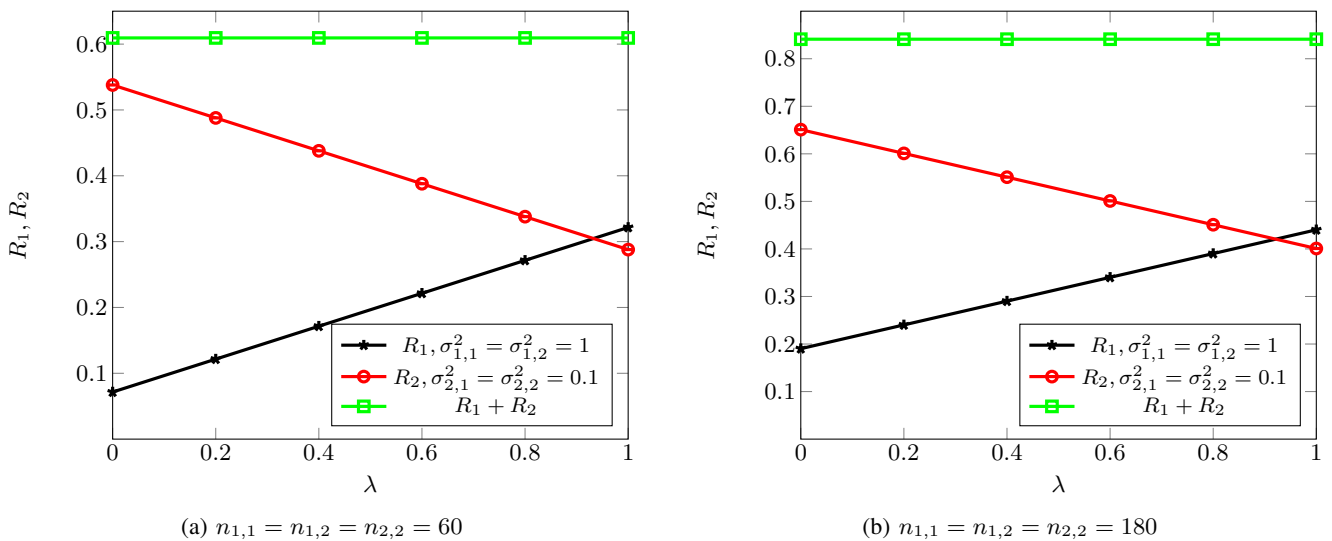

An important difference between our scheme and the time-sharing scheme appears when the channels to one receiver are stronger  than the channels to another receiver. Under our scheme, it is possible to adjust the design parameters $L_1$ and $L_2$ such that the rate to the weaker receiver increases while keeping the sum-rate approximately constant.  
In Theorem~\ref{Th1:bounds}, it is required that parameters $L_1$ and $L_2$ to be chosen such the condition in \eqref{eq:Ls} is satisfied. It is clear that,  large values of $L_1$ and $L_2$ reduce the transmission rates $R_1$ and $R_2$ and vice versa. In the analysis provided in the previous Section~\ref{sec:Marton},  $L_1$ and $L_2$ are set to be equal. In this section, we focus on the effect of changing the values of $L_1$ and $L_2$ on the achievable rates $R_1$ and $R_2$.  To this end, we introduce a parameter $\lambda$ as 
 \begin{equation}
 \lambda := \frac{1}{\ns}\log \left (\frac{L_2}{L_1}\right ).
\end{equation} 
By \eqref{eq:Ls}, 
\begin{IEEEeqnarray}{rCl}
\log L_1 &\ge&\frac{1}{2} \log \frac{ \log \left( \epsilon_{1,2} \right )}{ \log \left(1 - \left (1-\rho^2 \right )^{n_{1,2}-1} \right )} - \frac{\ns \lambda}{2}, 
\end{IEEEeqnarray}
and
\begin{IEEEeqnarray}{rCl}
\log L_2 &\ge&\frac{1}{2} \log \frac{ \log \left( \epsilon_{1,2} \right )}{ \log \left(1 - \left (1-\rho^2 \right )^{n_{1,2}-1} \right )} + \frac{\ns \lambda}{2}.
\end{IEEEeqnarray}
Therefore, increasing $\lambda$, increases $L_2$ (decreases $L_1$) which results in decreasing $R_2$ (increasing $R_1$).  

To numerically evaluate the effect of $\lambda$, we consider a case where the channels of the first receiver are weaker than the channels of the second receiver. In Fig.~\ref{fig3new}, we assume that $\sigff = \sigfs=1$ and $\sigsf = \sigss = 0.1$. In Fig.~\ref{fig3a}, we set $\nf = \ns = \nt = 60$ and in Fig.~\ref{fig3b} we set $\nf = \ns = \nt = 180$. In this figure,  $\P = 1$, $\epsilon_1 =\epsilon_2 = 10^{-5}$, $\epsilon_{12} = 10^{-7}$, $K_1 = K_2 = 0.5$, $\rho = 0.8$ and the values of power coefficients $\bef, \bes, \bet, \ba, \bb$ are set such that the sum-rate $R_1$ and $R_2$ is maximized.  We then increase $\lambda$ from $0$ (i.e., $L_1 = L_2$) to $1$ with step sizes of $0.2$. As can be seen from this figure, by increasing $L_2$ and consequently decreasing $L_1$, it is possible to increase the rate of $R_1$ while keeping the sum-rate $R_1 + R_2$ constant. For example, under our scheme,  it is possible to achieve $R_1 = R_2 = 0.31$ by setting $\lambda$ at approximately $0.95$.

\section{Conclusions} \label{sec:conclusion}
We have considered a broadcast setting in which a transmitter sends two different messages to two receivers. Messages are considered to have different arrival times and decoding deadlines such that their transmission windows overlap. For this setting, we have proposed a coding scheme that exploits  Marton's coding strategy.  We have derived rigorous bounds on the achievable rate regions for broadcast setting and point-to-point settings with one or multiple parallel channels. \hnr{ In the point-to-point setting with one channel and more parallel channels, the proposed achievability scheme was seen to be consistent with the normal approximation.}  In the broadcast setting, our scheme agreed with Marton's strategy for sufficiently large numbers of channel uses. Our numerical analysis  have shown  significant performance improvements over standard approaches based on time sharing for transmission of short packets. 
\appendices

\section{Proof of Theorem~\ref{Th1:bounds}} \label{sec:proofTh1}
In the following subsections we analyze the probability that events $\mathcal E_{1,2}$ and  $\mathcal E_{1}| \mathcal E_{1,2}^c$  occur.  The analysis related to $\Pr [\mathcal E_{2}| \mathcal E_{1,2}^c]$ is similar to that of $ \Pr [\mathcal E_{1}| \mathcal E_{1,2}^c]$. 

\subsection{Analyzing $\Pr[\mathcal E_{1,2}]$} \label{sec:VII-A}
Recall the definition of the error event $\mathcal E_{1,2}$ from \eqref{eq:e12}. Let 
\begin{align}
	\cos(\theta) = \frac{\langle \xfs(m_1,\ell_1), \xsf(m_2,\ell_2) \rangle}{n_{1,2}\sqrt{\beta_{1,2}\beta_{2,1}}\P}.
\end{align} 
Then,
\begin{IEEEeqnarray}{rCl}
\lefteqn{\Pr[\mathcal E_{1,2}]}\notag \\
 &=& \Pr \left[ \langle \xfs , \xsf \rangle \notin \mathcal D \right] \\
& = & \Pr  \left[ \langle \xfs , \xsf \rangle \notin\left [\ns \sqrt{\ba \bb  }\P  \rho :  \ns \sqrt{\ba \bb  }\P  \right ] \right] \notag \\ \\
& = & \prod_{\ell_1 = 1}^{L_1}\prod_{\ell_2 = 1}^{L_2} 1-  \Pr  \Big[ \langle \xfs(m_1,\ell_1), \xsf(m_2,\ell_2) \rangle \notag \\
&& \hspace{1.5cm}\in \left [\ns \sqrt{\ba \bb  }\P  \rho  :  \ns \sqrt{\ba \bb  }\P  \right ] \Big] \\
&= & \Bigg(1 - \Pr  \Big[ \langle \xfs(m_1,1), \xsf(m_2,1) \rangle \notag \\
&&\hspace{1cm} \in \left [\ns \sqrt{\ba \bb  }\P \rho  :  \ns \sqrt{\ba \bb  }\P  \right ] \Big]\Bigg)^{L_1L_2} \\
& = & \left(1- \Pr \Bigg [\rho   \le  \cos (\theta) \le 1  \Bigg ]\right)^{L_1L_2} \\
& = & \left(1+ \Pr \left [ \cos (\theta) \le \rho  \right ] - \Pr \left [ \cos (\theta) \le 1  \right ]\right)^{L_1L_2}  \\
& = & \left(1+ F_{\cos(\theta)}  (\rho ) - F_{\cos (\theta)} \left (1  \right )\right)^{L_1L_2} \label{eq:30}\\
& = & \left (F_{\cos^2(\theta)}  ((\rho)^2 )\right)^{L_1L_2}  \label{eq:31}\\
& = & \left( I_{(\rho)^2} (1, \ns-1) \right)^{L_1L_2}\label{eq:32}\\
& = & \left(1 - \left(1-\rho^2 \right)^{\ns-1}\right)^{L_1L_2}. \label{eq:34}
\end{IEEEeqnarray}
Note that in \eqref{eq:30}, $\cos^2 (\theta)\sim \text{Beta} (1, \ns-1)$ \hn{\cite{Jindal2006}}. As such, it follows that
\begin{IEEEeqnarray}{rCl}
F_{\cos^2(\theta)}  (x ) = I_x(1, \ns-1) =1- (1-x)^{\ns-1},
\end{IEEEeqnarray}
where $I_x(\cdot, \cdot)$ is regularized incomplete beta function. 
It then follows that
\begin{IEEEeqnarray}{rCl} \label{eq: e12f}
\Pr[\mathcal E_{1,2}]
=  \left(1 - \left(1-\rho^2 \right)^{\ns-1}\right)^{L_1L_2}.
\end{IEEEeqnarray}
In order to upper bound this error probability by a given threshold $\epsilon_{1,2}$, $L_1$ and $L_2$ should be chosen such that 
\begin{IEEEeqnarray}{rCl} 
L_1 \cdot L_2 \ge \frac{\log (\epsilon_{1,2})}{\log (1 - \left (1-\rho^2 \right )^{\ns-1})}.
\end{IEEEeqnarray}
This proves the bound in \eqref{eq:Ls}.

\subsection{Analyzing $\Pr [\mathcal E_{1}| \mathcal E_{1,2}^c]$}
Define the following Gaussian distributions: 
\begin{IEEEeqnarray}{rCl}
&&Q_{\vect Y_{1,j}} (\vect y_{1,j}) \sim \mathcal N(\vect y_{1,j}: \vect 0, \sigma_{y_{1,j}}^2 \mathrm I_{\nf}), \quad \text{for} \; j =1,2  \\
&&Q_{\hn{\yfs}| \xfs} (\vect y_{1,2}|\vect x_{1,2}) \sim \mathcal N(\hn{\yfs}: \vect \mu_{1,2}, \syuf \mathrm I_{\ns}), \IEEEeqnarraynumspace
\end{IEEEeqnarray}
where
\begin{subequations}\label{eq:sigmas}
\begin{IEEEeqnarray}{rCl}
\syf & = &  \bef \P + \sigff, \\
\sys &  = & \alpha^2 \bes \P + \sigfs, \\
  \syuf & = &\alpha^2 (1 - \rho^2) \bb \P + \sigfs, \label{eq:99c} \\
  \vect \mu_{1,2} &=& h \cdot \xfs, \quad  h := \alpha \left (1+ \rho  \sqrt{\frac{\bb}{\ba}}\right ). \label{eq:99d}
\end{IEEEeqnarray}
\end{subequations}
Recall the definition of $\alpha$ from \eqref{eq:alphan}. Note that $\sqrt{\frac{\bes }{\tilde \bes}} < \alpha < 1$. Therefore, all the parameters that are a function of $\alpha$  can also be upper and lower bounded accordingly. 
We then  introduce 
\begin{IEEEeqnarray}{rCl} \label{eq:tilde-i}
\lefteqn{\tilde i_1(\xf, \xfs; \yff, \yfs) }\notag \\
& \triangleq& \log \frac{f_{\yff|\xf} (\vect y_{1,1}|\vect x_{1,1})  Q_{\yfs| \xfs} (\vect y_{1,2}|\vect x_{1,2})}{\prod_{j=1}^2 Q_{\vect Y_{1,j}} (\vect y_{1,j})  }.
\end{IEEEeqnarray}
Following Remark~\ref{rem1}, we now continue our analysis based on this modified version of information density. 

To analyze $\Pr [\mathcal E_{1}| \mathcal E_{1,2}^c]$, we use the threshold-based metric bound \cite{Yuri2012}. Let $\gamma_1 \in \mathbb{R}$, since the first decoder selects among $M_1L_1$ codewords, thus
\begin{IEEEeqnarray}{rCl} \label{eq:85}
\Pr [\mathcal E_{1}| \mathcal E_{1,2}^c] &\le& \Pr [\tilde i_1(\xf, \xfs; \yff, \hn{\yfs}) \le \gamma_1] \notag \\
&&+ M_1 L_1 \Pr [\tilde i_1(\bar{\vect X}_{1,1}, \bar{\vect X}_{1,2}; \yff, \hn{\yfs}) \ge \gamma_1], \IEEEeqnarraynumspace
\end{IEEEeqnarray}
 where $\bar {\vect X}_{1,1} \sim f_{\xf} (\vect x_{1,1})$ and $\bar {\vect X}_{1,2} \sim f_{\xfs} (\vect x_{1,2})$ and are independent of $(\xf, \xfs; \yff, \yfs)$. Throughout our analysis, we interpret $ \Pr [\tilde i_1(\xf, \xfs; \yff, \yfs) \le \gamma_1]$ as the \emph{missed-detection} probability and $\Pr [\tilde i_1(\bar{\vect X}_{1,1}, \bar{\vect X}_{1,2}; \yff, \yfs) \ge \gamma_1]$ as the \emph{false alarm} probability.  

\subsubsection{Analyzing the missed-detection probability}
Note that since our inputs are non i.i.d, we cannot directly employ the Berry-Esseen theorem to bound  the missed-detection probability. As a result, we use the Berry-Esseen central limit theorem (CLT) for functions proposed in \cite[Proposition~1]{MolavianJaziArXiv}. To this end, we first need to show that the random variable $\tilde i_1(\xf, \xfs; \yff, \yfs)$ converges in distribution to a Gaussian distribution.   

\begin{lemma}\label{lemmanewi}
The following holds:
\begin{IEEEeqnarray}{rCl}
\lefteqn{ \tilde i_1(\xf, \xfs; \yff, \yfs)} \notag \\
 &&\sim \mathcal N( \nf C(\Omega_{1,1}) + \ns C(\Omega_{1,2}),  V_{1}),
 \end{IEEEeqnarray}
 where 
\begin{IEEEeqnarray}{rCl}
V_1 &: =& \nf V(\Omega_{1,1}) + \ns V(\Omega_{1,2}) + \ns \tilde V_1, 
\end{IEEEeqnarray}
with $V(x):= \frac{x(2+x)}{2(1+x)^2}$ and $\tilde V_1$ is defined in \eqref{eq:Vj}. 
\end{lemma}

\begin{IEEEproof}
See Appendix~\ref{Appnew}. 
\end{IEEEproof}


By the Berry-Esseen CLT for functions in \cite[Proposition~1]{MolavianJaziArXiv}, we have 
\begin{IEEEeqnarray}{rCl}
	\lefteqn{\Pr\left[\tilde{i}_1(\xf, \xfs; \yff, \yfs) \leq \gamma_1  \right] } \notag \\
	&\leq& \mathbb{Q}\left (\frac{\nf C(\Omega_{1,1}) + \ns C(\Omega_{1,2}) - \gamma_1 }{\sqrt{V_{1}}}\right) + \frac{6T_1}{\sqrt{V_1^3}}, \IEEEeqnarraynumspace
\end{IEEEeqnarray}
where $T_1$ is the  third central moment of $\tilde i_1(\xf, \xfs; \yff, \yfs)$.

%
Set
\begin{IEEEeqnarray}{rCl}
\gamma_1 & := & \log M_1 + \log L_1 -K_1\log (\nf+ \ns) 
\end{IEEEeqnarray}
  for some  $K_1$ and where  $C(x): = \frac{1}{2} \log(1 + x)$. 
\begin{lemma} \label{lemma2new}
The following inequality holds: 
\begin{IEEEeqnarray}{rCl}
T_1 &\le &T_{\max,1},
\end{IEEEeqnarray}
where  $V(x) := \frac{x(2+x)}{2(1+x)^2}$ and $T_{\max,1}$ is defined in \eqref{eq:Tmax}. 
\end{lemma}
\begin{IEEEproof}
See Appendix~\ref{app1}.
\end{IEEEproof}

We then have
\begin{IEEEeqnarray}{rCl}
\lefteqn{\Pr \left [  \tilde i_1(\xf, \xfs; \yff, \yfs)  \le  \gamma_1  \right ]} \notag\\
 && \le \frac{6T_1}{\sqrt{V_1^3}}  + \mathbb Q \left ( \frac{ \nf C(\Omega_{1,1}) + \ns C(\Omega_{1,2}) - \gamma_1 }{\sqrt{V_{1}}}\right ) \IEEEeqnarraynumspace\\
&& \stackrel{{(a)}}{\le} \frac{6T_{\max,1}}{\sqrt{(\nf V(\Omega_{1,1}) + \ns V(\Omega_{1,2}))^3}} \notag \\
&& + \mathbb Q \left ( \frac{ \nf C(\Omega_{1,1}) + \ns C(\Omega_{1,2})-\gamma_1  }{ \sqrt{V_{1}}}\right ), \IEEEeqnarraynumspace \label{eq:66}
\end{IEEEeqnarray}
where step $(a)$ follows by Lemma~\ref{lemma2new} and the fact that $\tilde V_1$ is positive. 

\subsubsection{Analyzing the false alarm probability}
To bound the false alarm probability, i.e., $\Pr [\tilde i_1(\bar{\vect X}_{1,1}, \bar{\vect X}_{1,2}; \yff, \yfs) \ge \gamma_1]$, we first use the following change of measure argument proposed by \cite[Eq.4]{Molavianjazi}
\begin{IEEEeqnarray}{rCl}
\lefteqn{\Pr [\tilde i_1(\bar{\vect X}_{1,1}, \bar{\vect X}_{1,2}; \yff, \yfs) \ge \gamma_1] }\\
& = & \int \int \int \int \mathbbm{1} \{ i_1(\bar{\vect X}_{1,1}, \bar{\vect X}_{1,2}; \yff, \yfs) \ge \gamma_1\} \notag \\
&&\hspace{0.5cm} f_{\xf} (\vect x_{1,1}) f_{\xfs}(\vect x_{1,2}) f_{\yff}(\vect y_{1,1}) f_{\yfs}(\vect y_{1,2})\notag \\
&& \hspace{1cm} d \vect x_{1,1} d \vect x_{1,2} d \vect y_{1,1} d \vect y_{1,2} \IEEEeqnarraynumspace \\
& = & \int \int \int \int \mathbbm{1} \{ i_1(\bar{\vect X}_{1,1}, \bar{\vect X}_{1,2}; \yff, \yfs) \ge \gamma_1\} \notag \\
&& f_{\xf} (\vect x_{1,1}) f_{\xfs}(\vect x_{1,2}) \frac{f_{\yff}(\vect y_{1,1})}{Q_{\yff}(\vect y_{1,1})} \frac{f_{\yfs}(\vect y_{1,2})}{Q_{\yfs}(\vect y_{1,2})}\notag \\
&& Q_{\yff}(\vect y_{1,1}) Q_{\yfs}(\vect y_{1,2}) d \vect x_{1,1} d \vect x_{1,2} d \vect y_{1,1} d \vect y_{1,2} \\
&\le& J_1 \Pr_{Q} [\tilde i_1(\bar{\vect X}_{1,1}, \bar{\vect X}_{1,2}; \yff, \yfs) \ge \gamma_1]
\end{IEEEeqnarray}
where 
\begin{IEEEeqnarray}{rCl} \label{eq:J1new}
\frac{f_{\yff}(\vect y_{1,1})}{Q_{\yff}(\vect y_{1,1})} \cdot \frac{f_{\yfs}(\vect y_{1,2})}{Q_{\yfs}(\vect y_{1,2})} &\le&J_1
\end{IEEEeqnarray}
and $J_1$ is defined in \eqref{eq:Ji}. 
See Appendix~\ref{app2} for the proof of \eqref{eq:J1new}.
We then  use \cite[Lemma 47]{Yuri2012} and the proof of this lemma in \cite[Appendix G]{Yuri2012}. Based on this lemma, we can bound the false alarm probability by
 \begin{IEEEeqnarray}{rCl} \label{eq:107}
	\lefteqn{\Pr [\tilde i_1(\bar{\vect X}_{1,1}, \bar{\vect X}_{1,2}; \yff, \yfs) \ge \gamma_1]}\\
	&\le&  \Pr_{Q} [\tilde i_1(\bar{\vect X}_{1,1}, \bar{\vect X}_{1,2}; \yff, \yfs) \ge \gamma_1] \notag \\
	 &\leq& J_1\frac{2^{\delta_1}}{2^{\delta_1}-1}\left(\frac{\delta_1}{\sqrt{2 \pi V_1}} + \frac{12T_{1}}{\sqrt{V_1^3}}\right)2^{-\gamma_1} ,
 \end{IEEEeqnarray}
 for any $\delta_1$. 
 Note that in the proof of \cite[Lemma 47]{Yuri2012} $\delta_1$ is set at  $\log 2$. See \cite[Appendix G]{Yuri2012} for the detailed proof.

By combining \eqref{eq:66} and \eqref{eq:107},  and the fact that $V_1 \ge \nf V(\Omega_{1,1}) + \ns V(\Omega_{1,2})$ we can bound the error probability in \eqref{eq:85} by 
\begin{IEEEeqnarray}{rCl}
\lefteqn{\Pr [\mathcal E_{1}| \mathcal E_{1,2}^c]}\notag \\ &\le& \frac{6 T_{\max,1} }{\sqrt{(\nf V(\Omega_{1,1}) + \ns V(\Omega_{1,2}))^3}}  \notag \\
&&+ \mathbb Q \left ( \frac{ \nf C(\Omega_{1,1}) + \ns C(\Omega_{1,2}) - \gamma_1   }{\sqrt{V_{1}}}\right )\notag  \\
&&+ \frac{J_1 2^{\delta_1}}{2^{\delta_1}-1}\Bigg(\frac{\delta_1}{\sqrt{2 \pi(\nf V(\Omega_{1,1})+\ns V(\Omega_{1,2}))}} \notag \\
&& \hspace{0cm} + \frac{12T_{\max,1}}{\sqrt{(\nf V(\Omega_{1,1}) + \ns V(\Omega_{1,2}))^3}}\Bigg)  (\nf+ \ns) ^{K_1}. \notag \\
\end{IEEEeqnarray}


As a result 
\begin{IEEEeqnarray}{rCl}
\lefteqn{\Pr[\hat{m}_1 \neq m_1] }\notag \\&\le& \frac{6 T_{\max,1} }{\sqrt{(\nf V(\Omega_{1,1}) + \ns V(\Omega_{1,2}))^3}}  \notag \\
&&+ \mathbb Q \left ( \frac{ \nf C(\Omega_{1,1}) + \ns C(\Omega_{1,2}) - \gamma_1  }{ \sqrt{V_{1}}}\right )\notag  \\
&&+  \frac{J_1 2^{\delta_1}}{2^{\delta_1}-1}\Bigg(\frac{\delta_1}{\sqrt{2 \pi(\nf V(\Omega_{1,1})+\ns V(\Omega_{1,2}))}} \notag \\
&& + \frac{12T_{\max,1}}{\sqrt{(\nf V(\Omega_{1,1}) + \ns V(\Omega_{1,2}))^3}}\Bigg)(\nf+ \ns) ^{K_1}\notag \\
&& +  \left (1 - \left (1-\rho^2 \right )^{\ns-1} \right )^{L_1 \cdot L_2}.
\end{IEEEeqnarray}
 
Given that the error probability $\Pr[\hat{m}_1 \neq m_1]$ should not exceed a given threshold $\epsilon_1$, thus  
\begin{IEEEeqnarray}{rCl}
\epsilon_1 &\ge& \frac{6 T_{\max,1} }{\sqrt{(\nf V(\Omega_{1,1}) + \ns V(\Omega_{1,2}))^3}}  \notag \\
&&+ \mathbb Q \left ( \frac{ \nf C(\Omega_{1,1}) + \ns C(\Omega_{1,2})-\gamma_1 }{ \sqrt{V_{1}}}\right )\notag  \\
&&+ \frac{J_12^{\delta_1}}{2^{\delta_1}-1}\Bigg(\frac{\delta_1}{\sqrt{2 \pi(\nf V(\Omega_{1,1})+\ns V(\Omega_{1,2}))}} \notag \\
&& + \frac{12T_{\max,1}}{\sqrt{(\nf V(\Omega_{1,1}) + \ns V(\Omega_{1,2}))^3}}\Bigg)(\nf+ \ns) ^{K_1}  \notag \\
&& +  \left (1 - \left (1-\rho^2 \right )^{\ns-1} \right )^{L_1 \cdot L_2}.
\end{IEEEeqnarray}

Recall the definition of $\Delta_1$ from \eqref{eq:Deltai}. Hence
\begin{IEEEeqnarray}{rCl} \label{eq:114n}
\epsilon_1 - \Delta_1 &\ge&  \mathbb Q \left ( \frac{ \nf C(\Omega_{1,1}) + \ns C(\Omega_{1,2}) - \gamma_1}{ \sqrt{V_{1}}}\right ).\IEEEeqnarraynumspace
\end{IEEEeqnarray}
By taking $\mathbb Q^{-1}$ from the both sides of \eqref{eq:114n} we have 
\begin{IEEEeqnarray}{rCl} \label{eq:114n2}
\mathbb Q^{-1} \left (\epsilon_1 - \Delta_1\right ) &\le&   \frac{ \nf C(\Omega_{1,1}) + \ns C(\Omega_{1,2}) - \gamma_1 }{ \sqrt{V_{1}}}. \IEEEeqnarraynumspace
\end{IEEEeqnarray}
Note that the $\mathbb Q^{-1} (x)$ is a decreasing function of $x$ and thus changes the direction of the inequality.

Finally,
\begin{IEEEeqnarray}{rCl}
\lefteqn{\log M_1  + \log L_1} \notag \\
&\le&  \nf C(\Omega_{1,1}) + \ns C(\Omega_{1,2})  \notag \\
&& + K_1 \log (\nf+ \ns)  -  \sqrt{V_{1}} \mathbb Q^{-1} \left ( \epsilon_{1}  - \Delta_1\right ). \label{eq:100}
\end{IEEEeqnarray}
This concludes the proof of  Theorem~\ref{Th1:bounds}.

\section{Proof of Proposition~\ref{prop1}} \label{sec:prop1}
 For large values of $\nf$ and $\ns$, choose $K_1$ and $\delta_1$ such that the effect of $\Delta_1$ is negligible given the other parameters. 
This proves the bound in \eqref{eq:upp1asym}. 
To prove the bound in \eqref{eq:Lsasym}, define
\begin{IEEEeqnarray}{rCl}
A&: =&1 - \left (1-\rho^2 \right )^{\ns-1}.
\end{IEEEeqnarray}
We then use the approximation $\log(1-x) \approx -x$ which holds when $x$ is very small.
Hence, for large values of $\ns$
\begin{IEEEeqnarray}{rCl}
\log L_1 + \log L_2 \ge \log (-\log  \epsilon_{1,2}) - \ns \log(1- \rho^2), \IEEEeqnarraynumspace
\end{IEEEeqnarray}
which concludes the proof. 


\section{Proof of Lemma~\ref{lemmanewi}} \label{Appnew}
By \eqref{eq:tilde-i}
\begin{IEEEeqnarray}{rCl}
 \lefteqn{\tilde i_1(\xf, \xfs; \yff, \yfs)} \notag \\
 & =&  \log \frac{f_{\yff|\xf} (\vect y_{1,1}|\vect x_{1,1}) }{Q_{\vect Y_{1,1}} (\vect y_{1,1})} \notag \\
 && + \log \frac{Q_{\yfs| \xfs} (\vect y_{1,2}|\vect x_{1,2})}{ Q_{\vect Y_{1,2}} (\vect y_{1,2})  }.
 \end{IEEEeqnarray}
 Let 
 \begin{IEEEeqnarray}{rCl}
 I_1 &: =&  \log \frac{f_{\yff|\xf} (\vect y_{1,1}|\vect x_{1,1}) }{Q_{\vect Y_{1,1}} (\vect y_{1,1})}, \\ I_2 &: =& \log \frac{Q_{\yfs| \xfs} (\vect y_{1,2}|\vect x_{1,2})}{ Q_{\vect Y_{1,2}} (\vect y_{1,2})  }.
 \end{IEEEeqnarray}
 \begin{lemma}
 The following hold: 
 \begin{IEEEeqnarray}{rCl}
 I_1 &\sim& \mathcal N \left (\nf C(\Omega_{1,1}),  \nf V(\Omega_{1,1})\right),\label{eq:I1} \\ 
 I_2 &\sim& \mathcal N \left (\ns C(\Omega_{1,2}), \ns V(\Omega_{1,2}) + \ns \tilde V_1\right ), \label{eq:I2}
 \end{IEEEeqnarray}
 where $C(x) : = \frac{1}{2} \log(1+x)$, $V(x): = \frac{x(2+x)}{2(1+x)^2}$ and  $\tilde V_1$ is defined in  \eqref{eq:Vj}.
 \end{lemma}
 \begin{IEEEproof}
See \cite[Section~III-D-2]{MolavianJaziArXiv} for the proof of \eqref{eq:I1} . We follow the same argument as in \cite[Section~III-D-2]{MolavianJaziArXiv} for the proof of \eqref{eq:I2}. We consider $\rho^* = \rho$ which results in $\langle \xfs, \xsf \rangle = \ns \rho \sqrt{\ba \bb} \P$ and $\alpha = 1$. 
\begin{IEEEeqnarray}{rCl}
\lefteqn{I_2} \notag \\
 &=& \log \frac{Q_{\yfs| \xfs} (\vect y_{1,2}|\vect x_{1,2})}{ Q_{\yfs} (\vect y_{1,2})  } \\
&=& \frac{\ns}{2}\log(1+\Omega_{1,2}) \notag \\
&& + \frac{1}{2} \left [ \frac{||\yfs||^2}{\sys} - \frac{||\yfs - h\xfs||^2}{\syuf}\right]\\
& = & \ns C(\Omega_{1,2})  \notag \\
&& + \frac{1}{2} \Bigg [\left ( \frac{1}{\sys} - \frac{1}{\syuf}\right) ||\yfs||^2 \notag \\
&& \hspace{0.5cm} - \frac{h^2}{\syuf} ||\xfs||^2 + \frac{2h}{\syuf} \langle \yfs, \xfs \rangle \Bigg]\\
& = & \ns C(\Omega_{1,2})  \notag \\
&& + \frac{1}{2} \Bigg [\left ( \frac{1}{\sys} - \frac{1}{\syuf}\right)   \notag \\
&&\hspace{0.5cm}\cdot ||\alpha(\xfs + \xsf) + \zfs||^2- \frac{h^2}{\syuf} ||\xfs||^2 \notag \\
&& \hspace{0.8cm} + \frac{2h}{\syuf} \langle \alpha(\xfs+\xsf)+\zfs, \xfs \rangle \Bigg]\\
& = & \ns C(\Omega_{1,2})  \notag \\
&& + \frac{1}{2} \Bigg [\left ( \frac{1}{\sys} - \frac{1}{\syuf}\right) \notag \\
&&\hspace{0.75cm} \cdot  \Big ( \alpha^2 (||\xfs||^2 + ||\xsf||^2+2\langle \xfs, \xsf \rangle) \notag \\
&& \hspace{1.5cm}+ ||\zfs||^2 + 2\alpha \langle \xfs + \xsf, \zfs \ \rangle \Big)  \notag \\
&&\hspace{0.5cm}- \frac{h^2}{\syuf} ||\xfs||^2 + \frac{2h}{\syuf}  \notag \\
&& \hspace{0.2cm}\Big(  \alpha ||\xfs||^2+ \alpha  \langle \xsf, \xfs \rangle + \langle \zfs, \xfs \rangle \Big) \Bigg]\\
& = & \ns C(\Omega_{1,2})  \notag \\
&& + \frac{1}{2} \Bigg [\left ( \frac{1}{\sys} - \frac{1}{\syuf}\right) \big (\ns \bes \P + ||\zfs||^2 \notag \\
&& \hspace{0.75cm}  + 2 \langle \xfs, \zfs  \rangle + 2 \langle \xsf, \zfs  \rangle \big)  \notag \\
&&\hspace{1cm}- \frac{h^2}{\syuf} \ns \P \ba + \frac{2h}{\syuf} \big (  \ns \P \ba \notag \\
&& \hspace{1.5cm}+ \ns \P \rho \sqrt{\ba \bb} + \langle \zfs, \xfs \rangle \big) \Bigg]\\
&\stackrel{(i)}{ \ge}& \ns C(\Omega_{1,2})  \notag \\
&&+ \frac{1}{2} \Bigg [ -\frac{\P (\rho \sqrt{\bb} + \sqrt{\ba})^2}{ \sys \syuf} \big(\ns \bes \P   \notag \\
&& \hspace{0.3cm}+ ||\zfs||^2+ 2 \langle \xfs, \zfs  \rangle + 2 \langle \xsf, \zfs  \rangle \big)  \notag \\
&&\hspace{0.4cm}- \frac{\ns \P (\sqrt{\ba} + \rho \sqrt{\bb})^2}{\syuf}  + \frac{2(1+ \rho \sqrt{\frac{\bb}{\ba}})}{\syuf} \notag \\
&& \hspace{-0.1cm} \cdot \left (  \ns \P \ba + \ns \P \rho \sqrt{\ba \bb} + \langle \zfs, \xfs \rangle \right) \Bigg] \\
& = & \ns C(\Omega_{1,2})  \notag \\
&& + \frac{1}{2} \Bigg [ -\frac{\P (\rho \sqrt{\bb} + \sqrt{\ba})^2}{ \sys \syuf} \notag \\
&& \hspace{0.5cm}  \big ( ||\zfs||^2+ 2 \langle \xfs, \zfs  \rangle + 2 \langle \xsf, \zfs  \rangle \big)  \nonumber \\
&& \hspace{1cm}+ \frac{2(1+ \rho \sqrt{\frac{\bb}{\ba}})}{\syuf} \langle \zfs, \xfs \rangle  \notag \\
&&\hspace{1.5cm}- \frac{\ns \P (\sqrt{\ba} + \rho \sqrt{\bb})^2}{\syuf} \left (1 + \frac{\bes \P }{\sys} \right) \notag \\
&&  + \frac{2(1+ \rho \sqrt{\frac{\bb}{\ba}})}{\syuf} \left (  \ns \P \ba + \ns \P \rho \sqrt{\ba \bb} \right) \Bigg] \notag \\ \\
& = & \ns C(\Omega_{1,2})  \notag \\
&& + \frac{1}{2} \Bigg [ -\frac{\P (\rho \sqrt{\bb} + \sqrt{\ba})^2}{ \sys \syuf} \notag \\
&& \hspace{0.6cm}\cdot \left ( ||\zfs||^2 + 2 \langle \xfs, \zfs  \rangle + 2 \langle \xsf, \zfs  \rangle \right)  \notag \\
&& \hspace{1cm}+ \frac{2(1+ \rho \sqrt{\frac{\bb}{\ba}})}{\syuf} \langle \zfs, \xfs \rangle  \notag \\
&&\hspace{1.3cm}- \frac{\ns \P (\sqrt{\ba} + \rho \sqrt{\bb})^2}{\syuf} \left (1 + \frac{\bes \P }{\sys} \right)  \notag \\
&& \hspace{2.9cm}+ \frac{2\ns \P (\sqrt{\ba} + \rho \sqrt{\bb})^2}{\syuf}\Bigg]\\
& = & \ns C(\Omega_{1,2})  \notag \\
&& + \frac{1}{2} \Bigg [ -\frac{\P (\rho \sqrt{\bb} + \sqrt{\ba})^2}{ \sys \syuf} \notag \\
&&\hspace{0.7cm} \cdot \left ( ||\zfs||^2 + 2 \langle \xfs, \zfs  \rangle + 2 \langle \xsf, \zfs  \rangle \right)  \notag \\
&& \hspace{1cm}+ \frac{2(1+ \rho \sqrt{\frac{\bb}{\ba}})}{\syuf} \langle \zfs, \xfs \rangle  \notag \\
&&\hspace{2.5cm} + \frac{\ns \P (\sqrt{\ba} + \rho \sqrt{\bb})^2 \sigfs }{\syuf \sys } \Bigg]\\
& = & \ns C(\Omega_{1,2})  \notag \\
&& + \frac{1}{2} \Bigg [ \frac{\P (\rho \sqrt{\bb} + \sqrt{\ba})^2}{ \sys \syuf}  \big(\ns \sigfs \notag \\
&& \hspace{0.7cm} -  ||\zfs||^2 - 2 \langle \xfs, \zfs  \rangle - 2 \langle \xsf, \zfs  \rangle \big)  \notag \\
&&\hspace{2.5cm} + \frac{2(1+ \rho \sqrt{\frac{\bb}{\ba}})}{\syuf} \langle \zfs, \xfs \rangle \Bigg ] \\
&=& \ns C(\Omega_{1,2}) + \frac{1}{2} \big[c_1(\ns \sigfs -||\zfs||^2)\notag \\
&& \hspace{2cm} + c_2 \langle \xfs, \zfs \rangle + c_3\langle \xsf, \zfs \rangle \big ], \label{eq:135}
\end{IEEEeqnarray}
where 
\begin{IEEEeqnarray}{rCl}
c_1 &: =& \frac{ \P (\rho \sqrt{\bb} + \sqrt{\ba})^2}{ \sys \syuf},\\
 c_2 &:=&   \frac{2}{\sys}, \quad c_3 := -2 c_1.
\end{IEEEeqnarray}
 \end{IEEEproof}
 The inequality in $(i)$ follows by considering $\alpha = 1$ and $\rho^* = \rho$. Note that 
 the summands in \eqref{eq:135} are not independent, since
  $\xfs$ and $\xsf$  are not independent across time. One can however express independent uniform random variables on the power shell as a functions of independent Gaussian random variables. To this end,  
 let $\vect W_1 \sim \mathcal N (0, I_{\ns})$ and $\vect W_2 \sim \mathcal N(0, I_{\ns})$ be i.i.d Gaussian random variables independent of the noise $\zfs$. Inputs $X_{12,t}$ and $X_{21,t}$ with $t \in \{1, \ldots, \ns\}$ thus can be expressed as
 \begin{IEEEeqnarray}{rCl}
 X_{12,t} &=& \sqrt{\ns \ba \P} \frac{W_{1,t}}{||\vect W_1||},  \\
  X_{21,t} &=& \sqrt{\ns \bb \P} \frac{W_{2,t}}{||\vect W_2||}, 
 \end{IEEEeqnarray} 
 To apply the CLT for functions proposed in \cite[Proposition~1]{MolavianJaziArXiv}, we consider the sequence $\{\vect U_t : = (U_{1,t}, \ldots, U_{5,t})\}_{t = 1}^{\infty}$ whose elements are defined as 
 \begin{IEEEeqnarray}{rCl}
 U_{1,t} &: =& \sigfs - Z_{12,t}^2, \quad U_{2,t}: = \sqrt{\ba \P} W_{1,t} Z_{12,t},  \\
  U_{3,t}&: =& \sqrt{\bb \P} W_{2,t} Z_{12,t}, \quad 
   U_{4,t}: = W_{1,t}^2 -1,  \\
     U_{5,t}&: =& W_{2,t}^2 -1.
 \end{IEEEeqnarray}
 Note that this random vector has an i.i.d. distribution across time $t = 1, \ldots, n$ and its moments can be easily verified to satisfy $\mathbb E [\vect U_1] = 0$ and $\mathbb E[||\vect U_t||_2^3] < \infty$. 
 The covariance matrix of this vector is given by 
 \begin{equation}
 \text{Cov} (\vect U) = \text{Diag} [2 \sigfs, \ba \P \sigfs , \bb \P \sigfs , 2, 2].
 \end{equation}

Define the function $f$ as 
\begin{IEEEeqnarray}{rCl}
f(\vect u) = c_1 u_1 + \frac{c_2 u_2}{\sqrt{1+u_4}} + \frac{c_3 u_3}{\sqrt{1+u_5}} . 
\end{IEEEeqnarray}
Notice that $f(\vect 0) = 0$, and all the first and second order partial derivatives of $f$ are continuous in a neighborhood of $\vect u = 0$. The Jacobian matrix $\{\pdv{f(\vect u)}{u_j}\}_{1\times 6}$ at $\vect u = 0$ thus is given by 
\begin{equation}
\mathrm{J} \big |_{\vect u = 0} = [c_1 \; c_2\;  c_3\;    0\;  0 ]. 
\end{equation}
Furthermore,  
\begin{IEEEeqnarray}{rCl}
\lefteqn{f\left (\frac{1}{\ns} \sum_{t = 1}^{\ns} \vect U_t \right) } \notag \\
 &= &\frac{c_1}{\ns} \sum_{t = 1}^{\ns} (\sigfs-Z_{12,t}^2)\notag \\
 &&  + \frac{\frac{c_2}{\ns}\sum_{t = 1}^{\ns} \sqrt{\ba \P} W_{1,t} Z_{12,t}}{\sqrt{1+\frac{1}{\ns} \sum_{t=1}^{\ns}(W_{1,t}^2-1)}}  \notag \\
 && \hspace{0.5cm}+ \frac{\frac{c_3}{\ns}\sum_{t = 1}^{\ns} \sqrt{\bb \P} W_{2,t} Z_{12,t}}{\sqrt{1+\frac{1}{\ns} \sum_{t=1}^{\ns}(W_{2,t}^2-1)}}    \\
& = & \frac{1}{\ns} \Big[c_1(\ns \sigfs - ||\zfs||^2) \notag \\
&& \hspace{1cm}+ c_2 \langle \xfs, \zfs \rangle + c_3\langle \xsf, \zfs \rangle  \Big ].
\end{IEEEeqnarray}
From the CLT in \cite[Proposition~1]{MolavianJaziArXiv}, we now conclude that the random variable $I_2$ converges in distribution to a Gaussian distribution with mean $\ns C(\Omega_{1,2})$ and variance 
\begin{IEEEeqnarray}{rCl}
\lefteqn{\frac{1}{2\ns} [c_1 \; c_2\;  c_3\;  0\;  0 ]  \text{Cov} (\vect U) [c_1 \; c_2\;  c_3\;  0\;  0 ]^T } \notag \\
&=& \frac{1}{2\ns} \left [2c_1^2 \sigfs + c_2^2 \sigfs \ba \P + c_3^2 \sigfs \bb \P  \right]. \IEEEeqnarraynumspace
\end{IEEEeqnarray}
This concludes the proof. 


\section{Proof of  Lemma~\ref{lemma2new}} \label{app1}

In this section, we upper bound the third moment. To this end, we employ the following inequality: 
\begin{IEEEeqnarray}{rCl}
	\lefteqn{\mathbb{E}\big[|\tilde{i}_1(\xf, \xfs; \yff, \yfs) }\notag \\
	&& - \mathbb{E}[\tilde{i}_1(\xf, \xfs; \yff, \yfs)]|^3\big] \notag \\
	&& \leq 2^3\mathbb{E}[|\tilde{i}_1(\xf, \xfs; \yff, \yfs)|^3],
\end{IEEEeqnarray}
which follows from the Minkowski inequality \cite{Mulholland1949}.

Let $f_{\tilde{i}}(\cdot)$ be the probability density function of $\tilde{i}_1(\xf, \xfs; \yff, \yfs)$ with $F_i$ as its cumulative density function. For simplicity, define $Z:=\tilde{i}_1(\xf, \xfs; \yff, \yfs)$. For any $\kappa_1 >1$ we have
\begin{align}
	&\mathbb{E}[|Z|^3]\notag\\
	&= \int_{-\infty}^{\infty} |z|^3f_{\tilde{i}}(z)\mathrm{d}z\notag\\
	&\leq \kappa_1 + \int_\kappa^{\infty} z^3 f_{\tilde{i}}(z)\mathrm{d}z + \int_{-\infty}^{-\kappa_1} |z|^3f_{\tilde{i}}(z)\mathrm{d}z\notag\\
	&= \kappa_1 + \int_\kappa^{\infty} z^3(f_{\tilde{i}}(z) - f_{\tilde{i}}(-z))\mathrm{d}z\notag\\
	&= \kappa_1 + \sum_{j=0}^{\infty} \int_{\kappa_1+j}^{\kappa_1+ j + 1} z^3(f_{\tilde{i}}(z) - f_{\tilde{i}}(-z))\mathrm{d}z\notag\\
	&\leq \kappa_1 + \sum_{j=0}^{\infty} (\kappa_1+ j + 1)^3 \int_{\kappa_1+j}^{\kappa_1+j+1} (f_{\tilde{i}}(z) - f_{\tilde{i}}(-z))
	\mathrm{d}z\notag\\
	&= \kappa_1 + \sum_{j=0}^{\infty} (\kappa_1+j+1)^3\big(F_{\tilde{i}}(\kappa_1 + j + 1) - F_{\tilde{i}}(\kappa_1 + j) \notag \\
	&   \hspace{2.5cm}+ F_{\tilde{i}}(-\kappa_1 - j) - F_{\tilde{i}}(-\kappa_1 - j - 1)\big) \notag \\
	&\le \kappa_1 + \sum_{j=0}^{\infty} (\kappa_1 + j + 1)^3 (1-  F_{\tilde{i}}(\kappa_1 + j) + F_{\tilde{i}}(-\kappa_1 - j) ).
\end{align}
Notice that 
\begin{IEEEeqnarray}{rCl}
\lefteqn{1-  F_{\tilde{i}}(\kappa_1 + j) } \notag \\
& =& \Pr [\tilde{i}_1(\xf, \xfs; \yff, \yfs) > \kappa_1 + j]
\end{IEEEeqnarray}
and
\begin{IEEEeqnarray}{rCl}
\lefteqn{F_{\tilde{i}}(-\kappa_1 - j)}\notag \\
 & = & \Pr [\tilde{i}_1(\xf, \xfs; \yff, \yfs) \le -\kappa_1 - j].
\end{IEEEeqnarray}
Hence, 
\begin{IEEEeqnarray}{rCl}
\lefteqn{\mathbb{E}[|\tilde{i}_1(\xf, \xfs; \yff, \yfs)|^3]} \notag \\
& \le&  \kappa_1 + 2 \sum_{j=0}^{\infty} (\kappa_1+j+1)^3 \notag \\
&& \hspace{0.5cm}\cdot \Pr \left [|\tilde{i}_1(\xf, \xfs; \yff, \yfs)| > \kappa_1 + j \right ].
\end{IEEEeqnarray}
\begin{lemma} \label{lemma2}
The following inequality holds:
\begin{IEEEeqnarray}{rCl}
\lefteqn{\Pr [|\tilde{i}(\xf, \xfs; \yff, \yfs) | > \kappa_1 + j ]} \notag \\
 &\le& \max \left \{ \frac{\zeta_1 \ell_1^{\frac{\nf}{2} -1}e^{-\ell_1} }{\Gamma \left (\frac{\nf}{2} \right)} ,  \frac{\tilde \zeta_1 \tilde \ell_1^{\frac{\ns}{2} -1}e^{-\tilde \ell_1} }{\Gamma \left (\frac{\ns}{2} \right)}\right \},\end{IEEEeqnarray}
 where
 \begin{IEEEeqnarray}{rCl}\label{eq:ltilde}
\ell_1&: =& \frac{1}{2\sigff}  \left (\sqrt{\frac{\tilde  \kappa_1 + \kappa_1 +j }{2k_1}} - \frac{b_1}{2} \right)^2, \\
 \tilde \ell_1 &: =&  \frac{1}{2\sigfs}  \left (\sqrt{\frac{\tilde \kappa_1 + \kappa_1 +j }{2\tilde k_1}} - \frac{\tilde b_1}{2} \right)^2,
\end{IEEEeqnarray}
 for $\zeta_1 > 1$ and $\tilde \zeta_1 >1$ satisfying
\begin{IEEEeqnarray}{rCl}
  \frac{\zeta_1}{(\zeta_1 - 1)(\frac{\nf}{2} - 1)} &<& \ell_1, \\
 \frac{\tilde \zeta_1}{(\tilde \zeta_1 - 1)(\frac{\ns}{2} - 1)} &<&\tilde \ell_1,
\end{IEEEeqnarray}
where $k_1, \tilde k_1, b_1, \tilde b_1$ and $\tilde \kappa_1$ are defined in \eqref{eq:37}. 
\end{lemma}
\begin{IEEEproof}
See Appendix~\ref{app4}. 
\end{IEEEproof}
\begin{lemma}
It holds that 
\begin{IEEEeqnarray}{rCl}
\lefteqn{\sum_{j = 0}^{\infty} (\kappa + j + 1)^3}\notag \\
&& \cdot \Pr \left [|\tilde{i}_1(\xf, \xfs; \yff, \yfs)| > \kappa + j \right ]\\ \notag 
&\le& \max \Bigg\{ \frac{\zeta_1 e^{-c_1}}{\Gamma \left (\frac{\nf}{2} \right)} A(\nf, k_1, b_1, c_1),\notag \\
&&  \hspace{1.5cm}\frac{\tilde \zeta_1 e^{-\tilde c_1}}{\Gamma \left (\frac{\ns}{2} \right)} A(\ns, \tilde k_1, \tilde b_1,\tilde c_1)\Bigg\},
\end{IEEEeqnarray}
where $k_1, \tilde k_1, b_1, \tilde b_1, c_1, \tilde c_1$ and  $A(\cdot, \cdot, \cdot, \cdot)$ are defined in \eqref{eq:37}. 
\end{lemma}
\begin{IEEEproof}
The proof is based on the following equality \cite{Srivastava2011}: 
\begin{IEEEeqnarray}{rCl}
\sum_{j = c}^{\infty} j^n e^{-j} = e^{-c} \Phi (e^{-1}, -n, c),
\end{IEEEeqnarray}
 where $\Phi (\cdot, \cdot, \cdot)$ is the Hurwitz Lerch transcendent. 
 \end{IEEEproof}
 Hence,
\begin{IEEEeqnarray}{rCl}						
\lefteqn{\mathbb{E}[|\tilde{i}_1(\xf, \xfs; \yff, \yfs)|^3] } \\
&\le&  \kappa_1 +  2\max \Bigg\{ \frac{\zeta_1 e^{-c_1}}{\Gamma \left (\frac{\nf}{2} \right)} A(\nf, k_1, b_1, c_1), \notag \\
&& \hspace{2cm} \frac{\tilde \zeta_1 e^{-\tilde c_1}}{\Gamma \left (\frac{\ns}{2} \right)} A(\ns, \tilde k_1, \tilde b_1,\tilde c_1)\Bigg\}. 
\end{IEEEeqnarray}

As a result
\begin{IEEEeqnarray}{rCl}
	\lefteqn{\mathbb{E}[|\tilde{i}_1(\xf, \xfs; \yff, \yfs)} \notag \\
	&& - \mathbb{E}[\tilde{i}_1(\xf, \xfs; \yff, \yfs)]|^3]  \notag \\
&\leq& 2^3 \kappa_1 + 2^4 \max \Bigg \{ \frac{\zeta_1 e^{-c_1}}{\Gamma \left (\frac{\nf}{2} \right)} A(\nf, k_1, b_1, c_1), \notag \\
&&\hspace{2.5cm}  \frac{\tilde \zeta_1 e^{-\tilde c_1}}{\Gamma \left (\frac{\ns}{2} \right)} A(\ns, \tilde k_1, \tilde b_1,\tilde c_1)\Bigg\} 
\end{IEEEeqnarray}
for any $\kappa_1 >1$. This concludes the proof. 



\section{Proof of Equation~\eqref{eq:J1new}}\label{app2}
By \cite[Proposition 2 and Appendix B]{Molavianjazi}, we have the following bounds:
\begin{IEEEeqnarray}{rCl}
\lefteqn{\frac{f_{\yff}(\yff)}{Q_{\yff}(\vect y_{1,1})}} \notag \\
 &\le& \hnr{e^{\frac{1}{6\nf}}\left (\frac{\nf}{\nf-2} \right)^{\frac{\nf+1}{2}} \frac{\pi }{2} \frac{1 + \Omega_{1,1}}{\sqrt{1 + 2\Omega_{1,1}}} } \label{eq:168n}
\end{IEEEeqnarray}
\begin{IEEEeqnarray}{rCl}
\lefteqn{\frac{f_{\yfs}(\vect y_{1,2})}{Q_{\yfs}(\vect y_{1,2})}} \notag \\
 &\le&  \frac{(\ba  + \bb  )}{2\sqrt{\pi}\sqrt{\ba \bb  }}e^{\frac{1}{6\ns} - \frac{1}{2}}  \notag \\
 && \hspace{0.1cm}\cdot \left ( \frac{\ns}{\ns-2}\right)^{\frac{\ns+1}{2}} \left ( \frac{\ns}{\ns-1}\right)^{\frac{\ns-2}{2}}. \label{eq:169n}
\end{IEEEeqnarray}
\hnr{Note that the bounds in \eqref{eq:168n} and \eqref{eq:169n} are different from the bounds in \cite[Proposition 2]{Molavianjazi} as we have removed the assumption that the blocklength is sufficiently large. 
The techniques are the same as in \cite[Appendix B]{Molavianjazi}. Combining \eqref{eq:168n} and \eqref{eq:169n} proves $J_1$ that is defined in \eqref{eq:Ji}.}

\section{Proof of Lemma~\ref{lemma2}} \label{app4}
Notice that 
\begin{IEEEeqnarray}{rCl}
\lefteqn{\tilde{i}(\xf, \xfs; \yff, \yfs)} \notag \\
 &=& \frac{\nf}{2} \log \frac{\syf}{\sigff} + \frac{\ns}{2} \log \frac{\sys}{\syuf} \notag \\
 && \hspace{0.5cm}- \frac{||\zff||^2 }{2 \sigff}+ \frac{||\xf + \zff||^2 }{2\syf} \notag \\
& & \hspace{1cm} - \frac{||\alpha(\xfs+\xsf) + \zfs -\vect \mu_{1,2} ||^2}{2\syuf} \notag \\
&& \hspace{1.5cm}+ \frac{||\alpha(\xfs  + \xsf)+ \zfs||^2 }{2\sys},
\end{IEEEeqnarray}
with $\alpha$ being defined in \eqref{eq:alphan}. By \eqref{eq:sigmas},
\begin{IEEEeqnarray}{rCl}
\lefteqn{|\tilde{i}(\xf, \xfs; \yff, \yfs)| } \notag \\&\le& \frac{\nf}{2} \log \frac{\syf}{\sigff} + \frac{\ns}{2} \log \frac{\sys}{\syuf} \notag \\
&& \hspace{0.5cm} + \frac{||\zff||^2 }{2 \sigff}+ \frac{||\xf + \zff||^2 }{2\syf} \notag \\
& & \hspace{1cm} + \frac{||\alpha \xsf + \zfs +\left( \alpha -h  \right)\xfs ||^2}{2\syuf}  \notag \\
&& \hspace{1.5cm}+ \frac{|| \alpha \xfs  +  \alpha \xsf+ \zfs||^2 }{2\sys} \\
& = & \frac{\nf}{2} \log \frac{\syf}{\sigff} + \frac{\ns}{2} \log \frac{\sys}{\syuf} \notag \\
&& \hspace{0cm}  + \frac{||\zff||^2 }{2 \sigff}+ \frac{||\xf ||^2+ ||\zff||^2 + 2 \langle \xf, \zff \rangle }{2\syf} \notag \\
& & \hspace{0.5cm} + \frac{||\alpha \xsf -\alpha  \rho \sqrt{\frac{\bb}{\ba}} \xfs ||^2 + ||\zfs||^2 }{2\syuf} \notag \\
&&\hspace{1cm}+ \frac{ 2 \langle \alpha \xsf -  \alpha \rho \sqrt{\frac{\bb}{\ba}} \xfs, \zfs \rangle}{2\syuf} \notag \\
&& \hspace{1.5cm}+ \frac{|| \alpha \xfs  +  \alpha \xsf||^2 + ||\zfs||^2  }{2\sys}\notag \\
&& \hspace{2cm}+ \frac{ 2 \alpha \langle  \xfs  +   \xsf, \zfs \rangle  }{2\sys} \\
& = & \frac{\nf}{2} \log \frac{\syf}{\sigff} + \frac{\ns}{2} \log \frac{\sys}{\syuf} + \frac{||\zff||^2 }{2 \sigff}\notag \\
&& + \frac{\nf \bef \P+ ||\zff||^2 + 2 \langle \xf, \zff \rangle }{2\syf} \notag \\
& & \hspace{0cm} + \frac{\alpha^2 \left (\ns \bb \P (1+ \rho^2) -2  \rho \sqrt{\frac{\bb}{\ba}} \langle \xfs,  \xsf \rangle \right)  }{2\syuf} \notag \\
& & \hspace{0cm} + \frac{ ||\zfs||^2 + 2 \alpha \langle \xsf, \zfs \rangle  - 2 \alpha \rho \sqrt{\frac{\bb}{\ba}} \langle \xfs, \zfs \rangle}{2\syuf} \notag \\
&& + \frac{\alpha^2 ( \ns \P ( \ba + \bb)  +  2\langle \xfs, \xsf \rangle )  }{2\sys} \\
&& + \frac{||\zfs||^2 + 2 \alpha \langle  \xfs , \zfs \rangle + 2\alpha \langle \xsf, \zfs \rangle  }{2\sys} \\
&\stackrel{(a)}{ \le} &  \nf C(\Omega_{1,1})+ \ns C(\Omega_{1,2}) + \frac{||\zff||^2 }{2 \sigff}\notag \\
&&  + \frac{\nf \bef \P+ ||\zff||^2 + 2 \sqrt{\nf  \bef \P} || \zff|| }{2\syf} \notag \\
& & \hspace{0.5cm} + \frac{ \ns \bb \P (1- \rho^2)   + ||\zfs||^2  }{2((1-\rho^2)\bb \P + \sigfs)} \notag \\
& & \hspace{1cm} + \frac{  2 \sqrt{ \ns \bb \P} (1+\rho) ||\zfs|| }{2((1-\rho^2)\bb \P + \sigfs)} \notag \\
&& \hspace{1.5cm}+ \frac{ \bes \ns \P  + ||\zfs||^2   }{2(\bes \P + \sigfs)},\notag \\
&& \hspace{2cm}+ \frac{ 2 \sqrt{\ns \P} (\sqrt{\ba} + \sqrt{\bb}) ||\zfs||   }{2(\bes \P + \sigfs)}\\
& = & k_1\left (\|\zff\| + \frac{b_1}{2}\right)^2 + \tilde k_1\left (\|\zfs\| + \frac{\tilde b_1}{2}\right)^2 - \tilde \kappa_1,
\end{IEEEeqnarray}

where $(a)$ is by \eqref{eq:con}, and by the fact that $\alpha < 1$ and $-||x||\cdot ||y|| < \langle x,y\rangle < ||x|| \cdot ||y||$. The parameters  $k_1, \tilde k_1, b_1, \tilde b_1$ and $\tilde \kappa_1$ are defined in \eqref{eq:37}. Thus,
\begin{IEEEeqnarray}{rCl}
\lefteqn{\Pr [	|\tilde{i}(\xf, \xfs; \yff, \yfs)| > \kappa + j]} \\
&\le &\Pr \Bigg[k_1\left (\|\zff\| + \frac{b_1}{2}\right)^2 \notag \\
&&\hspace{0.5cm} + \tilde k_1\left (\|\zfs\| + \frac{\tilde b_1}{2}\right)^2 > \tilde \kappa_1 + \kappa_1 +j \Bigg] \\
&\le &\max \Bigg \{\Pr \left [k_1(\|\zff\| + \frac{b_1}{2})^2  > \frac{\tilde \kappa_1 + \kappa_1 +j }{2}\right],\notag \\
&& \hspace{0.75cm} \Pr \left [ \tilde k_1(\|\zfs\| + \frac{\tilde b_1}{2})^2 >\frac{\tilde \kappa_1 + \kappa_1 +j }{2}\right] \Bigg \}\\
&\le &\max \Bigg \{\Pr \left [k_1(\|\zff\| + \frac{b_1}{2})^2  > \frac{\tilde \kappa_1 + \kappa_1 +j}{2}\right], \notag \\
&& \hspace{0.75cm} \Pr \left [ \tilde k_1(\|\zfs\| + \frac{\tilde b_1}{2})^2 >\frac{\tilde \kappa_1 + \kappa_1 +j }{2}\right] \Bigg \}\\
& = & 1- \min \Bigg \{ \Pr \left [ ||\zff|| \le \sqrt{\frac{\tilde \kappa_1 + \kappa_1 +j }{2k_1}} - \frac{b_1}{2} \right ],  \notag \\
&& \hspace{1cm} \Pr \left [ ||\zfs|| \le \sqrt{\frac{\tilde \kappa_1 + \kappa_1 +j}{2\tilde k_1}} - \frac{\tilde b_1}{2} \right ]\Bigg \} \\
& \stackrel{(i)}{=} & 1- \min \left \{ 1- \frac{\Gamma \left (\frac{\nf}{2}, \ell_1\right )}{\Gamma \left (\frac{\nf}{2} \right)},1- \frac{\Gamma \left (\frac{\ns}{2}, \tilde \ell_1 \right )}{\Gamma \left (\frac{\ns}{2} \right)}\right \} \\
& = & \max \left \{ \frac{\Gamma \left (\frac{\nf}{2}, \ell_1 \right )}{\Gamma \left (\frac{\nf}{2} \right)},\frac{\Gamma \left (\frac{\ns}{2}, \tilde \ell_1 \right )}{\Gamma \left (\frac{\ns}{2} \right)}\right \}\\
&\stackrel{(ii)}{\le}&\max \left \{ \frac{\zeta_1 \ell_1^{\frac{\nf}{2} -1}e^{-\ell_1} }{\Gamma \left (\frac{\nf}{2} \right)} ,  \frac{\tilde \zeta_1 \tilde \ell_1^{\frac{\ns}{2} -1}e^{-\tilde \ell_1} }{\Gamma \left (\frac{\ns}{2} \right)}\right \}
\end{IEEEeqnarray}
where in $(i)$ we use the fact that $||\zff||/\sigma_{1,1}$ and $||\zfs||/\sigma_{1,2}$ follow a central chi-distribution of degree $\nf$ and $\ns$, respectively, and 
\begin{IEEEeqnarray}{rCl}
\ell_1&: =& \frac{1}{2\sigff}  \left (\sqrt{\frac{\tilde  \kappa_1 + \kappa_1 +j }{2k_1}} - \frac{b_1}{2} \right)^2, \\
 \tilde \ell_1 &: =&  \frac{1}{2\sigfs}  \left (\sqrt{\frac{\tilde \kappa_1 + \kappa_1 +j }{2\tilde k_1}} - \frac{\tilde b_1}{2} \right)^2.
\end{IEEEeqnarray}
In $(ii)$, we use the following bound \cite{Pinelis2020}:
\begin{align}
	\Gamma(a,x) < \zeta x^{a-1}e^{-x},
\end{align}
for $a > 1, \zeta > 1, x > \zeta/(\zeta - 1)(a - 1)$. 
This concludes the proof.

     \begin{IEEEbiographynophoto}{Homa Nikbakht} (Member, IEEE) received the M.Sc. degree in Electrical Engineering from the University of Tehran in 2015 and the Ph.D. degree in Electrical Engineering from Télécom Paris in 2020. From January 2021 to June 2022, she was a Postdoctoral Fellow at the CITI Laboratory, INRIA, France. Since October 2022, she has been
  a Postdoctoral Research Associate at Princeton University. Her research interests are in emerging technologies in 5G and 6G wireless cellular networks, multi-terminal information theory and machine learning applications in wireless communications. 
 \end{IEEEbiographynophoto}
 
 \begin{IEEEbiographynophoto}
 {Malcolm Egan} received the Ph.D. in Electrical Engineering in 2014 from the University of Sydney, Australia. He is currently a Chargé de Recherche (Tenured Research Scientist) in Inria and a member of CITI, a joint laboratory between Inria, INSA Lyon and Université de Lyon, France. Previously he was an assistant professor in INSA Lyon, and a postdoctoral researcher with the Laboratoire de Mathématiques, Université Blaise Pascal, France and the Department of Computer Science, Czech Technical University in Prague, Czech Republic. He has also held visiting positions at the University of California Santa Barbara, Princeton University and the University of Bristol. His research interests are in the areas of information theory, statistical signal processing and machine learning with applications in communications.
  \end{IEEEbiographynophoto}
 
 \begin{IEEEbiographynophoto}{Jean-Marie Gorce} (Senior Member, IEEE) received the M.Sc. and Ph.D. degrees in electrical engineering from the Institut National des Sciences Appliquées (INSA), Lyon, France, in 1993 and 1998, respectively. He was a Co-Founder of the Centre for Innovation, Telecommunications and Integration of Services (CITI Lab), in 2001. He was a Visiting Scholar with Princeton University, Princeton, NJ, USA, from 2013 to 2014. He has been the Principal Investigator of several French and European sponsored projects related to wireless communications and networking. He is a Professor with INSA de Lyon. He is currently the Scientific Coordinator for the Experimental Facility CorteXlab affiliated to the SLICES European infarstructure. He has co-published more than 200 conference and journal articles. His research focuses on multi-user communicating systems, with approaches combining information theory, coding, distributed algorithms, signal processing and machine learning.
\end{IEEEbiographynophoto}

\begin{IEEEbiographynophoto}{H. Vincent Poor} (S’72, M’77, SM’82, F’87) received the Ph.D. degree in EECS from Princeton University in 1977.  From 1977 until 1990, he was on the faculty of the University of Illinois at Urbana-Champaign. Since 1990 he has been on the faculty at Princeton, where he is currently the Michael Henry Strater University Professor. During 2006 to 2016, he served as the dean of Princeton’s School of Engineering and Applied Science, and he has also held visiting appointments at several other universities, including most recently at Berkeley and Caltech. His research interests are in the areas of information theory, machine learning and network science, and their applications in wireless networks, energy systems and related fields. Among his publications in these areas is the book Machine Learning and Wireless Communications. (Cambridge University Press, 2022). Dr. Poor is a member of the National Academy of Engineering and the National Academy of Sciences and is a foreign member of the Royal Society and other national and international academies. He received the IEEE Alexander Graham Bell Medal in 2017.
\end{IEEEbiographynophoto}
 

\begin{thebibliography}{20}

%
%


\bibitem{Bennis2018}
M. Bennis, M. Debbah, and H. V. Poor, ``Ultrareliable and low-latency wireless communication: Tail, risk, and scale,'' \emph{Proceedings of the IEEE}, vol.~106, no. 10, pp. 1834--1853, Oct. 2018.



\bibitem{Wang2022}
Y.~Wang, W.~Chen, and H.~V.~Poor, ``Ultra-reliable and low-latency wireless communications in the high SNR regime: A cross-layer tradeoff,'' \emph{IEEE Transactions on Communications}, vol.~70, no.~1, pp. 149--162, Jan.~2022.

\bibitem{HomaITW2020}
H.~Nikbakht, M.~Wigger, and S.~Shamai (Shitz), ``Random user activity with mixed delay traffic,'' in  \emph{Proceedings of the IEEE Information Theory Workshop}, April, 11--14, 2021.

\bibitem{Yao2022}
J. Yao, Q. Zhang, and J. Qin, ``Joint decoding in downlink NOMA systems with finite blocklength transmissions for ultrareliable low-latency tasks,'' \emph{IEEE Internet of Things Journal}, vol.~9, no.~18, pp. 17705--17713, Sept.~ 2022.

\bibitem{HomaEntropy2022}
H.~Nikbakht, M.~Wigger, M.~Egan, S.~Shamai (Shitz), J-M.~Gorce, and  H.~V.~Poor, ``An information-theoretic view of mixed-delay traffic in 5G and 6G,'' \emph{ Entropy}, vol.~24, no.~5, article 637, 2022. 

\bibitem{Khan2022}
N. Khan and S. Coleri, ``Resource allocation for ultra-reliable low-latency vehicular networks in finite blocklength regime,'' in \emph{Proceedings of the  IEEE International Mediterranean Conference on Communications and Networking}, pp. 322-327, Athens, Greece, 2022.  




\bibitem{Liu2018}
Y. Liu, P. M. Olmos, and D. G. M. Mitchell, ``Generalized LDPC codes for ultra reliable low latency communication in 5G and beyond,'' \emph{ IEEE Access}, vol. 6, pp. 72002--72014, 2018.

\bibitem{Sharma2019}
A. Sharma and M. Salim, ``Polar code appropriateness for ultra-reliable and low-latency use cases of 5G systems,’’ \emph{International Journal of Networked and Distributed Computing}, vol.~7, no. 3, pp. 93--99, 2019.
 \bibitem{Yuri2012}
  Y. Polyanskiy, H. V. Poor and S. Verdu, ``Channel coding rate in the finite blocklength regime,'' \emph{IEEE Transactions on Information Theory}, vol. 56, no. 5, pp. 2307--2359, May, 2010.

\bibitem{Hayashi2009}
M. Hayashi, ``Information spectrum approach to second-order coding rate in channel coding,'' \emph{IEEE Transactions on Information Theory}, vol. 55, no. 11, pp. 4947-4966, Nov. 2009.
 
\bibitem{Strassen}
V. Strassen, ``Asymptotic estimates in Shannon’s information theory [Asymptotische Absch\"atzungen in Shannon’s Informationstheorie],'' in \emph{Transactions of the Third Prague Conference on Information Theory, Statistical Decision Functions, Random Processes},  pp. 689–723, Prague, Czechoslovak Academy of Sciences, June 1962, translated from German by Peter Luthy. [Online]. Available: http://pi.math.cornell.edu/pmlut/strassen.pdf.

\bibitem{Mahmood2023}
N. H.~Mahmood, I.~Atzeni, E.~A.~Jorswieck, and O.~L.~A.~López, ``Ultra-reliable low-latency communications: Foundations, enablers, system design, and evolution towards 6G,'' \emph{Foundations and Trends® in Communications and Information Theory}, vol. 20, no. 5-6, pp. 512--747, 2023.

\bibitem{Zhang2023}
H. Zhang and W. Tong, ``Channel coding for 6G extreme connectivity - requirements, capabilities and fundamental tradeoffs,'' \emph{IEEE BITS the Information Theory Magazine}, pp. 1--12, Oct. 2023.

\bibitem{Fujiwara2024}
S. Fujiwara and H. Ochiai, ``A flexible polar decoding architecture with adjustable latency and reliability,'' \emph{IEEE Open Journal of the Communications Society}, 2024. 


\bibitem{Lucas2022}
M. C. Lucas-Estañ and J. Gozalvez, ``Sensing-based grant-free scheduling for ultra reliable low latency and deterministic beyond 5G networks,'' \emph{IEEE Transactions on Vehicular Technology}, vol.~71, no. 4, pp. 4171--4183, April 2022.

\bibitem{Shannon1967}
C. E. Shannon, R. G. Gallager, and E. R. Berlekamp, ``Lower bounds to error probability for coding on discrete memoryless channels. I,'' \emph{Information Control}, vol.~10, no. 1, pp. 65--103, Feb.~1967.

  

\bibitem{Tan2015}
V. Y. F. Tan and M. Tomamichel, ``The third-order term in the normal
approximation for the AWGN channel,'' \emph{IEEE Transactions on Information Theory}, vol. 61, no. 5, pp. 2430–2438, May 2015.

\bibitem{Erseghe2016}
  T. Erseghe, ``Coding in the finite-blocklength regime: Bounds based on Laplace integrals and their asymptotic approximations,'' \emph{IEEE Transactions on Information Theory}, vol. 62, no. 12, pp. 6854--6883, Dec. 2016.

  \bibitem{Shulman2000}
  N.~Shulman and M. Feder. ``Static broadcasting,''  in \emph{Proceedings of the IEEE International Symposium on Information Theory (ISIT)},  Sorrento, Italy, 2000.
  
  
\bibitem{Langberg2021}
M.~Langberg and M.~Effros,``Beyond capacity: The joint time-rate region,'' {arXiv:2101.12236v1}, Jan.~2021. 

\bibitem{Lin2021}	
	P. -H. Lin, S. -C. Lin, P. -W. Chen, M. A. Mross, and E. A. Jorswieck, ``Second order rate regions of Gaussian broadcast channels under heterogeneous blocklength constraints,'' \emph{IEEE Transactions on Communications}, 2023.  

\bibitem{Mross2022}
M. A. Mross, P. -H. Lin,  and E. A. Jorswieck, ``Gaussian broadcast channels with heterogeneous finite blocklength constraints: Inner and outer bounds,'' \emph{IEEE Transactions on Communications}, 2024. 


\bibitem{HomaWCNC2022}
H. Nikbakht, M. Egan, and J. -M. Gorce, ``Joint channel coding of consecutive messages with heterogeneous decoding deadlines in the finite blocklength regime,'' in \emph{Proceedings of the IEEE Wireless Communications and Networking Conference}, pp. 2423--2428, Austin, TX, USA, 2022.

\bibitem{HomaISIT2022}
H. Nikbakht, M. Egan, and J. -M. Gorce, ``Dirty paper coding for consecutive messages with heterogeneous decoding deadlines in the finite blocklength regime,'' in \emph{Proceedings of the IEEE International Symposium on Information Theory}, pp. 2100--2105,  Espoo, Finland, 2022.

\bibitem{Costa1983}
M. H. M. Costa, ``Writing on dirty paper (Corresp.),'' \emph{IEEE Transactions on Information Theory}, vol. 29, no. 3, pp. 439–441, May 1983.
  \bibitem{Scarlett2015}
  J. Scarlett, ``On the dispersions of the Gel'fand - Pinsker channel and dirty paper coding,''  \emph{IEEE Transactions on Information Theory}, vol. 61, no. 9, pp. 4569--4586, Sept. 2015.

\bibitem{Caire2003}
G. Caire and S. Shamai, ``On the achievable throughput of a multiantenna Gaussian broadcast channel,''  \emph{IEEE Transactions on Information Theory}, vol. 49, no. 7, pp. 1691-1706, July 2003.

\bibitem{Nikbakht2023Globecom}
H.~Nikbakht, E.~Ruzomberka, M.~Wigger, S.~Shamai (Shitz), and H.~V.~Poor, ``Joint coding of eMBB and URLLC in vehicle- to-everything (V2X) communications,'' in \emph{Proceedings of the IEEE Global Communications Conference}, pp.~1--6, Kuala Lumpur, Malaysia,  2023.
\bibitem{Tajer2021}
A.~Tajer, A.~Steiner, and S.~Shamai (Shitz), ``The broadcast approach in communication networks,'' \emph{Entropy}, vol.~23, no.~1, article~120, 2021. 

\bibitem{Sheldon2021}
P. Sheldon, D. Tuninetti, and B. Smida, ``The Gaussian broadcast channels with a hard deadline and a global reliability constraint,'' in \emph{Proceedings of the IEEE International Conference on Communications}, pp. 1–6,  Montreal, QC, Canada, 2021.

\bibitem{Sheldon2022}
P. Sheldon, B. Smida, N. Devroye, and D. Tuninetti, ``Achievable rate regions for the Gaussian broadcast channel with fixed blocklength and per user reliability,'' in \emph{Proceedings of the 2022 58th Annual Allerton Conference on Communication, Control, and Computing (Allerton)}, pp. 1-6, Monticello, IL, USA, 2022.   

\bibitem{Tuninetti2022}
D Tuninetti, P Sheldon, B Smida, and N Devroye, ``On second order rate regions for the static scalar Gaussian broadcast channel,'' \emph{IEEE Journal on Selected Areas in Communications}, vol. 41, no. 7, pp. 1982-1999, July 2023. 

\bibitem{Qiu2023}
M. Qiu, Y. -C. Huang, and J. Yuan, ``Downlink transmission with heterogeneous URLLC services: Discrete signaling with single-user decoding,'' \emph{IEEE Journal on Selected Areas in Communications}, vol.~41, no.~7, pp. 2261--2277, July 2023.

\bibitem{Tuninetti2018}
D. Tuninetti, B. Smida, N. Devroye, and H. Seferoglu, ``Scheduling on the Gaussian broadcast channel with hard deadlines,'' in \emph{Proceedings of the IEEE International Conference on Communications (ICC)}, pp. 1--7, Kansas City, MO, USA, 2018.

\bibitem{Marton1979} 
K. Marton, ``A coding theorem for the discrete memoryless broadcast channel,'' \emph{IEEE Transactions on Information Theory}, vol. 25, no. 3, pp. 306--311, May 1979.
  
\bibitem{Jindal2006}  
  N. Jindal, ``MIMO broadcast channels with finite-rate feedback,'' \emph{IEEE Transactions on Information Theory,} vol. 52, no. 11, pp. 5045--5060, Nov. 2006.


\bibitem{Shannon1959}
C.~E.~Shannon, ``Probability of error for optimal codes in a Gaussian channel,'' \emph{The Bell System Technical Journal}, vol. 38, no. 3, pp. 611--656, 1959.





\bibitem{Molavianjazi}
E. MolavianJazi and J. N. Laneman, ``A second-order achievable rate region for Gaussian multi-access channels via a central limit theorem for functions,''  \emph{IEEE Transactions on Information Theory}, vol. 61, no. 12, pp. 6719--6733, Dec. 2015. 

\bibitem{MolavianJaziArXiv}
E.~MolavianJazi and J. N.~Laneman, ``A finite-blocklength perspective on Gaussian multi-access channels,'' arXiv:1309.2343, sep.~2013.  

\bibitem{Nikbakht2022}
H.~Nikbakht, M.~Wigger, S.~Shamai, J.~M.~Gorce, and H.~V.~Poor, ``Joint coding of URLLC and eMBB in Wyner's soft-handoff network in the finite blocklength regime,'' in \emph{Proceedings of the IEEE Global Communications Conference}, Rio de Janeiro, Brazil, pp.~1--6, 2022.


\bibitem{Mulholland1949}
 H.~P.~Mulholland,  ``On generalizations of Minkowski's inequality in the form of a triangle inequality,'' in \emph{Proceedings of the London Mathematical Society}, s2-51 (1): 294–307, 1949.

\bibitem{Srivastava2011}
H.~M.~Srivastava, R.~K.~Saxena, T.~K.~Pogány, and R.~Saxena,  ``Integral and computational representations of the extended Hurwitz-Lerch Zeta function,'' \emph{Integral Transforms and Special Functions}, vol. 22, no. 7,  pp.~487--506, July 2011.

\bibitem{Pinelis2020}
I.~Pinelis, ``Exact lower and upper bounds on the incomplete Gamma function,'' \emph{Journal of Mathematical Inequalities and Applications}, vol.~23, no.~4, pp. 1261--1278, 2020.  

\bibitem{Amos1974}
D. E. Amos, ``Computation of modified Bessel functions and their ratios'', \emph{Mathematics of Computation}, no. 125, pp.~239–251, 1974.


  


  
  
  
  


  



    \end{thebibliography}
\end{document}